\documentclass{new_tlp}

\usepackage{paper}

\title[Fifty Years of Prolog and Beyond]{%
 Fifty Years of Prolog and Beyond
  $^{\thanks{
      Ver\'{o}nica Dahl is thankful for the support provided for this work by
      NSERC's Discovery grant 31611024.
      José Francisco Morales and Manuel Hermenegildo are partially
      funded by MINECO MICINN PID2019-108528RB-C21 \emph{ProCode}
      project, by the Madrid P2018/TCS-4339 \emph{BLOQUES-CM} program,
      and by the Tezos foundation.
      Salvador Abreu acknowledges support from FCT through
      strategic project UIDB/04516/2020 (NOVA LINCS).
      }}$
}%
\author[Philipp Körner, Jo\~{a}o Barbosa et. al]
 {PHILIPP KÖRNER, MICHAEL LEUSCHEL\\
 Institut f\"{u}r Informatik, Universit\"{a}t D\"{u}sseldorf, Universit\"{a}tsstr. 1, D-40225 D\"{u}sseldorf\\
 \email{\{p.koerner,leuschel\}@uni-duesseldorf.de}
 \and
 JO\~{A}O BARBOSA, V\'{I}TOR SANTOS COSTA\\
 Department of Computer Science, Faculty of Science of the University of Porto\\
 \email{\{joao.barbosa,vscosta\}@fc.up.pt}
 \and
 VER\'{O}NICA DAHL\\
 Computing Sciences Department, Simon Fraser University \\
 \email{veronica\_dahl@sfu.ca}
 \and
 MANUEL V. HERMENEGILDO, JOSE F. MORALES\\ %
 IMDEA Software Institute and Universidad Polit\'{e}cnica de Madrid (UPM)\\
 \email{\{manuel.hermenegildo,josef.morales\}@imdea.org} %
 \and
 JAN WIELEMAKER\\
 Centrum voor Wiskunder en Informatica (CWI), Amsterdam\\
  \email{J.Wielemaker@cwi.nl}
 \and
 DANIEL DIAZ\\
 Centre de Recherche en Informatique, University Paris-1 \\
 \email{daniel.diaz@univ-paris1.fr}
 \and
 SALVADOR ABREU \\
 NOVA-LINCS, University of \'Evora \\
 \email{spa@uevora.pt}
 \and
 GIOVANNI CIATTO
 \\
 Dept. of Computer Science and Engineering,
 Alma Mater Studiorum---Univerità di Bologna
 \\
 \email{giovanni.ciatto@unibo.it}
}

\begin{document}
\renewcommand*{\thefootnote}{\arabic{footnote}}

\maketitle

\pagebreak
\vspace*{-6mm}
\begin{abstract}
\vspace*{-1mm}
    Both logic programming in general, and Prolog in particular, have a long and
    fascinating history, intermingled with that of many
    disciplines they inherited from or catalyzed.
    A large body of research has been gathered over the last 50 years,
    supported by many Prolog implementations. Many implementations are still actively
    developed, while new ones keep appearing.
    Often, the features added by different systems were
     motivated by the interdisciplinary needs of programmers and implementors,
    yielding systems that, while sharing the ``classic'' core
    language, in particular, the main aspects of the ISO-Prolog standard,
    also depart from each other in other aspects. This obviously
    poses challenges for code portability.
    The field has also inspired many related, but quite different
    languages that have created their own communities.

    This article aims at integrating and applying the main lessons
    learned in the process of evolution of Prolog.
    It is structured into three major parts. Firstly, we overview the evolution of
    Prolog systems and the community approximately up to the ISO standard,
    considering both the main historic developments and the motivations
    behind several Prolog implementations, as well as
    other logic programming languages influenced by Prolog.
    Then, we discuss the Prolog implementations that are most active after the appearance of the standard:
    their visions, goals, commonalities, and incompatibilities.
    Finally, we perform a SWOT analysis in order to
    better identify the potential of Prolog, and
    propose future directions along which
    Prolog might continue to add
    useful features, interfaces, libraries, and tools, while at the
    same time improving compatibility between implementations.
\end{abstract}

\begin{keywords}
\vspace*{-3mm}
  Prolog, %
  logic programming systems, portability, rationale, evolution, vision.
\vspace*{-2mm}
\end{keywords}

\enlargethispage{3\baselineskip{}}
\renewcommand\contentsname{Contents\\ [-2mm]}
\vspace*{-9mm}
\tableofcontents{}
\section{Introduction}

Logic programming languages in general, and Prolog in particular, have
a long and fascinating history, having catapulted computing sciences
from its old number-crunching, algorithm-focused and mostly imperative
paradigm into the new, unique paradigm of inferential engines.  Rather
than measuring performance through the number of calculations per
second, we can now do so through inferences per second -- a
qualitative leap, with import well beyond the natural language
processing uses for which Prolog had been first conceived.

Logic programming's truly novel
    characteristics distinguish it not only from   traditional
    imperative programming, but also from functional programming, some of whose aims and
    techniques it shares.
    The year TPLP celebrates its 20-year anniversary also marks the
    milestone of 50 years of evolution since the first steps
    towards Prolog, the first version of which was completed in 1972.
    Logic programming and Prolog have progressed over the years deeply
    intermingled with the evolution of the
    different areas they both resulted from, as well as those that
    they enabled.

    The Prolog language in particular has attracted sustained
    academic and practical interest since its origins, yielding a
    large body of research. %
    The language has been supported by numerous Prolog implementations, many of which are still
    in active development, while new ones keep appearing all the time.
    The large number of features added by different systems during this evolution
     were often motivated by the diverging needs of respective implementors.
    As a result, while sharing a core language including the main aspects of the ISO-Prolog
    standard, most Prolog systems also depart from each other in significant ways.
    This fertile evolution has also spawned many other new languages
    and paradigms that have created their own
    communities.

    This article is structured in three major parts.
    In the first part, \cref{sec:history},
    we outline the evolution of
    Prolog systems and the community approximately up to the development of the ISO standard in 1995,
    focusing on historic developments and scientific milestones. %
    This provides a condensed description of the history of Prolog including the steps that got us to the first standard, along with the main motivations behind each step.

    In the second part, \cref{sec:current-state}, we discuss
    the need for the ISO standard and analyze, with this standard in mind, how the Prolog implementations
    and community have evolved since.
    The section aims at
    documenting the vision, research and development focus for each implementation.
    Since most systems have incorporated
    significant functionality beyond the Prolog
    ISO-standard, we survey these non-standard features, with a special emphasis on portability.
    That section gathers very diffused information in one place;
     tables and small paragraphs %
     allow for convenient comparison of implementations.
    We also consider logic programming languages which have considerably departed from Prolog,
    but were obviously strongly influenced by it.

    In the third and last part, \cref{sec:future}, we gaze into the crystal
    ball and answer a few questions such as: How might Prolog and its community evolve in the future?
    Can we better unify the new aspects that are offered by different implementations?
    How should efforts for increased portability be organized?
    Does it make sense to aim for a unified language?
    And, what tools could be provided to ease development?
    Furthermore, we propose a plan for future steps that need to be taken to evolve Prolog as a language.
    The plan is founded on needs expressed by the community in a consultation %
     and on a comparison with successful evolution of other languages. %
    Our goal is to achieve further standardization and easier portability of code among implementations.
    We argue for why this is important and discuss why previous similar attempts have failed.
\section{Part I: The Early Steps of Prolog}%
\label{sec:history}
\begin{figure}[tb]
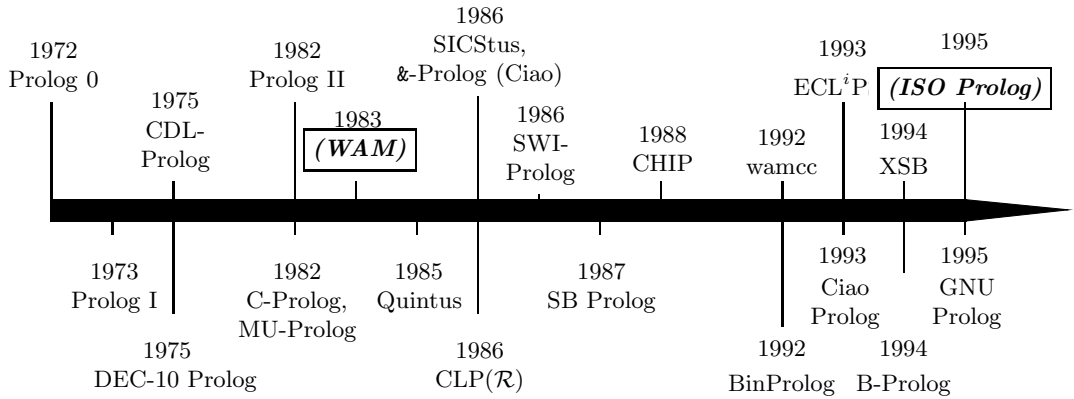

\centering
\catcode`\@=11
\def\chron@selectmonth#1{\ifcase#1\or January\or February\or
March\or April\or May\or June\or
July\or August\or September\or
October\or November\or December\fi}
\startchronology[startyear=1980,startdate=false,stopdate=false,stopyear=1995]
\chronoevent[markdepth=-50pt,year=false,textwidth=5em]{1980}{1972 ~ Prolog 0}
\chronoevent[markdepth=5pt,year=false,textwidth=5em]{1981}{1973 ~ Prolog I}
\chronoevent[markdepth=-20pt,year=false,textwidth=5em]{1982}{1975 ~ CDL-Prolog}
\chronoevent[markdepth=35pt,year=false,textwidth=7em]{1982}{1975 ~ \linebreak DEC-10 Prolog}
\chronoevent[markdepth=-50pt,year=false,textwidth=4.2em]{1984}{1982 ~ Prolog II}
\chronoevent[markdepth=5pt,year=false,textwidth=5em]{1984}{1982 ~ C-Prolog,\linebreak MU-Prolog}
\chronoevent[markdepth=-20pt,year=false,textwidth=4.5em]{1985}{1983 ~ \fbox{\textbf{\emph{(WAM)}}}}
\chronoevent[markdepth=5pt,year=false,textwidth=4em]{1986}{1985 ~ Quintus}
\chronoevent[markdepth=-52pt,year=false,textwidth=7em]{1987}{1986 \linebreak SICStus,\linebreak {\tt \&}-Prolog (Ciao)}
\chronoevent[markdepth=35pt,year=false,textwidth=5em]{1987}{1986 ~ CLP($\cal R$)} %
\chronoevent[markdepth=-15pt,year=false,textwidth=3em]{1988}{1986 ~ SWI-Prolog} %
\chronoevent[markdepth=5pt,year=false,textwidth=5em]{1989}{1987 \linebreak SB Prolog}
\chronoevent[markdepth=-20pt,year=false,textwidth=3em]{1990}{1988 ~ CHIP}
\chronoevent[markdepth=40pt]{1992}{BinProlog}
\chronoevent[markdepth=-20pt]{1992}{wamcc}
\chronoevent[markdepth=5pt,textwidth=4em]{1993}{Ciao Prolog}
\chronoevent[markdepth=-50pt]{1993}{\eclipse{}}
\chronoevent[markdepth=-20pt]{1994}{XSB}
\chronoevent[markdepth=40pt]{1994}{B-Prolog}
\chronoevent[markdepth=5pt,textwidth=5em]{1995}{GNU Prolog}
\chronoevent[markdepth=-50pt]{1995}{\fbox{\textbf{\emph{(ISO Prolog)}}}}
\stopchronology
\caption{Approximate timeline of some early Prolog systems (%
  up to the ISO Standard).}
\label{fig:timeline}
\end{figure}

This first part of the paper provides two major contributions.
First, it provides a general %
definition of what ``Prolog'' is and what a current Prolog
implementation %
generally looks like.
This allows us to focus in the paper on this class of systems,
although we also mention briefly other related ones.
The second contribution of this section is a description of the
evolution of Prolog systems approximately up to and including the
appearance and gradual adoption the ISO standard. %

Given the high variability of features and technologies which characterized pre-standard Prolog systems, some early systems may not perfectly fit a modern definition of Prolog.
Nevertheless, we choose to include these also in our historical discussion --- and consider them in any case Prolog systems --- because of their importance in shaping Prolog as we know it today. More concretely, we consider Prolog systems essentially all those discussed in this paper, except of course the ``systems influenced by Prolog'' of \cref{pl-inspired-otherlangs}.

Note that it is beyond the scope of our article to reconstruct all
the initial steps and theoretical developments that led to Prolog.
Fortunately, good recollections can be found for example by
Robert Kowalski\founder{} (\citeyear{Kowalski1988,10.1007/978-3-642-40564-8_2}), 
Jacques Cohen\founder{} (\citeyear{Cohen1988}),
Peter \cite{vanroy-survey}, 
Alain Colmerauer\founder{} (\citeyear{colmerauer1996birth}) and 
Maarten van Emden\founder{} (\citeyear{vanEmden2006}).\footnote{The asterisk by the name indicates people who are recognized for their impactful research as \emph{Founders of Logic Programming} by the Association of Logic Programming (ALP): \url{http://www.cs.nmsu.edu/ALP/the-association-for-logic-programming/alp-awards/}}
Here, we discuss those first steps that are useful for understanding the origin and evolution of the Prolog systems that have survived to the present day.

We also note that this paper is not aimed at providing a completely
exhaustive list of Prolog systems: the list is very large and
constantly changing, and many other implementations, such as Waterloo
Prolog \citep{roberts_1977}, UNSW Prolog \citep{SammutS83a}, or the
recently discontinued Jekejeke Prolog, to name just a few, have helped
to spread Prolog throughout the world, but we simply cannot cover them
all.  Instead, we have tried to concentrate on implementations that
constitute a milestone in the evolution of the language or offer some
specially interesting characteristics.  We redirect interested readers
to the historical sources archive maintained by \cite{preservation} to
learn about many of the earlier systems.

\cref{fig:timeline} provides a timeline overview of some of the most
impactful of the early Prolog systems treated in this section, i.e.,
approximately up to the development of the ISO standard.

Throughout the paper we attempt to assign meaningful dates to the
different Prolog systems covered. This is not always straightforward,
and the dates should not be given too much significance. The strategy
that we have followed is as follows:
we first looked for some authoritative source explicitly stating the date the Prolog system was developed or made publicly available.
We consider as ``authoritative source'' any paper from any logic programming-related conference or journal, as well as the Prolog system's home page, or any technical report or manual of the system.
In absence of these sources, we looked for resources on the web mentioning the Prolog system along with a date, and we selected the earliest date amongst all resources.

\subsection{Defining Prolog} \label{sec:prologdef}

Prolog is arguably the most prominent language under the logic programming (LP) umbrella. %
However, as we elaborate in the remainder of this paper, the evolution of Prolog did not follow a linear path.
Many contributions have been presented in the history of LP as implementations, extensions, variants, or subsets of Prolog.
Interestingly, while in some other
programming paradigms the custom is to create new language names when
making modifications or extensions to a given language, the Prolog
tradition has been instead to keep the name Prolog across this long
history of very substantial evolution.

In the following, we attempt to draw a line between what can be considered a Prolog implementation and what not.
We do so by defining Prolog from several perspectives.
We firstly provide a conceptual and minimalist definition of the essential features of Prolog (in a post-ISO-standard world).
We then overview a number of important (yet non-essential) features which any full-fledged implementation of Prolog should include.
Finally, we present a technical test users may perform to verify whether a technology can be considered as Prolog or not.

The objective of our definition is in any case inclusive, in the sense
that we aim at encompassing all systems that preserve the essence that
is generally recognized as Prolog, while allowing the many extensions
that have taken place and hopefully those that may be adopted in the
future.

\paragraph{Conceptual Definition}

Any Prolog implementation must \emph{at least} support:
\begin{enumerate}
    \item\label{feature:horn-clauses} Horn clauses with variables in the terms and arbitrarily nested function symbols as the basic knowledge representation means for both programs (a.k.a. knowledge bases) and queries;
    \item\label{feature:terms-syntax}
     the ability to manipulate predicates and clauses as terms,
     so that meta-predicates can be written as ordinary predicates;
    \item\label{feature:sld} SLD-resolution~\citep{DBLP:conf/ifip/Kowalski74} %
        based on Robinson's\founder{} principle (\citeyear{Robinson1965}) and Kowalski's procedural semantics \citep{DBLP:conf/ifip/Kowalski74} as the basic execution mechanism;
    \item\label{feature:unification}
     unification of arbitrary terms which may contain logic variables at any position, %
     both during SLD-resolution steps and as an explicit mechanism (e.g., via the built-in {\tt =/2});
   \item\label{feature:proof-tree} the automatic depth-first exploration of the proof tree for each logic query.
\end{enumerate}

Notably, item \ref{feature:horn-clauses} aims at excluding strict subsets of Prolog which do not support function symbols or require knowledge bases to be ground.
Item \ref{feature:terms-syntax} rules out custom rule engines for Horn clauses which do not support meta-programming, while requiring Prolog implementations to support meta-predicates.
In other words, real Prolog systems must at least support an efficient mechanism such as \texttt{call/1}, enabling programmers to write predicates accepting terms as arguments, to be interpreted as goals.
ISO-compliant implementations, for instance, employ meta-predicates to support negation, disjunction, implication, and other aspects which are not naturally supported by Horn clauses.
Item \ref{feature:unification} requires implementations to expose the unification mechanism to the users, and it cuts off subsets of Prolog employing weaker forms of pattern matching (e.g., where variables can only appear once and only at the top-level).
Items \ref{feature:sld} and \ref{feature:proof-tree} constrain Prolog solvers to a backward (goal-oriented) resolution strategy where a proof tree is explored via some traversal strategy.
ISO-compliant implementations support a sequential, depth-first, deterministic exploration of the proof tree, via backtracking. %
This is commonly achieved by selecting clauses in a top-down and sub-goals in a left-to-right fashion.
Other implementations may support further strategies: for instance tabled Prologs can deviate from pure depth-first traversal for tabled predicates.
Other Prologs may implement alternative search strategies in addition to depth-first, possibly for certain predicates.
The important issue here is to have at least a (default) mode in which the system is a true programming language, predictable in terms of cost in steps and memory.

\paragraph{Common Relevant Features}

Any Prolog implementation may also support:
\begin{enumerate}\setcounter{enumi}{5}
    \item\label{feature:control} some control mechanism aimed at letting programmers manage the aforementioned exploration;
    \item\label{feature:naf} negation as failure \citep{Clark1978}, and other logic aspects such as disjunction or implication;
    \item\label{feature:side-effects} the possibility to alter the execution context during resolution, via ad-hoc primitives;
    \item\label{feature:indexing} an efficient way of indexing clauses in the knowledge base, for both the read-only and read-write use cases;
    \item\label{feature:dcg} the possibility to express definite clause grammars (DCG) and parse strings using them;
    \item\label{feature:clp} constraint logic programming \citep{DBLP:conf/popl/JaffarL87} via ad-hoc predicates or specialized rules \citep{fruhwirth2009constraint};
    \item\label{feature:operators} the possibility to define custom infix, prefix, or postfix operators, with arbitrary priority and associativity.
\end{enumerate}

There, item \ref{feature:control} dictates that users should be provided with some mechanism to control the proof tree exploration.
ISO-compliant implementations provide the cut for this purpose, while other Prologs may expose further mechanisms.
For instance, in tabled Prologs users must explicitly specify which rules are subject to tabling, and in this way they retain some degree of control about the proof tree exploration. Similarly, delay declarations like {\tt when/2} allow one to influence the selection rule employed for SLD-resolution.
Item \ref{feature:naf} provides a practical way to realize negation on top of Horn clauses and SLD, which Keith Clark\founder{} gave a non-procedural semantics to --- namely, completion semantics --- showing that 
negation as failure is theoretically sound \citep{Clark1978}.
Furthermore, negation as well as other logic operators contribute to the perception of Prolog as a \emph{practical} programming language.
Item \ref{feature:side-effects} requires implementations to support, via \emph{side effects}, the dynamic modification of fundamental aspects which affect the resolution process, possibly as the resolution is going on.
These aspects may include the knowledge base (a.k.a. the \emph{dynamic} clause database), the flags, or the pool of currently open files, and their modification should be exposed to the user via ad-hoc meta-predicates.
For instance, ISO-compliant implementations rely upon built-in predicates like \texttt{assert/1}, \texttt{retract/1}, \texttt{set\_prolog\_flag/1}, etc. to serve this purpose.
In particular, to make both the access and modification of clauses efficient, item \ref{feature:indexing} plays a very important role: the satisfaction of this optional requirement is what distinguishes toy implementations from full-fledged Prolog systems.
Finally, while not strictly essential, items \ref{feature:dcg} and \ref{feature:clp} are two very successful features many modern Prolog system support.
Item \ref{feature:operators} is a nice-to-have feature which allows
a more natural notation when extending Prolog systems with custom functionalities,
without requiring a bare new language to be designed from scratch.

In particular, probabilistic extensions of Prolog such as ProbLog \citep{de-raedt-2007}, and \texttt{cplint} \citep{riguzzi-2007} benefit from custom operator definitions.

Of course, many other features may enrich (or be lacking from) a Prolog implementation.
Consider for instance full ISO library support, presence/lack of a module system, and so on.
While these are technical aspects which greatly affect the efficiency, effectiveness, and usability of Prolog implementations, we do not consider them as fundamental.

\paragraph{Technical Test}

As a rule of thumb, one can check if a logic solver can be considered as a Prolog system or not via the following test.
The test requires that the well-known \texttt{append/3} predicate can be written exactly as follows:
\begin{verbatim}
    append([], X, X).
    append([H | X], Y, [H | Z]) :- append(X, Y, Z).
\end{verbatim}
and can be queried in variety of ways, e.g., to append two lists: \texttt{append([1, 2], [c, d], R)}, deconstruct a list as in: \texttt{append(A, B, [1, 2])}, or with any arbitrary instantiation of the arguments, such as \texttt{append([X | T], [c], [Z, Z, Z])}.

Note that the above test excludes some logic programming languages, such as Datalog \citep{inbookdatalog}, as it does not support functors (just constants);
traditional ASP (Answer Set Programming) \citep{brewka2011answer}, as it does not cater for fully recursive first-order terms with no bound; or Mercury~\citep{SOMOGYI199617}, as it is based on pattern-matching and not on unification and only caters for linear terms.
CORAL~\citep{ramakrishnan1994coral}, on the other hand, is an interesting edge case:
while its default proof tree exploration strategy does not meet our definition, it can be instantiated to behave like Prolog.
Thus, we would consider the entire system not to be a Prolog (but it would qualify as an extension).
Other systems we choose not to consider as Prolog systems (but rather as systems derived from Prolog) are Gödel~\citep{hill1994godel}, Curry~\citep{HanusKuchenMoreno-Navarro95ILPS}, and Picat~\citep{zhou2015constraint}. %
Nevertheless, we do discuss these systems in some detail in Section~\ref{pl-inspired-langs}, where we discuss Prolog derivatives.
However, as mentioned before, we consider all other systems that are discussed in the paper to be Prologs.

\subsection{Ancestors of Prolog}

Prolog descends from three main branches of research: AI programming, automatic theorem proving, and language processing.

The field of AI was born around 1956 and quickly gave rise to the functional programming language LISP \citep{McCarthy1960}. %
A host of other AI languages followed, sometimes grouped under the denomination of \emph{Very High Level Languages}.
These languages had features such as symbolic processing and abstraction, that set them apart from more mundane languages.

Automatic theorem proving made a big step forward in a seminal paper by Alan Robinson introducing the resolution inference rule \citep{Robinson1965}.
Resolution extends modus ponens and modus tollens and includes unification.
Resolution can be used to obtain a semi-decision procedure for predicate logic %
and is at the heart of most inference procedures in logic programming.
In the wake of these advances,
an early visionary in the development of the logic programming field
was
Cordell Green, who already in the late 60s envisioned how to extend
resolution to automatically construct problem solutions, and
implemented this vision in particular for automatically answering
questions based on first-order logic, illustrating it as well for plan
formation, program synthesis, and program simulation, thus
presaging the possibility of moving symbolic programming beyond functions and into logic~\citep{DBLP:conf/ijcai/Green69,DBLP:books/garland/Green69}.
This represented perhaps the first zenith of logic in AI~\citep{Kowalski1988}.
Also notable is Ted Elcock, whose 1967 Aberdeen System, Absys, developed with Michael Foster, %
while not having directly influenced the development of Prolog,
was a declarative programming language that 
anticipated some of Prolog's features such as invertability, negation
as failure,
aggregation operators,
and %
the central role of backtracking~\citep{Elcock90}. 

Meanwhile, Alain Colmerauer was seeking to automate human-machine conversation, which led him to develop
 Q-systems \citep{Colmerauer1970LesSQ,colmerauer1996birth,10.1145/321556.321559},
a tree rewriting system that for many years served for English-to-French translation of Canadian meteorological reports. His aim of modifying Q-systems so that a complete question-answering system (rather than just the analyzer part of it) could be written in logic inspired him, amongst others, %
 to create Prolog.

Floyd's work on non-deterministic algorithms \citep{10.1145/321420.321422} (cf. the survey by Jacques Cohen (\citeyear{cohen1979non}))
 was another important influence, as was
Kowalski and Kuehner's SL-resolution (\citeyear{KOWALSKI1971227}).
SL-Resolution is a refinement of resolution which is still both sound and refutation complete for the full clausal form of first order logic, and underlies the procedural interpretation of Horn clauses
\citep{DBLP:conf/ifip/Kowalski74}.

A further simplification for the case of Horn clauses --- SLD-resolution \citep{DBLP:conf/ifip/Kowalski74} --- resulted from  Kowalski's efforts to reconcile the declarative nature of logic based representations of knowledge with PLANNER's procedural approach~\citep{DBLP:conf/ijcai/Hewitt69}.
The semantics of Horn clauses was explored by Kowalski and van Emden (\citeyear{KowVan1970}).
\subsection{The Birth of Prolog}

Colmerauer's aim of creating a human-machine communication system in logic had led him to further research French language analysis with
\cite{Pas1973}, and to numerous experiments
with Philippe Roussel and Jean Trudel on
automated theorem proving methods. Having learned about SL-resolution,
he invited Kowalski 
to visit Marseille in the summer of 1971. The visit led to Roussel's
use of SL-resolution in his thesis on formal equality in automated
theorem-proving (\citeyear{Rous1972}). In addition to its attractions as a
theorem-prover, SL-resolution had the additional attraction that its
stack-type operating mode was similar to the management of procedure
calls in a standard programming language, making it particularly
well-suited for implementation by
backtracking \`{a} la Floyd,
which Colmerauer adopted for efficiency, so as to avoid having to copy and save the resolvents. 
Yet, for language processing, Q-systems still seemed indispensable. During Kowalski's 1971 visit to Marseille, Kowalski and Colmerauer discovered that a certain way of representing formal grammars in clausal logic enables certain general-purpose proof procedures for first-order logic to behave as special-purpose parsing methods: SL-resolution as top-down parsing, hyper-resolution as bottom-up parsing, similar to Q-systems.

Then, Colmerauer defined a way to encode grammar rules in clauses, known today as the difference-list technique, and introduced extra parameters into the non-terminals to propagate and compute information, through which the analyzer could extract, as in Q-systems, a formula representing the information contained in a sentence.

Colmerauer and Kowalski's collaboration led in 1972 to a discovery analogous, for programs, to that made previously for grammars: that a certain style for representing programs in the clausal form of logic 
enables SL resolution to behave as a computational procedure for executing computer programs.

For this to happen, though, a simplification for efficiency of Kowalski's SL-resolution was implemented at the cost of incompleteness: linear resolution was constrained to unify only between the head literals of ordered clauses with ordered literals\footnote{"head literals" are not to be confused with what we now call the "head" of a clause: at the time, conditions and conclusions were distinguished by a "+" or -" sign preceding the literal in question, not by which side of Prolog's "if" symbol they were in- there was no such symbol in their clauses.}.  This made Colmerauer's aim of creating a human-machine communication system possible. The result was not only the first Natural Language (NL) application of what we now know as Prolog, but most importantly, the basis of Prolog itself: a linear resolution system restricted to Horn clauses that could answer questions (i.e., solve problems) non-deterministically in the problem domain described by the clauses input \citep{Col1972}.

However, the Marseille group was unaware of Horn clauses at the time. But Kowalski recognized that Marseille's principal "heresy" (in Colmerauer' words), a strategy of linear demonstration with unifications only at the heads of clauses, was justified for Horn clauses. Kowalski also clarified further simplifications that so far were only implicit: the elimination of "ancestor resolution" (which only works with non-Horn clauses) and the elimination of the "factoring" rule. Together with Maarten van Emden, he went on to define the modern semantics of Horn-clause programming (\cite{KowVan1970}).

\subsection{The Early Prolog Systems}

Prolog implementations evolved in interaction with ad-hoc, initially meta-programmed extensions of the language itself, created for the often interdisciplinary needs of applications.
In time, these extensions became, or evolved into, standard features of the language.
In this section we chronicle such early developments.

\subsubsection{Prolog 0, Prolog I (1972--1973)}

\paragraph{Basic Features:} As reported by \cite{Cohen1988} and later by \cite{colmerauer1996birth}, the first system (``Prolog 0'') was written in Algol-W by Roussel in 1972.
Practical experience with this system lead to a much more refined second implementation (``Prolog I'') at the end of 1973 by Battani, Meloni, and Bazzoli, in Fortran.
This system already had the same operational semantics and most of the built-ins that later became part of the ISO standard, such as the search space pruning operator (the ``cut''), relevant for Prolog to become a practical AI language. %
Efficiency was greatly improved by adopting the structure-sharing technique by \cite{boyer1972sharing} to represent the clauses generated during a deduction.

\paragraph{Higher-order logic extensions:}
Basic facilities for meta-programming higher-order logic extensions were present in Prolog systems from the very beginning, and many later systems include extended higher-order capabilities beyond the basic \predicate{call/1} predicate---e.g., $\lambda$Prolog~\citep{10.5555/868509}, BinProlog~\citep{tarau1992binprolog,tarau2012binprolog}, Hyprolog~\citep{christiansen2005hyprolog}.
Some of the most influential early extensions are discussed into the following sub-paragraphs.

\subparagraph{Constraints:}
Interestingly, Prolog 0 already included the \texttt{dif/2} ($\neq$) predicate, as a result of Roussel's thesis (\citeyear{Rous1972}).
The predicate sets up a constraint that succeeds if both of its arguments are different terms, but \emph{delays the execution} of the goal if they are not sufficiently instantiated.
\subparagraph{Coroutining:}%
\label{subpar:coroutining}
Although \predicate{dif/2} was neither retained in Prolog I nor became part of the ISO standard, it meant a first step towards the extension of unification to handle constraints: while it introduced the negation of unification, it also allowed an early form of coroutining.
Building on this work,  Ver\'{o}nica Dahl\founder{} introduced a \predicate{delay} meta-predicate  serving to dynamically reorder the execution of a query's elements by delaying a predicate's execution until statically defined conditions on it become true, and used it to extend Prolog with full coroutining --- i.e., the ability to execute either a list of goals or a first-order logic formula representing a goal, by proving them in efficient rather than sequential order. With Roland Sambuc, she developed the first Prolog automatic configuration system, which exploited coroutining, for the SOLAR 16 series of computers ~\citep{dahl1976systeme}.

\subparagraph{Safe Negation as Failure:}%
\label{subpar:naf}

Dahl also used \predicate{delay/2} to make negation-as-failure (NaF) (the efficient but generally unsafe built-in predicate of Prolog I which consists of assuming \predicate{not(p)} if every proof of \predicate{p} fails) safe, simply by delaying the execution of a negated goal until all its variables have been grounded.
This approach to safe negation and to coroutining  made its way into many NL consultable systems, the best known being perhaps Chat 80~\citep{warren1982efficient}, and more importantly, into later Prologs, as we discuss later.

\subparagraph{Deductive Databases:}

Dahl then ushered in the deductive database field by developing the first relational database system written in Prolog~\citep{dahl1977systeme, Dahl1982}.
Other higher-order extensions to Prolog included in this system or in the one by \cite{dahl1976systeme}, such as \texttt{list/3}  (now called \texttt{setof/3}), have become standard in Prolog.

\paragraph{Metamorphosis Grammars:} \label{sect-metamorphosis-grammars}

Metamorphosis Grammars (MGs)~\citep{Col75} were Colmerauer's language processing formalism for Prolog.
They constituted at the time a linguist's dream,
since they elegantly circumvented the single-head restriction of Prolog's Horn clauses, thus achieving the expressive power of transformational grammars in linguistics, which, as type-0 formal grammars, allow more than one symbol in their left-hand side.
This allowed for fairly direct, while also executable, renditions of the linguistic constraints then in vogue: a single rule could capture a complete parsing state through unification with its left-hand (multi-head) side, in order to enforce linguistic constraints through specifying, in its right-hand side, how to re-write it.

The first applications of MGs were compilation~\citep{Col75};
 French consultation of automatic configuration systems~\citep{dahl1976systeme}, where a full first-order logic interlingua was evaluated through coroutining;
 and Spanish consultation of database systems~\citep{dahl1977systeme,Dahl1979}, where a set-oriented, three valued logic interlingua~\citep{Col1979,Dahl1979} was evaluated, allowing amongst other things for the detection of failed presuppositions~\citep{dahl1977systeme,Dahl1982}.
Coroutining was used in the system by \cite{dahl1976systeme} not only for feasibility and efficiency, as earlier described, but also  to permit different paraphrases of a same NL request to be reordered into a single, optimal execution sequence.

A simplification of MGs, Definite Clause Grammars (DCGs), was then developed by Fernando Pereira\founder{} and David H.D.\ Warren\founder{},\footnote{Not to be confused with David S.\ Warren, see later.} in which rules must be single-headed like in Prolog, while syntactic movement is achieved through threading syntactic gap arguments explicitly.
DCGs were popularized in 1980~\citep{pereira1980definite} and became a standard feature of Prolog.
It is worth highlighting that the ``DCG'' name does not refer to the fact that they can translate to definite clauses (since all four subsets of MGs can, just as a side effect of being included in MGs), but to their restriction to single heads, which makes them similar in shape to definite clauses.

More specialized Prolog-based grammars started to emerge.
Their uses to accommodate linguistic theories, in particular Chomskyan, were studied as early as 1984~\citep{DahlRL}, leading to the new research area of ``logic grammars'' \citep{Abramson1989}.

\paragraph{Further Theoretical Underpinnings:}

In 1978, Keith Clark published a paper that showed NaF to be correct with respect to the logic program's completion \citep{Clark1978}.
Simultaneously, Ray Reiter\founder{} provided a logical formalization of the related  ``Closed World Assumption'' \citep{Reiter1978}, which underlies NaF's sanctioning as false of anything that cannot be proved to be true: since in a closed world, every statement that is true is also known to be true, it is safe - in such worlds- to assume that what is not known to be true is false.
This then led to substantial research on non-monotonic reasoning in logic programming, and to inspiring foundational work on deductive databases by Reiter himself, as well as Herve Gallaire\founder{}, Jack Minker\founder{} and Jean-Marie Nicolas (\citeyear{Gallaire84logicand}).
The work of \cite{cohen1979non} on non-determinism in programming languages was also influential in these early stages.

\subsubsection{CDL-Prolog (1975)}

As discussed by Peter Szeredi\founder{} (\citeyear{szeredi2004early}), a group at NIM IG\"USZI in Hungary was trying to port the Marseille system to the locally available machine in 1975.
At the same time, Szeredi, who was part of another group at NIM IG\"USZI, completed his first (unnamed) Prolog implementation using the Compiler Definition Language (CDL) developed by Cornelis Koster, one of the authors of the Algol 68 report, marking the beginning of a series of substantial contributions to Prolog.

\subsubsection{DEC-10 Prolog (1975)} \label{sec:DEC10}

In 1974, David H.D.\ Warren visited Marseille and developed a plan generation system in Prolog, called Warplan \citep{Warren1974}.
He then took Prolog with him as a big deck of punched cards and installed it on a DEC-10 in Edinburgh,  %
where he enhanced it with an alternative ``front-end'' (or ``superviser'') written in Prolog,
to better tailor it to the Edinburgh computing environment and the wider character set available
(the Marseille group had been restricted by a primitive, upper-case-only, teletype connection to a mainframe in Grenoble).
He distributed this version to many groups around the world.

He then set out to address what he perceived as a limitation of the
implementations of Prolog up to that point in time: they were
comparatively slower and more memory hungry than other high-level
AI-languages of the time and, in particular, than LISP.  In what would
eventually become his PhD thesis work~\citep{warren-phd}, David H.D.\ Warren pieced away
on one side the elements of Prolog that could be implemented in the
same way as the most efficient symbolic programming languages
(activation records, argument passing, stack- and heap-based memory
management, etc.), and applied to them well-established compilation
techniques.  Then, for those elements of Prolog that were more novel,
such as unification and backtracking, he developed or applied specific
compilation and run-time techniques such as optimization of
unification by clause head pre-compilation, fast recovery of space on
backtracking, trailing, or structure sharing-based term
representation~\citep{boyer1972sharing} (the latter already present in
Marseille Prolog).  He used as target again the DEC-10 with the
TOPS-10 operating system, and exploited architectural features of the
DEC-10 such as arbitrary-depth indirect memory access, particularly
suited for the structure-sharing technique.
The product of this effort was the first compiler from Prolog to
machine code.  This resulted in a large leap in performance for
Prolog, both in terms of speed and memory efficiency, rivaling that of
Lisp systems.  This was documented in~\citep{WaPePe77}, in what was to
be a landmark publication on Prolog in a mainstream Computer
Science venue.
A version of this compiler dated 1975 is part of the archive maintained by \cite{preservation}.

Fernando Pereira and Lu\'\i{}s Moniz Pereira\founder{}, both at LNEC in Lisbon,
also made major contributions to
the %
development of the complete DEC-10 Prolog system, which also included
now ``classic'' built-ins such as \texttt{setof/3} and
\texttt{bagof/3}. %
A significant element in DEC-10 Prolog's popularity was the
availability of an example-rich user
guide~\citep{warrenDEC10,PePeWa78}.  %
All these features, coupled with the improved syntax and performance,
and the fact that the DEC-10 (and later DEC-20) were the machines of
choice at the top AI departments worldwide, made DEC-10 Prolog
available to (and used by) all these departments, and in general by
the AI research community. This led to DEC-10 Prolog becoming very
popular and it spread widely from about 1976 onward.
By 1980, the system also featured a garbage collector and last-call
optimization~\citep{Warren1980AnIP}.  Also in 1980, David H.D.\ Warren
and Fernando Pereira adapted it to TENEX/TOPS-20, which had then
become the operating system(s) and machines most widely used for AI
research.

The contributions made by the authors of DEC-10 Prolog were
fundamental for the coming of age of Prolog: they proved that Prolog
could not only be elegant and powerful, but it could also come with
the usability, speed, and efficiency of a conventional programming
language.
As a result, DEC-10 Prolog had a large influence on most Prologs after
it, and its syntax, now known also as the ``Edinburgh syntax,'' and
many of its features %
constitute a fundamental component of the current Prolog ISO standard.

However, for all its merits, the one drawback of DEC-10 
Prolog %
was that it was deeply tied to its %
computer architecture and thereby inherently not portable to new
machines, in particular to the
then-emerging 32-bit computer architectures with virtual memory.
This prompted the development of other, more portable Prologs, described below, and eventually the Warren Abstract Machine (see \cref{sec:wam}).

\subsubsection{Unix Prolog (1979)}

As discussed by Chris \cite{Mellish1979}, there were a number of Prolog interpreters at the time that used the DEC-10 syntax but were internally quite different.

The objective of these other systems was to develop a portable, yet still reasonably-performing Prolog system, written in a mainstream source language, and that could be compiled on more mainstream, 32-bit machines (including later Unix systems such as, for example, the DEC VAX family, which became ubiquitous).

The first system to achieve portability was Unix Prolog by \cite{Mellish1979}, written for PDP-11 computers running Unix, which was also ported to the RT-11 operating system.
Unlike Marseille Prolog and DEC-10 Prolog, it used structure-copying rather than structure-sharing.
It led to Clocksin and Mellish writing an influential textbook (\citeyear{ClocksinMellish1981}) which describes a standard ``core'' Prolog, compatible with both DEC-10 Prolog and Unix Prolog.

\subsubsection{LPA Prolog (1980)}

Logic Programming Associates (LPA) was founded in 1980 out of the
group of Kowalski at the Department of Computing and Control
at Imperial College London, including, amongst others, Clive Spenser,
Keith Clark, and Frank McCabe~\cite{lpa-homepage}.  LPA distributed micro-PROLOG which ran
on popular 8-bit home computers of the time such as the Sinclair
Spectrum and the Apple II, and evolved to be one of the first Prolog
implementations for MS-DOS. LPA Prolog evolved to support
the Edinburgh syntax around 1991, and is still delivered today as a compiler and
development system for the Microsoft Windows platform.

\subsubsection{MU-Prolog (1982)}

In 1982, another implementation named MU-Prolog~\citep{naish:report:82,naish:thesis:86} was developed by Lee Naish at Melbourne University.
It was initially a simple interpreter to understand the workings of Prolog, as the author could not find a Prolog system for his hardware.

The system offered efficient coroutining facilities and a delay mechanism similar to those discussed in \cref{subpar:coroutining} to automatically delay calls to negation and if-then-else constructs, as well as meta-logical (e.g., \predicate{functor/3}) and arithmetic predicates.
Its rendition of the \predicate{delay} predicate, here called  \predicate{wait}, allows for declarations to be provided manually but also generated automatically.

MU-Prolog was one of the first shipping database connections, module systems, and dynamic loading of shared C libraries, as well as sound negation (through \predicate{delay/2}, as in \cref{subpar:naf}) %
and a logically pure \predicate{findall/3} predicate, a consequence of its variable binding-controlled delayed goal execution.
MU-Prolog was later succeeded by NU-Prolog~\citep{NU-Prolog}, bringing
MU-Prolog's features to the Warren Abstract Machine (see \cref{sec:wam}).

 \subsubsection{C-Prolog (1982)}

As a first foray into getting Edinburgh Prolog on 32-bit address machines, Lu\'{i}s Damas %
created an Edinburgh-syntax Prolog interpreter for an ICL mainframe with Edinburgh-specific time sharing system (EMAS) and systems programming language (IMP).
This interpreter used the structure sharing approach by \cite{boyer1972sharing} and copied as far as possible the built-in predicates of DEC-10 Prolog.
When Fernando Pereira got access to a 32-bit DEC VAX 11/750 at EdCAAD in Edinburgh in 1981, he rewrote EMAS Prolog in C for BSD 4.1 Unix.
This required many adaptations from the untyped IMP into the typed C, and he also made it even closer to DEC-10 Prolog in syntax and built-in predicates.
The whole project became known as C-Prolog later on \citep{pereira1983c}.
The archive maintained by \cite{preservation} contains a readme file from 1982.

Although implemented as an interpreter, C-Prolog was reasonably efficient, portable and overall a very usable system.
Thus it quickly became influential amongst the Edinburgh implementations, helping to establish ``Edinburgh Prolog'' as the standard.
It contributed greatly to creating a wider Prolog community, and remained extensively used for many years.

\subsection{From Prolog Compilation to the WAM}\label{sec:wam}

Following David H.D.\ Warren's first Prolog compiler, described in \ref{sec:DEC10}, there were a number of other compiled systems up until 1983, including Prolog-X~\citep{bowen1983portable} and later NIP, the ``New Implementation of Prolog'' (for details, cf. the survey by \cite{vanroy-survey}).

In 1983, %
 funded by DEC in SRI, who wanted to have the Prolog performance of the DEC-10/20 implementation ported to the VAX line,
David H.D.\ Warren devised an \emph{abstract machine}, i.e., a memory architecture and an instruction set that greatly clarified the process of implementing a high-performance Prolog system~\citep{Warren1983AnAP}.
This machine became widely known as the Warren Abstract Machine, the WAM.
The proposal was basically a reformulation of the ideas of the DEC-10
compiler, which translated Prolog source to a set of abstract
operations which were then expanded to machine code~\citep{warren-phd},
but expressed in a more accessible way.
In particular, it was described in legible pseudo-code, as opposed to DEC-10 machine code.
Warren made some changes with respect to the DEC-10 system, such as passing parameters through registers instead of the stack.
Also, instead of the structure sharing approach used in the DEC-10 work, the WAM used the structure copying approach by Maurice Bruynooghe\founder{} (\citeyear{BruynoogheMaurice1976Aifp}).
The WAM also included the idea of compiling to intermediate code (bytecode), as introduced by the programming language Pascal and its p-code~\citep{nori1974pascal}, which made compiled code very compact and portable, an approach that is still advantageous
today with respect to native code in some contexts.
The first software implementation of the WAM was for the Motorola 68000 implemented for Quintus by David H.D. Warren, which he also adapted to the VAX line.
Evan Tick, later %
designed a pipelined microprocessor organization for Prolog machines based on the WAM \citep{Tick1984}.

Copies of the SRI technical report describing the WAM were passed around widely among those who had an interest in Prolog implementations, 
and the WAM became the standard blueprint for Prolog compilers and continues to be today.
The WAM was made even more widely accessible and easier to understand with the publication of A\"it-Kaci's \emph{Tutorial Reconstruction}~(\citeyear{DBLP:books/mit/AitKaci91}), building on earlier tutorials by~\cite{iclp89-tutorial} and Nasr.

Much work was done after that on further optimization techniques for
WAM-based Prologs, achieving very high levels of sequential
performance. This very interesting topic is outside the scope of
this paper, but is covered in detail in the excellent survey
by~\cite{vanroy-survey}, and much of this work is by Van Roy himself.
Further work, beyond the survey, includes, e.g., 
dynamic compilation~\citep{vitor07:dynamic_compilation},
instruction merging~\citep{insmerging-sicstus} (pioneered by Quintus),
advanced indexing~\citep{vitor07:indexing,DBLP:conf/iclp/VazCF09},
optimized compilation~\citep{van-roy-computer,morales04:p-to-c-padl,carro06:stream_interpreter_cases},
optimized tagging~\citep{tagschemes-ppdp08},
etc.
Also, the compilation of Prolog programs to WAM code was proven mathematically correct by~\cite{Brger1995TheW}, and the proof was machine verified by~\cite{schellhorn1998wam,schellhorn1999verification}.

\subsection{The FGCS Initiative}\label{sec:fgcs}

In 1982 Japan's Ministry of International Trade and Industry (MITI)
started the Fifth Generation Computer Systems (FGCS) initiative in
order to boost Japan's computer industry.  The technical objective was
to build large parallel computers and apply them in artificial
intelligence tasks, with logic programming as the basis, and in
particular Prolog.  The research was conducted across Japanese computer
industries and at a specific research center, ICOT.
Among the first results were hardware sequential Prolog machines
called PSI (for Personal Sequential Inference), similar to those
developed for Lisp at the time by companies such as Lambda Machines,
Thinking Machines, Xerox,
Borroughs, etc.
A series of parallel machines were also developed in the project.

However, at the point of combining parallelism and logic programming,
a language shift occurred.
During a visit to ICOT, Ehud Shapiro developed
what he defined as a \emph{subset} of concurrent
Prolog~\citep{shapiro83,10.5555/535468}.  This referred to the fact
that, in order to reduce the implementation complexity stemming from
the interactions between concurrency and Prolog's backtracking, the
latter was left out in this initial design.  As in other concurrent
logic programming languages %
at the time, such as Parlog~\citep{Clark1986}, \emph{committed choice}
was supported instead, where only the first clause whose \emph{guard}
succeeds is executed.  This guard part consists of a marked set of
literals (normally built-ins) at the beginning of the clause.  This
inspired the Guarded Horn Clauses (GHC) language
of~\cite{DBLP:conf/lp/Ueda85}, as the Kernel Language 1
(KL1)~\citep{ueda1990design}, which became the core language of the
FGCS project.
While the ``kernel'' denomination of KL1 indicated a desire to
eventually recover
the declarative search capabilities of Prolog,
the basic characteristics of KL1
remained throughout the FGCS project.
With the departure from Prolog, an essential part of the language's
elegance and functionality was lost, and this arguably detracted from
the potential impact of the FGCS.

It can be argued that the fifth generation project was successful in a
number of ways.  From the technical point of view, in addition to the
programming language work, it produced many results in parallel
machines, scheduling, parallel databases, parallel automated theorem
proving, or parallel reasoning systems.  Perhaps most importantly, it
accelerated much work elsewhere.  This included a significant line of
research into concurrent (constraint) logic languages
(see~\cref{pl-inspired-lplangs}).  But, more relevant herein, all the
work on parallel implementation of Prolog, which in the end was
done at other centers throughout the world rather than in Japan (we
return to this briefly in \cref{sec:parallelism}).

Beyond the technical part, %
the FGCS project
developed very valuable expertise in computer architecture,
parallelism, languages, software, etc. and nurtured a whole generation
of researchers in areas that were hitherto not so well covered
in Japan.
Furthermore, the FGCS project spurred a number of similar initiatives
around the world that led to important legislative changes and funding
schemes
that last until today.  For example %
laws were developed that allowed companies to collaborate on
``pre-competitive'' research.  This gave rise %
to
the Microelectronics Computer and Technology Corporation (MCC)
in the US and to the European Computer Research Center (ERC) in
Europe, where hardware Prolog machines were also developed, and, most
importantly, the EU ESPRIT program that has continued to the present
day in the form of the current framework programs.  An
account of the outcomes of the FGCS project was presented
by~\cite{DBLP:journals/cacm/ShapiroWFKFUKCT93}.

\subsection{Parallelism}%
\label{sec:parallelism}

In parallel to the FGCS project, logic programming and Prolog were
soon recognized widely as providing good opportunities for parallel
execution, largely because of their %
clean semantics and potentially flexible control.
This spurred a fertile specialized research and development topic,
and several parallel implementations of Prolog or derivatives thereof
were developed, targeting both shared-memory multiprocessors and
distributed systems. As mentioned before, many concurrent Prolog derivatives were also developed.
Going over this very large and fruitful field of research is beyond the scope of this paper; good accounts may be found in the articles by~\cite{partut-toplas}, \cite{DBLP:journals/csur/KergommeauxC94} and \cite{kacsuk1992implementations}.
There is also a survey on this topic in this same special issue of the TPLP
  journal~\citep{DBLP:journals/corr/abs-2111-11218}.
However, it is worth mentioning that two of the current Prolog systems, SICStus and Ciao, have their origins in this body of work on parallelism.

\paragraph{Or-Parallelism: SICStus, Aurora, MUSE (1985)}
Around 1985 the Swedish Institute of Computer Science (SICS) was founded and Mats Carlsson joined SICS to develop a Prolog engine that would be a platform for research in or-parallelization of Prolog, i.e., the parallel exploration of alternative paths in the
execution.
This work was performed in the context of the informal ``Gigalips'' project, involving David H.D.\ Warren at SRI and researchers from Manchester and Argonne National Laboratory, as well as and-parallel efforts (described below).
This resulted in quite mature or-parallel Prologs,
such as Aurora~\citep{lusk1990aurora} and MUSE~\citep{ali1990muse}.
The objective of these Prologs was to achieve effective speedups through or-parallel execution transparently for the programmer and supporting full Prolog.
This led to SICS distributing SICStus Prolog, which
quickly became popular in the academic environment.

\paragraph{And-Parallelism: RAP-WAM and \&-Prolog (1986), a.k.a.\  Ciao Prolog}\label{sec:andpar}
Since 1983, the University of Texas at Austin conducted research on and-parallelization of Prolog,
i.e., executing in parallel steps within an execution path,
complementary to or-parallelism.
The appearance of
the WAM led to \&-Prolog's abstract machine, the
RAP-WAM \citep{Hampaper}, which extended the WAM with parallel
instructions, lightweight workers, multiple stack sets, task stealing,
etc.
Richard Warren, Kalyan Muthukumar, and Roger Nasr joined the project,
which continued
now also at
MCC (also funded by DEC). The RAP-WAM %
was recoded using early versions
of SICStus, also becoming part of the ``Gigalips'' effort.
\&-Prolog
extended Prolog with constructs for
parallelism and concurrency, and incorporated a parallelizing compiler~\citep{iclp90-annotation-short,annotators-jlp-short} %
which performed global
analysis using the \emph{ProLog Abstract Interpreter}, PLAI~\citep{pracabsin-short,abs-int-naclp89}, %
based on abstract interpretation~\citep{Cousot77-short}.
This allowed the exploitation of parallelism transparently to the user, while supporting full Prolog, and, on shared-memory multiprocessors, was the first proposed WAM extension to achieve effective parallel speedups~\citep{effofai-toplas-short}.
This %
infrastructure was later extended to support
constraint logic programs~\citep{clppar-plilp96-short,consind-toplas-short}.
\&-Prolog evolved into Ciao Prolog (cf. Section~\ref{sec:ciao}).

\subsection{Constraints}\label{sec:conspast}

As discussed by \cite{Colmerauer1984EquationsAI}, in 1982, a new version of Prolog, Prolog II~\citep{PrologInfTrees,VANEMDEN1984143}, was developed in Marseille by Alain Colmerauer, Henri Kanoui, and Michel van Caneghem, for which they shared in 1982 the award \emph{Pomme d'Or du Logiciel Fran\c{c}ais}.
This release brought two major contributions to the future paradigm of Constraint Logic Programming (CLP)~\citep{DBLP:conf/popl/JaffarL87,JaM94-clpsurvey,marriot-stuckey-98-wdoi}: moving from unification to equations and inequations over rational trees, %
and innovative extensions to constraint solving and its semantic underpinnings driving into richer domains.%

\subsubsection{The CLP Scheme and its Early Instantiations}
CLP was presented by \cite{DBLP:conf/popl/JaffarL87} in their landmark paper as a language framework, parameterized by the \emph{constraint domain}.
The fundamental insight behind the CLP scheme is that new classes of
languages can be defined by replacing the unification procedure
 in the resolution steps by
 a more general process for solving \emph{constraints}
 over specific \emph{domains}.
Jaffar and Lassez proved that, provided certain conditions are met by
the constraint domain, the fundamental results regarding correctness
and (refutation) completeness of resolution are preserved.
Traditional LP languages and Prolog are particular cases of the scheme
in which the constraints are equalities over the domain of Herbrand
terms, and can be represented as CLP(${\cal H}$).
The CLP framework was first instantiated as the CLP($\cal R$) system~\citep{10.1145/129393.129398}, which implemented linear equations and inequations over real numbers, using incremental versions of Gaussian elimination and the Simplex algorithm.
CLP($\cal R$) was widely distributed, becoming a popular system.
In the meantime, the research group at ECRC (the European Computer Research Centre)\footnote{The European counterpart of MCC, see~\cref{sec:andpar}.} developed CHIP~\citep{chip} (for \emph{Constraint Handling in Prolog}) over the late 1980s, which interfaced Prolog to domain-specific solvers stemming from operations research and successfully introduced constraints over finite domains, CLP($\fd{}$).
CHIP also introduced the concept of \emph{global constraints}~\citep{beldiceanu1994introducing}, which is arguably a defining feature of CLP and Constraint Programming.
Other instances of the CLP scheme supported constraints over
\emph{intervals}, as implemented by
BNR-Prolog~\citep{older-programming},
and constraints over booleans, which are usually implemented as a
specialization of finite domains and are useful to express
\emph{disjunctive constraints}, whereby a set of constraints may be
placed which encode multiple alternatives, without resorting to
Prolog-level backtracking.

\subsubsection{Later Marseille Prologs}

\paragraph{Prolog III (1990)}

\cite{colmerauer1990introduction} focused on improving some limitations of Prolog II.
It now included the operations of addition, multiplication and subtraction as well as the relations $\leq,<,\geq,$ and $>$.
It also improved on the manipulation of trees, together with a specific treatment of lists, a complete treatment of two-valued Boolean algebras, and the general processing of the relation $\neq$.
By doing so, the concept of unification was replaced by the concept of constraint solving in a chosen mathematical structure.
By mathematical structure, we mean here a domain equipped with operations and relations, the operations being not necessarily defined everywhere.

\paragraph{Prolog IV (1996)}

\cite{colmerauer1996bases} generalized to discrete and continuous domains the technique of constraint solving by enclosure methods.
The solving of an elementary constraint, often qualified local,
consists in narrowing %
the domain ranges of its variables, which generally are intervals.
In a system where numerous constraints interact, interval narrowing and propagation is performed iteratively, until a fixed point is reached.
It also moved closer to the ISO standard syntax.

\subsubsection{Opening the Box}\label{openbox}

While the early instantiations on the CLP scheme, such as
CLP($\cal R$), the CLP scheme predecessor Prolog II, %
BNR Prolog, Prolog III and IV, etc.\ were all specialized systems, new technology incorporated into
Prolog engines for supporting extensions to unification, such as
meta-structures~\citep{Neu90} and attributed
variables~\citep{holzbaur-plilp92}, enabled a \emph{library-based
  approach} to supporting embedded constraint satisfaction in standard
Prolog systems. This approach was first materialized in Holzbaur's libraries for supporting CLP
over reals, as in CLP($\cal R$), as well as the rationals,
CLP($\cal Q$)~\citep{holzbaur-clpqr}.
On the CLP($\fd{}$) side,
work progressed to replace the segregated ``black box'' architecture of CHIP by a transparent one~\citep{DBLP:journals/lncs/HentenryckSD94}, in which the underpinnings of the constraint solver are described in user-accessible form (\emph{indexicals}): such is the proposal discussed and implemented
by \cite{DBLP:conf/iclp/DiazC93}, \cite{DBLP:conf/iclp/CarlsonCD94}, and \cite{DBLP:journals/jlp/CodognetD96}.
Having elementary constraints to compile to is an approach which has largely been adopted by the attributed variable-based Prolog implementations of CLP($\fd$), present in most Prolog systems.
SICStus and GNU Prolog incorporate high-performance native implementations, which nevertheless follow this conceptual scheme.

\paragraph{Constraint Handling Rules (1991, cf. \citep[p. xxi]{fruhwirth2009constraint})}\label{chr}
On the trail of providing finer-grained control over the implementation of constraints, \cite{fruhwirth1992constraint,fruhwirth2009constraint} introduced \emph{Constraint Handling Rules} (CHR), in which syntactically enhanced Prolog clauses are used to describe and implement the progress of the constraint satisfaction process.
CHR is both a theoretical formalism related to first-order and linear logic, and a rule-based constraint programming language that can either stand alone or blend with the syntax of a host language.
When the host language is Prolog, CHR extends it with rule-based concurrency and constraint solving capabilities.
Its multi-headed rules allow expressing complex interactions succinctly, through rule applications that transform components of a shared data structure: the ``constraint store''.
A solid body of theoretical results guarantee best known time and space complexity, show that confluence of rule application and operational equivalence of programs are decidable for terminating CHR programs, and show that a terminating and confluent CHR program can be run in parallel without any modification and without harming correctness.
Applications are multiple, since CHR, rather than constituting a single constraint solver for a specific domain, allows programmers to develop constraint solvers in any given domain.

\bigskip
It should be noted that CLP has spurred the emergence of a very active research field and community, focusing on Constraints, with or without the Logic Programming part.

\subsection{Tabling}%
\label{lbl:tabling}

Tabling is a technique first developed for natural language processing, where it was called Earley parsing~\citep{Kay67,Earley70}.
It consists of storing in a table (a.k.a. chart in the context of parsing) partial successful analyses that might come in handy for future reuse

Its adaptation into a logic programming proof procedure, under the name of Earley deduction, dates from an unpublished note from 1975 by David H.D.\ Warren, as documented by \cite{PerShi87}.
An interpretation method based on tabling was later developed by \cite{tamaki1986old}, modeled as a refinement of SLD-resolution.

David %
S. Warren\footnote{Not to be confused with David H.D.\ Warren.} and his students adopted this technique with the motivation of changing Prolog's semantics from the completion semantics to the minimal model semantics.
Indeed, the completion semantics cannot faithfully capture important concepts such as the transitive closure of a graph or relation. %
The minimal model semantics is able to capture such concepts.
Moreover, tabled execution terminates for corresponding programs such as for the transitive closure of a cyclic graph.
This makes Prolog more declarative.

Tabling consists of maintaining a table of goals that are called during execution, along with their answers, and then using the answers directly when the same goal is subsequently called.
Tabling gives a guarantee of total correctness for any (pure) Prolog program without function symbols, which was one of the goals of that work.

\paragraph{XSB Prolog (1994)}\label{xsb}
The concept of tabled Prolog was introduced in XSB Prolog~\citep{sagonas1994xsb}.
This resulted in a complete implementation~\citep{rao1997xsb} of the \textit{well-founded semantics}~\citep{van1991well}, a three-valued semantics that represents values for true, false and unknown.

\subsection{Prolog Implementations after the WAM}

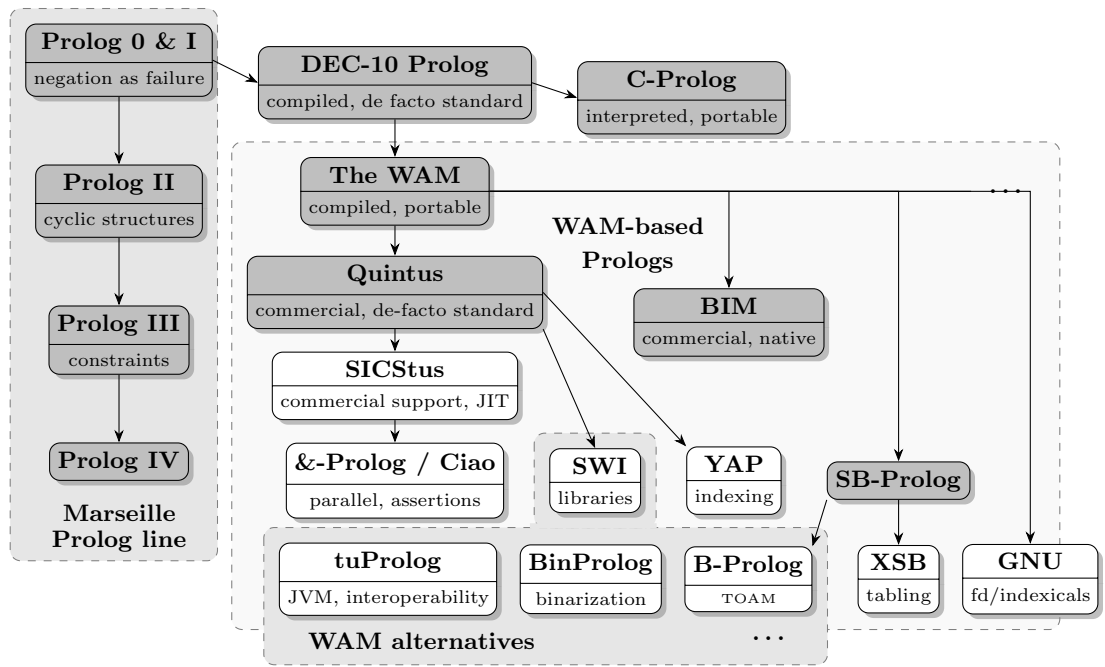
\begin{figure}
\tikzstyle{inactive}=[rectangle, draw=black, rounded corners, fill=gray!50, drop shadow,
        text centered, anchor=north, text=black]
\tikzstyle{active}=[rectangle, draw=black, rounded corners, fill=white, drop shadow,
        text centered, anchor=north, text=black]
\tikzstyle{myarrow}=[->, >=open triangle 90, thick]
\tikzstyle{line}=[-, thick]

\pgfdeclarelayer{background}
\pgfdeclarelayer{foreground}
\pgfsetlayers{background,main,foreground}
\usetikzlibrary{positioning}
\begin{tikzpicture}[node distance=.6cm]
    \node (Prolog1) [inactive, rectangle split, rectangle split parts=2]
        {
            \textbf{Prolog 0 \& I}
            \nodepart{second}{\scriptsize negation as failure}
        };

    \node (Prolog2) [inactive, rectangle split, rectangle split parts=2, below=of Prolog1, yshift=-3mm]
        {
            \textbf{Prolog II}
            \nodepart{second}{\scriptsize cyclic structures}
        };
    \draw[-Stealth] (Prolog1.south) -- (Prolog2.north) ;

    \node (Prolog3) [inactive, rectangle split, rectangle split parts=2, below=of Prolog2, yshift=-3mm]
        {
            \textbf{Prolog III}
            \nodepart{second}{\scriptsize constraints}
        };
    \draw[-Stealth] (Prolog2.south) -- (Prolog3.north) ;

    \node (Prolog4) [inactive,
    below=of Prolog3, yshift=-3mm]
        {
            \textbf{Prolog IV}
        };
    \draw[-Stealth] (Prolog3.south) -- (Prolog4.north) ;

    \node (DEC10) [inactive, rectangle split, rectangle split parts=2, right=of Prolog1, yshift=-3mm]
        {
            \textbf{DEC-10 Prolog}
            \nodepart{second}{\scriptsize compiled, de facto standard}
        };
    \draw[-Stealth] (Prolog1.east) -- (DEC10.west);

    \node (CProlog) [inactive, rectangle split, rectangle split parts=2, right=of DEC10, yshift=-2mm]
        {
            \textbf{C-Prolog}
            \nodepart{second}{\scriptsize interpreted, portable}
        };
    \draw[-Stealth] (DEC10.east) -- (CProlog.west);

    \node (WAM) [inactive, rectangle split, rectangle split parts=2, below=of DEC10, yshift=1mm]
        {
            \textbf{The WAM} %
            \nodepart{second}{\scriptsize compiled, portable}
        };
    \draw[-Stealth] (DEC10.south) -- (WAM.north);
    \node (Quintus) [inactive, rectangle split, rectangle split parts=2, below=of WAM, yshift=2mm]
        {
            \textbf{Quintus}
            \nodepart{second}{\scriptsize commercial, de-facto standard}
        };
    \draw[-Stealth] (WAM.south) -- (Quintus.north);

    \node (SICStus) [active, rectangle split, rectangle split parts=2, below=of Quintus, yshift=3mm]
        {
            \textbf{SICStus}
            \nodepart{second}{\scriptsize commercial support, JIT}
        };
    \draw[-Stealth] (Quintus.south) -- (SICStus.north);

    \node (BIM) [inactive, rectangle split, rectangle split parts=2, right=of Quintus, xshift=6mm, yshift=-4mm]
        {
            \textbf{BIM} %
            \nodepart{second}{\scriptsize commercial, native}
        };
    \draw[-Stealth] (WAM.east) -| (BIM.north);

    \node (Ciao) [active, rectangle split, rectangle split parts=2, below=of SICStus, yshift=3mm]
        {
            \textbf{\&-Prolog / Ciao}
            \nodepart{second}{\scriptsize parallel, assertions}
        };
    \draw[-Stealth] (SICStus.south) -- (Ciao.north);

    \node (SWI) [active, rectangle split, rectangle split parts=2, right=of Ciao]
        {
            \textbf{\ SWI\ } %
            \nodepart{second}{\scriptsize libraries}
        };
    \draw[-Stealth] (Quintus.south east) -- (SWI.north);

    \node (YAP) [active, rectangle split, rectangle split parts=2, right=of SWI]
        {
            \textbf{YAP} %
            \nodepart{second}{\scriptsize indexing}
        };
    \draw[-Stealth] (Quintus.east) -- (YAP.north west);

    \node (SB) [inactive,
    right=of YAP]
        {
            \textbf{SB-Prolog}
            \nodepart{second}
        };
    \draw[-Stealth] (WAM.east) -| (SB.north);

    \node (XSB) [active, rectangle split, rectangle split parts=2, below=of SB]
        {
            \textbf{XSB}
            \nodepart{second}{\scriptsize tabling}
        };
    \draw[-Stealth] (SB.south) -- (XSB.north);

    \node (GNU) [active, rectangle split, rectangle split parts=2, right=of XSB, xshift=-3mm]
        {
            \textbf{GNU} %
            \nodepart{second}{\scriptsize fd/indexicals}
        };
    \draw[-Stealth] (WAM.east) -| (GNU.north);

    \node (OtherW) [right=of WAM, xshift=58mm]
        {
            \textbf{\ \ldots\ }
        };
    \draw[] (WAM.east) -- (OtherW.west);

    \node (BProlog) [active, rectangle split, rectangle split parts=2, left=of XSB]
        {
            \textbf{B-Prolog}
            \nodepart{second}{\tiny TOAM}
        };
    \draw[-Stealth] (SB.south west) -- (BProlog.north east);

    \node (Bin) [active, rectangle split, rectangle split parts=2, left=of BProlog, xshift=3mm]
        {
            \textbf{BinProlog}
            \nodepart{second}{\scriptsize binarization}
        };

    \node (tuProlog) [active, rectangle split, rectangle split parts=2, left=of Bin, xshift=3mm]
        {
            \textbf{tuProlog}
            \nodepart{second}{\scriptsize JVM, interoperability} %
        };

    \node (OtherNW) [below=of BProlog, xshift=3mm, yshift=4mm]
        {
            \textbf{\ \ \ldots\ \ }
        };

    \node (Marseille1) [below= 2mm of Prolog4]
        {
            \textbf{Marseille}
        };
    \node (Marseille2) [below= -1mm of Marseille1]
        {
            \textbf{Prolog line}
        };

    \node (WAM-comment2) [right=of Quintus, xshift=-2mm, yshift=4mm]
        {
            \textbf{Prologs}
        };
    \node (WAM-comment1) [above=0mm of WAM-comment2]
        {
            \textbf{WAM-based}
        };

    \node (NWAM-comment) [below=1mm of tuProlog,align=right,xshift=5mm]
    {
      \textbf{WAM alternatives}
    };

    \begin{pgfonlayer}{background}
      \path (Prolog1.west |- Prolog1.north)+(-0.2,0.2) node (a) {};
      \path (Marseille2.east |- Marseille2.south)+(+0.3,0) node (c) {};
      \path[fill=gray!20,rounded corners, draw=black!50, dashed]
      (a) rectangle (c);

      \path (Quintus.west |- WAM.north)+(-0.2,0.2) node (wam-a) {};
      \path (OtherW.east |- GNU.south)+(+0.2,-0.2) node (wam-c) {};
      \path[fill=gray!5,rounded corners, draw=black!50, dashed]
      (wam-a) rectangle (wam-c);

      \path (tuProlog.west |- tuProlog.north)+(-0.2,0.2) node (nwam-a) {};
      \path (BProlog.east |- NWAM-comment.south)+(+0.3,-0.1) node (nwam-c) {};

      \path (tuProlog.west |- tuProlog.north)+(-0.2,0.2) node (nwam-a) {};
      \path (SWI.west |- tuProlog.north)+(-0.2,0.2) node (nwam-u) {};
      \path (SWI.west |- SWI.north)+(-0.2,0.2) node (nwam-r) {};
      \path (SWI.east |- SWI.north)+(0.2,0.2) node (nwam-d) {};
      \path (SWI.east |- tuProlog.north)+(0.2,0.2) node (nwam-e) {};
      \path (BProlog.east |- tuProlog.north)+(0.2,0.2) node (nwam-f) {};
      \path (BProlog.east |- NWAM-comment.south)+(+0.2,-0.1) node (nwam-c) {};
      \path (tuProlog.west |- NWAM-comment.south)+(-0.2,-0.1) node (nwam-g) {};
      \path[fill=gray!20,rounded corners,draw=black!50,dashed]
      (nwam-a) rectangle (nwam-c);
      \path[fill=gray!20,rounded corners,draw=black!50,dashed]
      (nwam-r) rectangle (nwam-e);
      \path[fill=gray!20,draw=gray!20]
      (nwam-u)+(-0.05,0.05) rectangle (nwam-e)+(0.05,-0.05);

    \end{pgfonlayer}
\end{tikzpicture}
\caption{Prolog Heritage.  Systems with a dark gray background are not
  supported any more.  Arrows denote influences and inspiration of
  systems. The bottom section of each block includes just some highlight(s); see the text
  for more details. Quick legend: JIT = ``Just in Time [Compiler]'', JVM = ``Java Virtual Machine'', TOAM = ``Tree-Oriented Abstract Machine''.
}
\label{fig:heritage}
\end{figure}

As mentioned before, the WAM became the standard for Prolog compilers and continues to be today.
In this section, we review how the main Prolog systems developed more
or less until the appearance of the ISO standard.
An overview of the most influential Prolog systems and their influence is given in~\cref{fig:heritage}.

\subsubsection{Early Proprietary Prologs}
\label{sec-early-commercialisation}

The WAM aroused much interest and many Prolog implementations started out as an exercise to properly understand it while others were aimed directly at commercialization.
Three of the early commercial Prolog systems were Quintus Prolog, BIM-Prolog and VM/Prolog by IBM.

\paragraph{Quintus Prolog (1984)}

Shortly after the WAM was proposed, \emph{Quintus} Computer Systems was founded by David H.D.\ Warren, William Kornfeld, Lawrence Byrd, Fernando Pereira and Cuthbert Hurd, with the goal of selling a high-performance Prolog system for the emerging 32-bit processors.
One of the earliest documents available about Quintus %
is a specifications note \citep{quintus-ps}. %
Quintus used the DEC-10 Prolog syntax and built-ins, and was based on the WAM.  %
Currently, Quintus is distributed by~\cite{quintus-homepage}. %
Quintus quickly became the \emph{de facto} standard at the time, influencing most Prolog systems that were created afterwards.
For many years, it offered the highest-performance implementation and was the standard in terms of syntax, built-ins, libraries, and language extensions.
Its success inspired many more Prolog systems to emerge, including the ones we discuss below.

\paragraph{BIM-Prolog (1984)}\label{sections-bim-prolog}

In 1984, BIM (a Belgian software house) in cooperation with the Katholieke Universiteit Leuven, %
and under the guidance of Maurice Bruynooghe,
started a project aiming at implementing a commercial Prolog system: BIM-Prolog.
A collection of documents is still available on the internet \citep{homepageBimProlog} and notable contributions were made
  amongst others by Bart Demoen, Gerda Janssens, Andr\'{e} Mari\"{e}n, Alain Callebaut and Raf Venken.
BIM Prolog was funded by the Belgian Ministry of Science Policy and was based on the WAM.
One of the earliest resources available is
an internal report~\citep{bim-ir}. %
BIM-Prolog developed into a system with the first WAM-based compiler to native code (as opposed to, e.g., threaded code by Quintus),
with interfaces to several database systems (Ingres, Unify, etc.),
a graphical debugger in the style of dbxtool, a bi-directional interface to C, decompilation even of static code, and multi-argument indexing of clauses---which overcame the common practice of indexing Prolog clauses via their head's first argument alone.
Its first release was on SUN machines, and later it was ported to Intel processors.
BIM was involved in the later ISO standardization effort for Prolog.
BIM went out of business in 1996.

\paragraph{IBM Prolog (1985)}
Several Prolog systems that ran on specific IBM hardware
remained unnamed and were referred to as IBM Prolog.
Here, we focus on Prolog systems distributed by IBM.
In 1985, IBM announced a tool named VM Programming in Logic or VM/Prolog~\citep{5387713},
which was its Prolog implementation for the 370,
focusing on AI research and development.
Its development started in 1983 by Marc Gillet at IBM Paris
according to~\cite{vanroy-survey}.
In 1990, a 16-bit Prolog system for OS/2 was announced,
including a database and dialog manager.
It was able to call programs written in other IBM languages,
such as macro assembler, C/2 and REXX scripts.
While its syntax was based on its predecessor,
it also provided support for the Edinburgh syntax
considering the ongoing ISO standard development.
It was maintained until 1992,
at which time it was succeeded to by the 32-bit implementation IBM SAA AD/Cycle Prolog/2~\citep{benichou1992prolog}.
IBM withdrew from the market in 1994.

\paragraph{SICStus Prolog (1986)}

A preliminary specification of SICStus existed in 1986~\citep{carlsson1986sicstus}, drawing inspiration from DEC-10 Prolog as well as from Quintus.
As already mentioned, SICStus was at first an open-source project aimed at supporting or-parallelism research, and became the basis of much other research, turning into an invaluable tool for other research groups as well as for commercial applications.
In addition to the open-source nature, powerful reasons for this popularity were the compatibility with the DEC-10 and Quintus Prolog de-facto standards, very good performance, and compact generated code.
Execution profiling and native code compilation were also added later.

At the end of the 80s, the Swedish Funding Agency and several companies funded the industrialization of SICStus, which eventually became a commercial product.
In 1998, SICS acquired Quintus Prolog and a number of its features made their way into newer SICStus Prolog versions.
SICStus is ISO-conforming and provides support for web-based applications.
It also supports several constraint domains, including a powerful finite domain solver.
Notably, SICStus is still alive and well as a commercial product, and its codebase is still actively maintained.

\subsubsection{Open-Source and Research-Driven Prolog Systems Based on the WAM}
\label{sssec:wam-based-prologs}

Further, generally open-source Prologs were developed featuring extensions and alternatives arising from the needs of specific application areas or from experimentation with issues such as control, efficiency, portability, global analysis and verification, and, more recently, interoperability and multi-paradigm support and interaction. This section examines some of these.

\paragraph{YAP Prolog (1985)}

As further discussed by \cite{costa2012yap}, the YAP Prolog project started in 1985.
In contrast to other systems discussed here, early versions of it were cast as a proprietary system which was later released as open source software.
Luís Damas, the main developer, wrote a Prolog compiler and parser in C (still used today).
Since the emulator was originally written in m68k assembly, the result was a system that was and felt fast.
As the 68k faded away, Damas developed a macro language that could be translated into VAX-11, MIPS, Sparc and HP-RISC.
Unfortunately, porting the emulator to the x86 was impossible, so a new one was designed in C, making it also easier for some excellent students to contribute.
Rocha implemented the first parallel tabling engine~\citep{DBLP:journals/tplp/RochaSC05} and Lopes the first Extended Andorra Model emulator~\citep{DBLP:journals/tplp/LopesCS12}.
This work was well-received by the community, but proved difficult to use in scaling up real applications.
The problem seemed to be that many YAP applications used Prolog as a declarative database manager.
In order to support them, the team developed JITI~\citep{costa2007demand}, a just-in-time multi-argument indexer that uses \emph{any} instantiated arguments to choose matching clauses, hoping to avoid shallow backtracking through thousands or millions of facts.
JITI's trade-off is extra space---the mega clause idea reduces overhead by compacting clauses of the same type into an array~\citep{DBLP:conf/padl/Costa07}, and the exo-emulation saves space by having a single ``smarter" instruction to represent a column of a table~\citep{DBLP:journals/tplp/CostaV13}.

\paragraph{Ciao Prolog (1993) a.k.a.\  \&-Prolog (1986)}\label{sec:ciao}
As mentioned before, \&-Prolog started in 1986, based initially on early versions of SICStus.
The early 90s brought much evolution, leading to its re-branding as Ciao Prolog~\citep{ciao-prolog-compulog}.
One of the main new aims %
was to point out future directions for Prolog, and to show how
features that previously required a departure from Prolog (such as those in,
e.g., Mercury, G\"{o}del, or AKL, and from other paradigms), could be brought to Prolog without losing Prolog's essence.
A new module system and code transformation facilities were added that allowed defining many language extensions (such as constraints, higher-order, objects, functional notation, other search and computation rules, etc.) as libraries in a \emph{modular} way~\citep{ciao-ppcp-short,CIAO,ciao-modules-cl2000-short}, and also facilitated global analysis.
Also, the progressively richer information inferred by the PLAI analyzers was %
applied to enhancing program development, leading to the Ciao assertion language and pre-processor,
CiaoPP~\citep{prog-glob-an-short,preproc-disciplbook-short,ciaopp-sas03-journal-scp-short}, which allowed \emph{optionally} specifying and checking many properties such as types, modes, determinacy, non-failure, or cost, as well as auto-documentation.
A native, optimizing compiler was also developed, and the abstract machine was rewritten in a restricted dialect of Prolog, ImProlog~\citep{morales05:generic_eff_AM_implem_iclp,morales15:improlog-tplp}.

\paragraph{SB-Prolog (1987)}

SB-Prolog was a Prolog system that, according to the \cite{SB-Prolog} %
became available  in 1987, and had been started as an exercise to understand the WAM.
It was made freely available in the hope that its source code would be of interest to other Prolog researchers for understanding, use, and extension.
Indeed, it became the foundation of two other Prolog systems, XSB (cf. \cref{xsb}) and B-Prolog (cf. \cref{bprolog}).
The goal of XSB Prolog~\citep{sagonas1994xsb} at its release in 1993 was to allow new application areas of Prolog.
As an example, a recent survey of its applications to NLP is given by \cite{christiansen2018natural}.

\paragraph{Andorra I  (1991)} \label{sec-andorra}
Sometimes also referred as \emph{Andorra Prolog}, Andorra I
is a Prolog system developed by \cite{10.1145/109625.109635}.
This system exploited both (deterministic) AND-parallelism and OR-parallelism, while
also providing a form of implicit coroutining, and ran on the
shared-memory multiprocessors of the time, the Sequent
Symmetry.
OR-parallelism was supported by using binding arrays to
access common variables and the Aurora
scheduler~\citep{lusk1990aurora}.
The implementation of AND-parallelism,
that (dynamically) identified which
goals in a clause are determinate and can be evaluated independently in parallel,
came to be known as the \emph{Andorra Principle} and is akin to
the concept of \emph{sidetracking}~\citep{PereiraP1979intelligent},
itself a form of coroutining.
Adherence to Prolog operational
semantics meant that subgoal order sometimes needs to remain fixed
and, also, that a cut may impact parallel execution.
Implementing an efficient Prolog system
which could exploit both forms of parallelism led to difficulties,
for which solutions
would follow in the guise of different computational models, namely
the Extended Andorra Model~\citep{eam-slides} and the Andorra Kernel
Language (AKL)~\citep{akl-paradigms}.

\paragraph{GNU Prolog (1999) a.k.a. Calypso (1996), and wamcc (1992)}

As stated on the GNU Prolog home page \citep{homepageGnuProlog}, the development of GNU Prolog started in January 1996 under the name Calypso.
A few years later, in 1999, the first official release of GNU Prolog saw the light \citep{diaz2012implementation}.

GNU Prolog is derived from wamcc~\citep{DBLP:conf/iclp/CodognetD95}, a system developed in 1992--1993 as a foundational framework for experiments on extensions to Prolog, such as intelligent backtracking techniques, coroutining, concurrency and constraints.
The \texttt{wamcc} Prolog system was designed to be easily maintainable, lightweight, portable and freely available, while still reasonably fast.
Its approach consisted in using the WAM as an intermediate representation in a multi-pass compilation process, producing C code which was subsequently compiled by GCC, to yield native code which was then linked to produce an executable.
\texttt{wamcc} was used as the basis for the development of
\text{CLP($\fd$)}~\citep{DBLP:journals/jlp/CodognetD96}, which
introduced transparent user-defined propagators for finite domain ($\fd$)
constraint solving (CLP).
In a later stage, when the ISO standard was being introduced in 1995,
the \text{CLP($\fd$)} system was redesigned to become more
standards-compliant and to increase its compile-time performance to
compete with that of interpreted Prolog systems.
As C was used as the intermediate language, compiled Prolog programs
would map to considerably larger C programs which were very slow to
compile using GCC, with little benefit as the code had very explicit
and low-level control flow (e.g., labels and gotos.)  This situation
led to the replacement of C by a much simpler specialized
mini-assembly language~\citep{DBLP:conf/padl/DiazC00} as the
intermediate language for compiling Prolog.
After it was approved by the Free Software Foundation,
this system became GNU Prolog.

\paragraph{\eclipse{} (1993)}

The earliest resource about the \emph{\eclipse{}} logic programming system \citep{schimpf_shen_2012} is the technical paper by \cite{WallaceV1993}.
Originally, it was an integration of ECRC's SEPIA, an extensible Prolog system, Mega-Log, an integration of Prolog and a database, and (parts of the) CHIP systems.
It was then further developed into a Constraint Logic Programming system with a focus on hybrid problem solving and solver integration.
It is data driven, allowing for array syntax and structures with field names.
The system also contains a logical iteration construct that eliminates the need for most of the basic recursion patterns.
Since September 2006, \eclipse{} is an open-source project owned by Cisco. %

\subsection{Alternatives to the WAM}

While most Prolog systems are based on the WAM, some alternatives were explored and are also used to date.
In the following, we briefly describe some impactful implementations.

\paragraph{SWI-Prolog (1986)}%

As stated on the project home page~\citep{swi-homepage}, SWI-Prolog \citep{WielemakerHM08} started in 1986 as a recreational project, though the main reason for its development was the lack of recursive calls between Prolog and C in Quintus.
Hence, it soon gained a significant user community (which may be the largest user base for a Prolog system today), partially because it spread over the academic ecosystem as it could be used by universities for teaching.

At its very core, SWI-Prolog is based on an extended version of the ZIP virtual machine by \cite{bowen1983portable}, that is, a minimal virtual machine for Prolog implementing a simple language consisting of only seven instructions.
SWI-Prolog-specific extensions aim at improving performance in several ways: ad-hoc instructions are introduced to support unification, predicate invocation, some frequently used built-in predicates, arithmetic, control flow, and negation-as-failure.
Prolog can easily be compiled into this language, and the abstract machine code is easily decompiled back into Prolog.
This feature is often exploited to interleave compiled and interpreted code execution---which may be needed, e.g., in debug mode.

In the past, SWI-Prolog has incorporated technologies first implemented in other systems, %
e.g., delimited continuations following the design by \cite{SchrijversDDW13},
BinProlog-like continuation passing style as designed by \cite{tarau1992binprolog},
or tabling based on XSB. %

\paragraph{LIFE (1991)}\label{sec:life}
LIFE (Logic, Inheritance, Functions, and
Equations) was an
experimental language developed by Hassan A\"it-Kaci and his group,
first at MCC and later at the DEC Paris Research Lab~\citep{Ait-KaciPodelski91PLILP,HASSAN93}.
It extended
Prolog with type inheritance and functions.  Functions were
implemented using \emph{residuation}, where they are delayed until
their arguments are sufficiently instantiated. Also, extensions to the
WAM were developed for implementing unification of feature terms in
order-sorted theories.

\paragraph{BinProlog (1992)}\label{binprolog}

Paul Tarau started work on BinProlog in 1991, researching alternatives to the then relatively mature WAM.
The first trace of BinProlog in the literature is the paper from \cite{tarau1992binprolog}.
In particular, Tarau was interested in a simpler WAM and in exploring what could be removed without too harsh performance losses.
He was also specializing the WAM for the efficient execution of binary programs by compiling a program's clauses into binary clauses, passing continuations explicitly as extra arguments.
This approach has advantages for implementing concurrency and distributed execution, and related mechanisms such as engines.
However, it also has a cost, since it conflicts with tail- and last-call optimization.
It is thus a veritable WAM alternative trading efficiency for flexibility.

BinProlog %
supports multi-threading and networking.
With human language processing needs in mind, hypothetical reasoning was built-in as well.
This took the form of intuitionistic and linear (affine) implication plus a novel form of implication (timeless assumptions, designed and first meta-programmed by Ver\'{o}nica Dahl for dealing with backward anaphora) \citep{dahl1998assumptions}, later incorporated as well into Constraint Handling Rule Grammars (CHRG) and into Hyprolog (which will be discussed in \cref{pl-inspired-lplangs}).

\paragraph{B-Prolog (1994)}\label{bprolog}

The first version of \emph{B-Prolog} \citep{zhou_2012} was released in 1994, as the reader may verify by scrolling the version history publicly available on the \cite{bprolog-updates}. %
It uses a modified version of the WAM named TOAM (Tree-Oriented Abstract Machine).
The TOAM differs from the WAM in three key aspects: firstly, arguments are passed through the stack rather than registers.
Secondly, it uses a single stack frame for a predicate call rather than two~\citep{zhou1996parameter}.
Lastly, the first-argument indexing of the WAM is improved by using matching trees inspired by Forgy's Rete algorithm~\citep{forgy1989rete}.

\paragraph{tuProlog (2001) and \textsc{2P-Kt} (2021)}

\cite{tuprolog-padl01} proposed tuProlog, another successful attempt at supporting Prolog without compiling to the WAM. %
It consists of a lightweight Prolog implementation targeting the Java Virtual Machine (JVM).
In particular, tuProlog relies on an original state-machine-based implementation of the SLDNF resolution mechanism for Horn Clauses, aimed at interpreting Prolog programs on the fly \citep{tuprolog-sac08}, without requiring any compilation step.

More recently, the whole project has been re-written by \cite{2pkt-swx16} as a Kotlin multi-platform project (codename \textsc{2P-Kt}) targeting both the JVM and JS platforms.
The Prolog state machine has been slightly extended as well to support the lazy enumeration of data streams via backtracking, as discussed by \cite{2pkt-jelia2021}.

\subsection{Early Steps in Building the Community}

After the first successes in Marseille, the availability of Prolog
systems such as DEC-10 Prolog and, especially, the explosion of widely
available Prolog systems that followed the appearance of the WAM, an
international community grew around Prolog and LP.

The First International Logic Programming Conference (ICLP) was held in
Marseille, in September 1982, and the second in Uppsala, Sweden in July 1984. %
There have also
been editions of the Symposium on Logic Programming and the North
American Conference of Logic Programming.

In 1984, the first issue of the Journal of Logic Programming (JLP) with Alan Robinson as founding Editor-in-Chief was
published, marking the solidification of the field of logic
programming within computer
science. %

The Association for Logic Programming (ALP) was founded at the 3rd
ICLP conference in 1986 and JLP became its official journal.
The ALP is the main coordination body within the community, sponsoring conferences
and workshops related to logic programming,
granting prizes and honors, providing support for attendance at
meetings by participants in financial need, etc.
The ALP was followed by other country-specific associations.

In 2001, the first issue of Theory and Practice of Logic Programming (TPLP)
was published. TPLP is a successor to JLP established in response to considerable price increases by Elsevier
 and to provide the logic
programming community with a more open-access journal \citep{DBLP:journals/cacm/Apt01},
and took over from JLP as the
official journal of the ALP.
In 2010 ICLP started publishing its proceedings directly as special
issues of TPLP, pioneering (together with VLDB and SIGGRAPH) a
tendency that is now followed by top conferences in many areas
\citep{dagstuhl-tplpmove}.

\section{Part II:  The Current State of Prolog}%
\label{sec:current-state}

    Currently, there are many Prolog implementations ---
    the \cite{complangprolog-faq} %
    lists 35 different free systems at the time of writing (last modified May 2021).
    While many of those are not maintained anymore or have even become unavailable,
    many others are actively maintained and extended regularly.
    Thanks to the ISO-standardized core of Prolog, significant core
    functionality is shared amongst the different Prolog systems.
    However, implementations also diverge from the standard in some
    details.  For the non-standardized interfaces and the additional
    libraries, differences become more marked, which leads to
    incompatibilities between different Prologs.  In addition, most
    systems have incorporated functionality that goes well
    beyond the Prolog ISO-standard.

    In the following, the background,
    benefits, and shortcomings of the ISO standard,
    as well as contributions based on it are discussed (\cref{sec:iso}).
    \Cref{sec:rationales} then discusses the more active current
    Prolog implementations and what renders them unique, documenting their visions for Prolog
    and their main development or research focus.
    \cref{sec:features} analyzes which non-standard
    features are available throughout the many current Prolog systems,
    and what the state of these features is with a special emphasis on
    the differences.
    \Cref{sec:takeaways-sec3}
    gives preliminary conclusions on the current state of Prolog features,
    whether they are important for portability, and if
    these differences can be easily reconciled.
    Finally, \cref{sec:influence-other-lang}  takes a look over the horizon to discuss
    which non-standard features have inspired
    other successful languages, as well as other interesting concepts
    that may be or become relevant for
    the Prolog community.

\subsection{The ISO Standard and Portability of Prolog Code}%
\label{sec:iso}

    As discussed in \cref{sec:history},
    the success of the WAM gave rise to many Prolog systems in the 80s and early 90s.
    Yet, at this point, the Prolog language was evolving without central stewardship.
    While originally the two traditional camps in Marseille and Edinburgh
    steered their developments at their respective location,
    many Prolog systems around the world aimed for extensions and new uses of the language.
    However, the Edinburgh/DEC-10 syntax and functionality became
    progressively accepted as the de-facto standard,
    helped by the wide dissemination of systems such as C-Prolog and
    later the very influential Quintus system. Many popular systems,
    such as SICStus, YAP, Ciao, SWI, XSB, etc.\ tried to maintain
    compatibility with this standard.

\paragraph{The Core ISO Standard:}%
\label{iso}

    Work on a Prolog standard started in 1984
    and was organized formally in 1987~\citep{neumerkelswi7iso}.
    Its major milestone was the ISO Prolog standard in 1995
    \citep{ISOProlog} that we will refer to as Part 1 or \emph{core}
    standard~\citep{iso-prolog}.
    It solidified the Edinburgh/Quintus
    de-facto standard, and greatly helped establish a common
    \emph{kernel}.
    Furthermore, it greatly increased the confidence of users in the portability of
    code between Prolog systems (especially important amongst industrial
    users), and the hope of having libraries that would be able to
    build on top of rich functionality and be shared as well.
    This was achieved to some extent, and indeed many libraries (for example,
    the excellent code contributed to the community by
    Richard O'Keefe~\citep{edinburghprologtools,paineprologlib}),
    are present today in almost identical form in many systems.

    However, adoption of the ISO standard was not a painless process.
    It was a compromise amongst many parties that did not describe
    any particular single system and, thus, compliance did force the
    different Prolog systems to make changes that, even if often
    minor, were not always backwards compatible.
    This
    made many Prolog vendors face the difficult decision of choosing
    between fully adopting the standard and breaking existing user
    code, or allowing slight deviations that made user code remain
    compatible.  Therefore, it took understandably some time for the
    standard to be adopted. At some point, the community even became
    concerned that the ISO standard might not be taken
    seriously, and some voices asked whether it \emph{should not}
    be taken seriously. In a post in the ALP newsletter,
    \cite{bagnara1999iso} pointed out that many implementations
    differed from the standard in at least some way, and that those
    differences were poorly documented.  Other shortcomings were
    pointed out by several Prolog system main contributors, such as, e.g., by
    \cite{CarlssonM12}, \cite{diaz2012implementation}, and \cite{wielemaker2011portability}.
    However, both aspects, adoption of the core standard and precise
    documentation of where each system departs from the standard, have
    improved progressively. In practice, most systems tend to follow
    the kernel part of the standard, and many systems continue to
    evolve even today to complete aspects in which they are not
    compliant.

    In addition to the battery of tests provided by~\cite{iso-prolog},
    a useful tool to analyze compliance
    with the core standard was developed in by
    \cite{szabo2006improving}, where a test suite of about 1000
    tests was developed and run on several Prolog systems, offering detailed test
    results.
    This suite is
    widely used for detecting areas of non-compliance in
    Prolog systems and was useful to improve compliance.  It has
    also been useful as a means for detecting areas in which the ISO
    standard can be improved, based on failing tests, ambiguities,
    typos, and inconsistencies that have been found in the standard
    document.
    These tests and others were coded as assertions and unit tests in
    the Ciao assertion language by \cite{testchecks-iclp09};
    encoding and testing compliance (both statically and
    dynamically) was one of the design
    objectives of the Ciao assertion language.
    Ulrich
    Neumerkel %
    has greatly contributed by testing and analyzing %
    a comprehensive set of features and programs over many 
    systems~\citep{Neu92,Neu93,HoaMes98,Mesnard2002,Triska2009}.  His
    attention to detail has been essential in bringing ISO standard
    compatibility to %
    many implementations.\footnote{On this
    topic, Neumerkel maintains the site
    \url{https://www.complang.tuwien.ac.at/ulrich/iso-prolog/}.}

    The fact that some aspects of the core standard still require
    additional work has been pointed out also by several authors.  For
    example, Paulo \cite{moura2005prolog}, in another issue of the ALP newsletter,
    points out that revisions of the standard are necessary to close
    gaps
    in the built-in predicates (as well as other issues like exceptions,
    scopes of declared operators, or meta-predicates).
    He demands a strong standard, and also requests mature libraries.
    In later reflections during a special session at ICLP 2008, \cite{moura2009uniting}
    reported on his interesting experiences when implementing
    Logtalk, which, as mentioned before, %
    is
    portable between many Prolog compilers.
    Over the course of a decade, Moura was able to locate hundreds of
    bugs and incompatibilities between all the targeted Prologs.
    However, Moura also asserts that
    Prolog developers\footnote{We will use the term Prolog \emph{developers} to refer to developers \emph{of} Prolog. Developers \emph{in} Prolog will be refered to as Prolog \emph{programmers}.} have generally addressed these issues, and there
    has been considerable improvement in portability.
    Another very valuable result of these efforts by Moura
    is a very comprehensive set of compatibility tests
    and libraries covering the target Prolog systems.
    
\paragraph{Beyond the Core Standard: The Module System}

    Despite leaving room for improvement, the adoption of the core
    standard can be considered a reasonable success. However, an important
    remaining shortcoming is that it does not address many features that modern
    Prolog systems offer, such as modules.
    The Prolog module system was addressed by the ISO
    standardization group in the second part of the standard \citep{ISOPrologPt2}
    which appeared five years later. However, while the core standard
    was wise enough to reflect the de-facto standards of the time,
    this second part proposed a module system that was a radical
    departure from any module system used by any Prolog at the
    time. Instead, by the time this part came out, the Prolog
    developer community had already settled for the de-facto module standard,
    which was the Quintus module system. Thus, this
    second part of the standard was largely ignored.
    Fortunately, the Quintus-like module system is still widely
    supported currently.
    Some systems include extensions to this de-facto standard, that while
    sometimes incompatible, generally preserve
    backwards compatibility. This has allowed the development %
    of some libraries that rely only on the Prolog ISO core, which are present in
    almost identical form in many systems.
    Only a few systems have module systems with radically differing
    semantics or no module system at all.

\paragraph{Beyond the Core Standard: Libraries and Extensions}

    Other aspects that are outside the core standard are libraries and
    many extensions such as coroutining, tabling, parallel
    execution, exceptions, constraints, etc.

    While, as mentioned before, many libraries are present in most systems,
    unfortunately the adoption of the de-facto
    module standard did not result in the establishment of a full
    set of common libraries across systems.
    Often, non-standard Prolog features are required by more involved libraries,
    and in other cases, similar yet independent developments were not reconciled.

    \cite{schrijvers2008uniting} published an article
    aptly named
    \emph{Uniting the Prolog Community}
    discussing issues related to portability
    and incompatibility, specially in the structure and content of
    libraries.
    This prompted two Prolog implementations, SWI-Prolog and YAP, to
    work together more closely on the issue.  A basic compatibility
    was established, that allowed writing Prolog programs in a
    portable manner by abstraction, emulation, and a small amount of
    conditional, dialect-specific
    code~\citep{wielemaker2011portability}.  The overall approach works
    fairly well and, as demonstrated in two case studies (reported in the same paper), large
    libraries can be ported between both Prologs with manageable
    effort.

    Some time later, in 2009, the~\cite{prolog-commons} %
    was established,
    with the objective of developing a common, public-domain set of
    libraries for systems that support the Prolog language.  The group
    met a number of times in Leuven~\citep{swift2009prologcommons}, with attendance from developers
    of most major Prolog systems, and some useful progress was made
    during this period.  Leading Prolog system developers worked
    towards a set of common libraries, a common mechanism for
    conditional code compilation, and closer documentation syntax.
    During early discussions other interesting topics were raised,
    such as the necessity of a portable bidirectional foreign language
    interface for C.  The work of the group also resulted in 17 libraries and 8 more in development.

    While the standard initiatives like the Prolog Commons Working
    Group have been taking major steps forward in the coordination of the Prolog
    developer community, there has been unfortunately less activity
    lately on standardization.
    Clearly, a higher involvement of the Prolog developer community
    in the evolution of the standard and/or
    alternative standardization efforts such as the Prolog Commons
    seems to be a necessity.

\subsection{Rationales and Unique Features of Prolog Implementations}%
\label{sec:rationales}

    Naturally, there are different interests between the Prolog systems
    that are maintained for commercial usage and those
    maintained for research: while the former generally seek maturity
    and stability, the latter generally concentrate on
    advancing the capabilities and uses of the language.
    This divergence of interests raises the question of what features \emph{do} work similarly
    between Prologs, what \emph{can} be adapted, and what can be considered
    as a ``de-facto'' standard that is valid today.
    Along this line, we argue that two interesting questions deserving attention are the following:
    \begin{itemize}
        \item can maintainers of Prolog systems agree on additional features and
        common interfaces?

        \item can the Prolog Commons endeavor or similar efforts be continued?
    \end{itemize}

    Accordingly, this section takes another look at currently active Prolog implementations.
    In contrast to \cref{sec:history}, it ignores their initial motivation
    and historic development, concentrating on their current development foci
    and unique features.
    A brief summary is presented in \cref{tbl:unique} and expanded in the following.

\begin{table}
    \caption{Unique Features and Foci of Prolog Systems.}%
    \label{tbl:unique}
\begin{tabularx}{\textwidth}{lX}
\hline
    System & Uniqueness \\ \hline
    B-Prolog & action rules, efficient CLP supporting many data structures \\
    Ciao & multi-paradigm, %
    module-level feature toggle, extensible language,
    static+dynamic verification of assertions (types, modes), 
    performance/scalabilty, language interfaces, parallelism \\
    \eclipse{} & focus on CLP, integration of MiniZinc and solvers, backward-compatible language evolution of Prolog \\
    GNU Prolog & extensible CLP($\fd{}$) solver, lightweight compiled programs \\
    JIProlog & semantic intelligence / NLP applications \\
    Scryer & new Prolog in development, aims at full ISO conformance \\ %
    SICStus & commercial Prolog, focus on performance and stability,
    sophisticated constraint system, advanced libraries, JIT compilation \\
    SWI-Prolog & general-purpose, focus on multi-threaded programming and support of protocols (e.g., HTTP) and data formats (e.g., RDF, XML, JSON, etc.), slight divergence from ISO, compatibility with YAP, \eclipse{} and XSB \\
    tuProlog & bi-directional multi-platform interoperability (JVM, .NET, Android, iOS), logic programming as a library \\
    XSB & commercial interests, tabled resolution, additional concepts (e.g., SLG resolution, HiLog programming) \\
    YAP & focus on scalability, advanced indexing, language integrations (Python, R), integration of databases \\\hline
\end{tabularx}
\end{table}

\oldtodo{EVERYONE: extend and give current focus of research}
    \paragraph{B-Prolog}
    \citep{zhou_2012,homepageBprolog} is a high-performance implementation of ISO-Prolog.
    It extends the language with several interesting concepts,
    such as action rules~\citep{zhou2006programming},
    which allow delayable sub-goals to be activated later.
    B-Prolog also offers an efficient system for constraint logic programming
    that supports many data structures.

    \paragraph{Ciao Prolog}
    \citep{DBLP:journals/tplp/HermenegildoBCLMMP12,homepageCiao}
    is a general-purpose, open
    source, high-performance Prolog system which supports the ISO-Prolog and other
    de-facto standards, while at the same time including many
    extensions. A characteristic feature is its set of %
    program development tools, which include static and dynamic verification
    of program assertions (see \cref{sec:typing}), testing,
    auto-documentation, source debuggers, execution visualization,
    partial evaluation, or automatic parallelization.  Another
    characteristic feature is extensibility, which has allowed the
    development of many LP and multi-paradigm language extensions which can be
    turned on and off at will for each program module, while
    maintaining full Prolog compatibility.  Other important foci
    are robustness, scalability, performance, and efficiency, with an incremental,
    highly-optimizing compiler, that produces fast and small
    executables.  Ciao also has numerous libraries and interfaces to
    many programming languages and data formats.

    \paragraph{\eclipse{}}
    \citep{DBLP:conf/padl/WallaceS99,Krzysztof2007,homepageEclipse} is a system that aims for backward compatibility with ISO Prolog (and, to some extent, compatibility with the dialects of Quintus, SICStus and SWI-Prolog),
    but also tries to evolve the language.
    Its research focus is constraint logic programming.
    The system integrates the popular MiniZinc constraint modeling language~\citep{nethercote2007minizinc}, by means of a library that allows users to run MiniZinc models,
    as well as third-party solvers.
    While \eclipse{} is open-source software, commercial support is available.
    \paragraph{GNU Prolog}
    \citep{diaz2012implementation,homepageGnuProlog} is an open-source
    Prolog compiler and extensible constraint solver over finite
    domains ($\fd$).  It compiles to native executable code in several
    architectures, by means of an intermediate platform-independent
    language which reduces compilation time.  GNU Prolog strives to be
    ISO-compliant and compiled programs are lightweight and efficient,
    as they do not require a run-time interpreter. The system's design eases the development of
    experimental extensions, and attains good performance,
    despite being built on a simple WAM architecture with few
    optimizations and a straightforward compiler.

    \paragraph{JIProlog}
    \citep{chirico2001jiprolog} was the first Prolog interpreter for a
    Java platform and, some years later, it was the first Prolog interpreter
    for a mobile platform with its implementation for J2ME.
    Its strengths include
    bidirectional Prolog-Java interoperability (meaning that Java programs can call Prolog and vice-versa),
    the possibility to let Prolog programs interoperate with JDBC-compliant data base management systems, and
    the possibility to run the Prolog interpreter on Android.
    Along the years, JIProlog has been exploited in the construction of expert systems, as well as semantic web or data mining applications.

    \paragraph{Scryer Prolog} (\citeyear{scryer-homepage})
    is a quite recent Prolog implementation effort whose
    WAM-based abstract machine is written in the Rust language.
    Scryer is open source and aims for full ISO compliance.
    Since Scryer Prolog is a new, from scratch implementation,
    it has the opportunity to select different trade-offs and
    implementation choices.
    Scryer is still heavily in development at the time of writing, and
    thus its features are in relative flux. We have thus not included
    it in the feature overview table~(\cref{tbl:feature-overview}).
    Nevertheless, it does already have at least preliminary support for
    features such as modules, tabling, constraint domains,
    indexing, attributed variables, and coroutines.
    The reader is directed to the evolving Scryer Prolog documentation to
    follow up on this system.

    \paragraph{SICStus Prolog}
    \citep{carlsson1986sicstus,CarlssonM12,homepageSicstus} is now a
    commercial %
    Prolog system.  It adheres to the ISO standard and has a strong focus on
    performance and stability.  An additional trait of the system is
    its sophisticated constraint system, with advanced libraries and
    many essentials for constraint solvers, such as coroutines,
    attributed variables, and unbounded integers.  The {\tt block}
    coroutining declaration is particularly
    efficient.
    It also incorporates many of the characteristics, features, and
    library modules of Quintus Prolog.
    Since release 4.3, SICStus also contains a JIT (just-in-time)
    compiler to native code, but currently has no multi-threading or
    tabling support.  SICStus is used in many commercial
    applications---cf. \citep[Section 6]{CarlssonM12}.

    \paragraph{SWI-Prolog}
    \citep{wielemaker_schrijvers_triska_lager_2012,homepageSwi} is a general-purpose Prolog system,
    intended for real-world applications.
    For this, it has to be able to interface with other (sub-)systems.
    Thus, the development focus lies on multi-threaded programming,
    implementations of communication network protocols such as HTTP,
    and on libraries that can read and write commonly used data formats,
    such as RDF, HTML, XML, JSON and YAML.
    Notably, for the last two formats, specific data structures need to be supported,
    which has motivated the divergence from the ISO standard in favor of real strings, dictionaries,
    distinguishing the atom \verb$'[]'$ from the empty list \verb$[]$, and non-normal floating-point numbers (Inf, NaN).
    Strings, extended floating-point numbers, and support for rational numbers have been synchronized with \eclipse{}.
    Its top priorities are robustness, scalability and compatibility with both
    older versions of SWI-Prolog and the ISO standard, as well as with YAP, \eclipse{} (data types), and XSB (tabling).

    \paragraph{tuProlog}%
    \citep{tuprolog-padl01,homepageTuProlog} is a relatively recent, research-oriented system which is the technological basis of several impactful works at the edge of the multi-agent systems and logic programming areas, such as \textsf{TuCSoN} \citep{tucson-jir98}, \textsf{ReSpecT} \citep{respect-scp41}, and LPaaS \citep{lpaas-bdcc2}.
    The main purpose of tuProlog is to make Prolog and LP ubiquitous \citep{2p-alpnews2013}.
    It provides basic mechanisms such as knowledge representation, unification, clause storage, and SLDNF resolution \emph{as a library} via multi-platform interoperability (e.g., to JVM, .NET, Android, and iOS) and multi-paradigm integration. %
    Further, tuProlog also supports the direct manipulation of objects from within Prolog.
	Recent research efforts are focused on widening the pool of
	\emph{(i)} logics and inferential procedures such as argumentation support \citep{arg2p-cilc2020}, and probabilistic LP,
	\emph{(ii)} platforms it currently runs upon, e.g., JavaScript \citep{homepageTuProlog}, and
	\emph{(iii)} programming paradigms and languages it is interoperable with, cf. \citep{kotlindsi4prolog-woa2020}.

    \paragraph{$\tau$Prolog}
    (\citeyear{homepageTau}), also referred to as Tau Prolog, is another noteworthy implementation of Prolog focusing on easing the use of logic programming in the Web.
    In particular, it provides a JavaScript-native library
    that facilitates the use of Prolog in web applications, both from the browser- and the server side.
    In other words, $\tau$Prolog pursues a similar intent with respect to tuProlog and JIProlog: bringing Prolog interpreters to high-level platforms and languages, except it focuses on another platform, JavaScript.
    Accordingly, $\tau$Prolog makes it very easy to run a Prolog interpreter in a web page, even without a server behind the scenes.

    \paragraph{XSB}
    \citep{DBLP:conf/deductive/SagonasSW93,sagonas1994xsb,DBLP:conf/procomet/Warren98,homepageXsb}
    is a research-oriented system but its development is also
    influenced by continued use in commercial applications.  Its most
    distinctive research contribution is tabled
    resolution~\citep{swift2012xsb} which has, since, been adopted in
    other systems.  In the XSB manual, the developers explicitly
    refrain from calling it a Prolog system, as it extends the latter
    with concepts such as SLG-resolution and HiLog programming.
    We return to tabling below in Section~\ref{sec:tabling}.
    \paragraph{YAP}
    \citep{costa2012yap} is a general-purpose Prolog system focused on scalability,
    mostly based on advanced clause indexing,
    and on integration with
    other languages, specifically Python and R.  There is a strong
    interest in trying to make Prolog as declarative as possible,
    looking at ways of specifying control (such as types), and in
    program scalability by considering modules.  There is also a
    long term goal of integrating databases into Prolog by having a
    driver that allows YAP to use a database as a Prolog predicate.
In the future, YAP's strength lies in the ability to write and maintain large applications.
Its team has worked on three key points toward this goal:
interfacing with other languages~\citep{DBLP:conf/padl/AngelopoulosCAWCW13},
interfaces to enable collaboration between Prolog dialects, %
and tools for declaring and inferring program properties, such as types~\citep{DBLP:conf/ppdp/Costa99}.

\subsection{Overview of Features}%
\label{sec:features}

In this section,
we survey the availability of features
that are often appreciated by Prolog programmers.
The goal is to find out whether there are commonalities and
even a ``de facto standard'' with respect to
features amongst most Prolog systems.
An overview of which features are available in what Prolog system
is given in~\cref{tbl:feature-overview}.
We give our conclusions regarding portability in \cref{sec:takeaways-sec3}. %
Due to the sheer number of existing Prolog systems,
we consider only the more actively developed and mainstream ones.
In the following, we will briefly discuss each surveyed feature, classifying them in four different groups:
core features that usually cannot be reasonably emulated on top of
simpler features;
extensions to the language semantics and the execution model;
libraries written on top of the core and (optionally, one or more)
extensions;
tools and facilities to debug, test, document, and perform static
analysis.

\paragraph{A Note on Portability:} The above classification
sheds some light onto the challenges of attaining portability of
sophisticated Prolog code. Compatibility at the core features (1) is
relatively easy, and this enables the sharing of a substantial number of
libraries (2). Extensions (3) represent a more complicated evolving
landscape, where some of them require deep changes in the system
architecture.  It stands to reason that the existence of multiple
Prolog implementations (or alternative \emph{core}s) might be a
\emph{necessary} good step, that should be regarded as a healthy sign
rather than an inconvenience. Nevertheless, this requires a periodic
revisit, dropping what did not work and promoting cross-fertilization
of ideas. On the other hand, tools (4), despite being more complex, have
also the advantage of being more flexible: sometimes they can run on
one system while still being usable with others (e.g., IDEs, documentation,
analysis, or refactoring tools).

\begin{table}[t!]
\clearpage%
\caption{Feature Overview of Several Maintained Prolog(-like) Systems.
    Constraint Logic Programming (CLP) abbreviations: $\fd$ = finite-domain, $\cal Q$ = rational numbers, $\cal R$ = real numbers, $\cal B$ = boolean variables.
    Indexing Strategy abbreviations: FA = first argument, N-FA = non-first argument, MA = multiple-argument, JIT = just-in-time, all = all aforementioned strategies.}%
\label{tbl:feature-overview}
\begin{tabular}{lcccc}
\hline
System            & Open Source & Modules  & Non-Standard Data Types  & Foreign Language Interfaces                                       \hfill \\\hline
B-Prolog          & \xmark      & \xmark   & arrays, sets, hashtables & C, Java                                                                  \\
Ciao              & \cmark      & \cmark   & \xmark                   & C, Java, Python, JavaScript \ignore{,Gecode, PPL, GSL, LLVM, SAT}        \\
ECLiPSe           & \cmark      & \cmark   & arrays, strings          & C, Java, Python, PHP                                                     \\
GNU Prolog        & \cmark      & \xmark   & arrays                   & C, Java, PHP                                                             \\
JIProlog          & \cmark      & \cmark   & \xmark                   & Java                                                                     \\
SICStus           & \xmark      & \cmark   & \xmark                   & C, Java, .NET, Tcl/Tk                                                    \\
SWI               & \cmark      & \cmark   & dicts, strings           & C, C++, Java                                                             \\
$\tau$Prolog      & \cmark      & \cmark   & \xmark                   & JavaScript                                                               \\
tuProlog          & \cmark      & \xmark   & arrays                   & Java, .NET, Android, iOS                                                 \\
XSB               & \cmark      & \cmark   & \xmark                   & C, Java, PERL                                                            \\
YAP               & \cmark      & \cmark   & \xmark                   & C, Python, R                                                             \\
\end{tabular}

\vspace{.4cm}

\begin{tabular}{lcccccc}
System            &  CLP                                          & CHR      & Tabling  & Parallelism  & Indexing       & Type / Mode  \hfill\\\hline
B-Prolog          &  $\fd{}$, $\cal B$, $\mathit{Set}$            & \cmark   & \cmark   & \xmark       & N-FA           & \xmark             \\
Ciao              &  $\fd{}$, $\cal Q$, $\cal R$                  & \cmark   & \cmark   & \cmark       & FA, MA         & \cmark             \\
ECLiPSe           &  $\fd{}$, $\cal Q$, $\cal R$, $\mathit{Set}$  & \cmark   & \xmark   & \cmark       & most suitable  & \xmark             \\
GNU Prolog        &  $\fd{}$, $\cal B$                            & \xmark   & \xmark   & \xmark       & FA             & \xmark             \\
JIProlog          &  \xmark                                       & \xmark   & \xmark   & \xmark       & undocumented   & \xmark             \\
SICStus           &  $\fd{}$, $\cal B$, $\cal Q$, $\cal R$        & \cmark   & \xmark   & \xmark       & FA             & \xmark             \\
SWI               &  $\fd{}$, $\cal B$, $\cal Q$, $\cal R$        & \cmark   & \cmark   & \cmark       & MA, deep, JIT  & \xmark             \\
$\tau$Prolog      &  \xmark                                       & \xmark   & \xmark   & \xmark       & undocumented   & \xmark             \\
tuProlog          &  \xmark                                       & \xmark   & \xmark   & \cmark       & FA             & \xmark             \\
XSB               &  $\cal R$                                     & \cmark   & \cmark   & \cmark       & all, trie      & \xmark             \\
YAP               &  $\fd{}$, $\cal Q$, $\cal R$                  & \cmark   & \cmark   & \xmark       & FA, MA, JIT    & \xmark             \\
\end{tabular}

\vspace{.4cm}

\begin{tabular}{lccccc}
System            & Coroutines & Testing      & Debugger          &   Global Variables  & Mutable Terms     \hfill  \\ \hline
B-Prolog          & \cmark     & \xmark       & trace             &   \cmark            & \xmark                     \\
Ciao              & \cmark     & \cmark       & trace / source    &   \cmark            & \cmark                     \\
ECLiPSe           & \cmark     & \cmark       & trace             &   \cmark            & \xmark                     \\
GNU Prolog        & \xmark     & \xmark       & trace             &   \cmark            & \cmark                     \\
JIProlog          & \xmark     & \xmark       & trace             &   \xmark            & \xmark                     \\
SICStus           & \cmark     & \cmark       & trace / source    &   \xmark            & \cmark                     \\
SWI               & \cmark     & \cmark       & trace / graphical &   \cmark            & \cmark                     \\
$\tau$Prolog      & \xmark     & \xmark       & \xmark            &   \xmark            & \xmark                     \\
tuProlog          & \xmark     & \xmark       & spy               &   \xmark            & \xmark                     \\
XSB               & \cmark     & \xmark       & trace             &   \xmark            & \xmark                     \\
YAP               & \xmark     & \xmark       & trace             &   \cmark            & \xmark                     \\\hline
\end{tabular}

 \clearpage
\end{table}

\subsubsection{Core Features}

\paragraph{Module System} \label{sec:modules}

As mentioned before, while
most Prolog systems support structuring the code into
modules, virtually no implementation adheres to the modules part of the ISO
standard.  Instead, most systems have decided to support as
\emph{de-facto} module standard the Quintus/SICStus module system.
However, further convenience predicates concerning modules
are provided by some implementations only,
and often have subtle differences in their semantics.

Interesting cases include GNU Prolog
which %
initially chose not to implement a module system at all, although a
similar functionality was later brought in by means of contexts and
dynamic native unit code loading~\citep{DBLP:conf/inap/AbreuN05};
Logtalk which
demonstrates that code reuse and isolation can be implemented on top
of ISO Prolog using source-to-source transformation \citep{de2003design}; 
Ciao which designed a strict module
system that, while being basically compatible with the \emph{de-facto}
standard used by other Prolog systems, is amenable to
precise static analysis, supports term hiding, and facilitates programming in the
large~\citep{ciao-modules-cl2000-short,termhide-padl2018};
and XSB, which offers an \emph{atom-based} module system~\citep{sagonas1994xsb}.
The latter two systems allow controlling the visibility of terms in
addition to that of predicates.

\oldtodo{JAN \& JOSE: complete merge of data structures and data types section}
\paragraph{Built-in Data Types}
The ISO Prolog standard requires support for atoms, integers,
floating-point numbers and compound terms with only little
specification on the representation limits, usually available
as Prolog flags.

In practice, most limits evolve with the hardware (word length,
floating-point units, available memory, raw performance), open
software libraries (e.g., multiple precision arithmetic), and system
maturity. Since the standard does not specify minimum requirements for
limits, special care must be taken in the following cases:
\begin{itemize}
\item \emph{Integers} may differ between Prolog systems.
  E.g., a given system may not support arbitrary precision arithmetic.
  Furthermore, the minimum and maximum values representable in standard precision
   may be smaller than implied by word length (due to tagging).
\item \emph{Maximum arity} of compound terms may be limited
  (\texttt{max\_arity} Prolog flag). Despite the arity of user terms
  usually falling within the limits, this is an issue with automatic
  program manipulation (e.g., analyzers) or libraries representing
  arrays as terms.
\item \emph{Atoms} have many implementation-defined aspects, such as their maximum length, number of character codes (such as
  ASCII, 8-bit or Unicode), text encoding (UTF-8 or other), whether
  the code-point 0 can be represented, etc.
\item \emph{Strings} are traditionally represented as list of codes or
  characters (depending on the value of \verb|double\_quotes| flag),
  or as dedicated data types. Although this can be regarderd as very
  minor issue, combining different pieces of code expecting different
  encodings is painful and error prone.
\item \emph{Garbage collection} of atoms may not be available. This
  may lead to non-portability due to resource exhaustion in programs
  that create an arbitrary number of atoms.
\item \emph{Floating point numbers} are not specified to be
  represented in a specific way. The IEEE double standard is most
  prevalent across all Prolog systems.  However, support for constants
  such as \verb|NaN|, \verb|-0.0|, \verb|Inf| as well as rounding
  behavior may differ.  \eclipse{}, for example, does interval
  arithmetic on the bounds.
\end{itemize}

For convenience, systems may offer other data types by means of
(non-standard) extension of the built-in data types, for example
\emph{rational numbers} (useful for the implementation of CLP($\cal Q$)),
key-value dictionaries, and compact representations of strings.  There
is no consensus on those extensions or portable implementation
mechanisms, thus more work is needed in this area.

\paragraph{Foreign (Host) Language Interface} \label{sect:fli}
Like any %
programming language, Prolog is more
suited for some problems than for others.
With a foreign language interface, it becomes easier
to embed it into a software system,
where it can be used to solve part of a problem
or access legacy software and libraries written in another language.
Since this is also a non-standard feature,
the interfaces themselves, as well as the targeted languages,
differ quite a lot amongst different systems.
An important case is interfacing Prolog with the \emph{host}
implementation language of the system (e.g., C, Java, JavaScript).
The main issues revolve around the external representation for Prolog
terms (usually in C or Java) and whether non-determinism is visible
to the host language or should be handled at the Prolog level.  The
latter aspect is usually resolved by hiding backtracking from the
foreign language program, except in where there is a natural
counterpart: such is the case when the language has consensual
built-in support for features like iterators.
A more detailed survey describing and comparing different features
is given by Bagnara and Carro~(\citeyear{bagnara2002foreign}).

\subsubsection{Libraries}

\paragraph{Constraint Satisfaction}
\label{section-clp}

As mentioned in Section~\ref{sec:conspast},
advances such as finite domain implementation based on indexicals and,
specially, progress in the underlying technology in Prolog engines for
supporting extensions to unification, such as
meta-structures~\citep{Neu90} and attributed
variables~\citep{holzbaur-plilp92}, enabled the \emph{library-based
  approach} to supporting embedded constraint satisfaction that is now
present in most Prolog systems.
Since, many constraint domains have been implemented as libraries,
such as $\cal R$ and $\cal Q$ (linear equations and
inequations over real or rational numbers), $\fd$ (finite
domains), $\cal B$ (booleans), etc.

Systems vary in how the constraint satisfaction process gets
implemented.
In SICStus and \eclipse{}, the constraint
library is partly implemented in C, in the case of GNU Prolog in a dedicated DSL designed to specify propagators.  Several other
systems, such as Ciao, SWI-Prolog, XSB, and YAP, use Prolog
implementations built on top of attributed variables (as
mentioned above), such as those of \cite{holzbaur-plilp92} %
or \cite{DBLP:conf/flops/Triska12}, or local ones.
While the C-based
implementations provide a performance edge, the Prolog implementations
are small, portable, and may use unbounded integer arithmetic when
provided by the host system.

CHR~\citep{fruhwirth2009constraint}, described also in
section~\ref{chr}, is available in several Prolog systems as a library
which, rather than working on a single, specific domain, enables the
writing of rule-based constraint solvers in arbitrary domains.
CHR provides a higher-level way of specifying propagation and
simplification rules for a constraint solver, although possibly at some
performance cost.

\paragraph{Data Structures} Different Prolog systems also ship
varying numbers of included libraries such as code for AVL trees, ordered sets, etc.
Because they are usually written purely in standard Prolog,
those implementations can usually be dropped in place without larger
modifications.

\subsubsection{Extensions}\label{sec:extension}

\paragraph{Tabling} \label{sec:tabling}
As discussed in \cref{lbl:tabling}, tabling can be used to improve the
efficiency of Prolog programs by reusing results of predicate calls
that have already been made, at the cost of additional memory.
It improves the termination properties of Prolog programs by
delaying \textit{self-recursive} calls.
Tabling was first implemented in XSB and currently a good number of other Prolog
implementations support it (e.g., B-Prolog, Ciao, SWI, YAP).
XSB and recent SWI-Prolog versions improve support for negation using
stratified negation and well-founded semantics.  Both systems also provide
\textit{incremental} tabling which automatically updates tables that depend on
the dynamic database, when the latter is modified.  Some systems (YAP, SWI-Prolog)
support \textit{shared} tabling which allows a thread to reuse answers that
are already computed by another thread.  Ciao supports a related
concept of concurrent facts for communication between
threads~\citep{shared-database}, combines tabling and
constraints~\citep{arias-ppdp2016}, and supports negation based on
stable model semantics~\citep{scasp-iclp2018}.

\paragraph{Parallelism} \label{subsec:parallelism}
Today, new hardware generations do not generally bring large
improvements in sequential performance, but they do often
bring increases in the number of CPU cores.
However,
ISO-Prolog only specifies semantics for single-threaded code and does
not specify built-ins for parallelism.
As already discussed in~\cref{sec:parallelism}, several parallel implementations of Prolog
or derivatives thereof have been developed, targeting both shared-memory
multiprocessors and distributed systems.
Support for or-parallelism is not ubiquitous nowadays, although
systems like SICStus were designed to support it and this feature
can possibly be easily recovered.
Ciao still has some native support for and-parallelism and concurrency,
and its preprocessor CiaoPP still includes auto-parallelization.
A different, more coarse-grained form of parallelism is
multi-threading (this is what the parallelism column
in~\cref{tbl:feature-overview} gathers).

\paragraph{Indexing} \label{sec:indexing}
Indexing strategies of Prolog facts and rules are vital for Prolog development as they
immediately influence how Prolog predicates are written with performance in mind.
The availability of different indexing strategies is an important issue
that affects portability of Prolog programs:
if a performance-critical predicate cannot be indexed efficiently,
run-time performance may be significantly affected.
There are several strategies for indexing:

\begin{itemize}
    \item First-argument (FA) indexing is the most common strategy where the first
        argument is used as index. It distinguishes atomic values and
	the principal functor of compound terms.
    \item Non-first argument indexing is
        a variation of first-argument indexing that uses the same or similar
	techniques as FA on one or more alternative arguments.
        E.g., if a predicate call uses variables for the first argument,
        the system may choose to use the second argument as the index instead.
    \item Multi-Argument (MA) indexing
        creates a combined index over multiple
        instantiated arguments if there is not a sufficiently selective
	single argument index.
    \item Deep indexing is
        used when multiple clauses use the same principal
        functor for some argument.  It recursively uses the same or similar
	indexing techniques on the arguments of the compound terms.
    \item Trie indexing
        uses a prefix tree to find applicable clauses.
\end{itemize}

In addition to the above indexing techniques, one can also distinguish
systems that use directives to specify the desired indexes and systems
that build the necessary indexes just-in-time based on actual
calls.  One should note that the first form of indexing (FA) is the
only one which may be effectively relied upon when designing portable
programs, for it is close to universal adoption.

\paragraph{Type and Mode Annotations} \label{sec:typing}
As Prolog is a dynamic language,
it can be hard to maintain larger code bases
without well-defined (and checkable) interfaces.
Several approaches have been proposed to achieve or enforce sound
typing in Prolog programs~\citep{mycroft1984polymorphic,%
dietrich1988polymorphic,gallagher2004abstract,schrijvers2008towards}.
While these approaches are closer to strong typing,
they have not caught on with
mainstream Prolog.
Many Prolog systems offer support, e.g., for mode annotations, yet the
directives usually have no effects except their usage for
documentation.
The few Prolog or Prolog-like systems that really address these issues and
incorporate a type and mode system include
Mercury~\citep{SOMOGYI199617} and Ciao~\citep{prog-glob-an-short}.
The former is rooted in the Prolog tradition, but departs from it in
several significant ways (see Sections~\ref{sec:prologdef} and
\ref{sec:influence-other-lang}).
The latter aims to bridge the static and dynamic language approaches,
while preserving full compatibility with traditional non-annotated
Prolog.
A fundamental component %
is its assertion language~\citep{assert-lang-disciplbook-short} that is processed by CiaoPP (cf. \cref{sec:ciao}).
CiaoPP then is capable of finding non-trivial bugs statically or
dynamically, and can statically verify that the program complies with
the specifications, even interactively \citep{verifly-2021-tplp-short}.
The Ciao model can be considered an antecedent of the now-popular
\emph{gradual-} and \emph{hybrid-typing}
approaches~\citep{DBLP:conf/popl/Flanagan06-hybrid-type-checking-short,Siek06gradualtyping,DBLP:conf/popl/RastogiSFBV15-short}
in other programming languages.

\paragraph{Coroutining} \label{sect:coroutines}
As discussed in \cref{subpar:coroutining}, coroutining was first
introduced into Prolog in order to influence its default left-to-right
selection rule. From a logical perspective, logic programs are
independent of the selection rule.  However, from a practical,
programming language perspective, procedural factors might influence
efficiency, termination, and even faithfulness to the intended
results. Consider for instance a program that tries to calculate the
price of an as yet unknown object X (this could result from, say, some
natural language interface's ordering of different paraphrases).
Coroutining can ensure that the program behaves as intended by
reordering these goals so that objects are instantiated before
attempting to calculate their prices.

Coroutining is an important feature of modern Prolog systems, allowing
programmers to write truly reversible predicates and improve their
performance.  It represents a step forward towards embodying the
equation of ``Algorithm = Logic + Control''
by~\cite{kowalski1979algorithm}.
Early mechanisms for coroutining focused on variations of the {\tt
  delay} primitive by \cite{dahl1976systeme}, which dynamically
reorders the execution of predicates according to statically,
user-defined conditions on them.
The {\tt freeze}
variation %
was present in Prolog
II~\citep{colme82:prolog2,DBLP:conf/iclp/Boizumault86}, which delays
the execution of its second argument (understood to be a goal) until
its first argument, a single variable occuring in the second argument,
is bound.
A more flexible variation is the {\tt wait} primitive present in
MU-Prolog~\citep{DBLP:conf/ijcai/Naish85}, which was subsequently
generalized to \texttt{when}.  This concern evolved into the {\tt
  block} declaration found in modern Prologs, which supports the
annotation of an entire predicate (rather than of each individual
call), resembling a mode declaration.  This approach thus leads to
more readable programs and more efficient code.
It can be argued that the coupling of goal evaluation to the binding
of variables ultimately led to the development of Constraint Logic
Programming (CLP)~(\ref{section-clp}).
\subsubsection{Tools}

\paragraph{Unit Testing Structures}
One of the most important tools in software development
is a proper testing facility.
Some Prolog systems ship a framework for unit testing,
and while the basic functionality is shared,
usually they don't adhere to the same interface.
SWI-Prolog ships a library named \verb|plunit|,
while SICStus uses a modified version of it.
Both versions are entirely written in Prolog,
yet they rely on system-specific code to function properly.
Ciao relies on the \texttt{test} assertions of its assertion language,
which also include test case generation.
\eclipse{} offers a library named \verb|test_unit|
with several test primitives
that assert whether calls should succeed, raise errors, etc.
Other systems seem to rely on the fact
that a small ad-hoc testing facility
is rather easy to implement.

\paragraph{Debugging}  \label{sect:debugging}         %
A good debugger is vital to understand and fix undesired behaviors of
programs.  Prolog control flow is different from that of most other
programming languages, because it has backtracking.  To address this,
most Prolog systems provide some form of tracing debugger based on the
4-port debugger introduced for DEC-10 Prolog
by~\cite{byrd1980understanding}, which allows for the tracing of
individual goals at their call, exit, redo, and failure ports (states).
Most systems allow setting of spy points (similar to breakpoints) and
some provide a very Prolog-specific and powerful debugging tool: the
\emph{retry} command which allows one to ``travel back in time'' to
the entry point of a call that, for some reason, misbehaved.  The
latter feature assumes Prolog programs without side effects.
A few systems, such as SWI, SICStus, and Ciao, additionally offer
source-level debugging that allows following the steps of execution
directly on the program text, thus providing a more conventional view.

\subsection{Takeaways}%
\label{sec:takeaways-sec3}

When considering \cref{tbl:feature-overview},
one can see that, despite undeniable differences amongst Prolog systems,
many Prolog systems offer similar features.

\subsubsection*{Available and Mostly Compatible Features}

Mostly compatible \emph{module} systems have been widely adopted,
even if they virtually all diverge from the ISO document.
The existence of a de-facto module standard makes it possible to
write production-quality libraries that are portable across most
Prolog systems.  %

Facilities for \emph{multi-threaded programming} are also common:
A number of systems offer predicates based on the corresponding
technical recommendation document~\citep{ISOProlog-parallel},
sometimes with some differences in semantics or syntax.

Most systems also offer libraries for \emph{constraint programming},
though they differ in performance and expressiveness.
Yet, no standard interface or even a proposal for one exists.

Almost all Prolog systems embrace dynamic typing,
and \emph{type and mode} annotations are used for documentation or
optimization, yet are not enforced or verified at all.  Ciao, with its
combination of the dynamic and static approaches
is the significant exception here.

While support for \emph{tabling} is present in various systems, the features
and interfaces can differ.
Programs that rely on it (beyond simple answer memoization
which can be implemented fairly simply using the dynamic database),
specially if they use special features,
can be hard to port. Thus, progress needs to be made towards better
portability in this area.

\subsubsection*{Discrepancies}
One gets a mixed result when considering support for other features:
some systems, such as \eclipse{} offer extensive library support of \emph{data structures},
whereas others remain rather basic, without a large standard library.

\emph{Coroutining} is not available on all systems, and
the various primitives ({\tt when}, {\tt block}, {\tt freeze}, {\tt
  dif}) sometimes have some variations
 amongst those systems that do support it.
A similar situation occurs for \emph{global variables} and \emph{mutable terms}.
Similarly, \emph{testing frameworks} are missing in several systems, but usually can be provided in form of a portable library.

Almost all Prolog systems support at least one \emph{foreign language interface},
in order to leverage existing libraries
and to widen the domains where logic programming can be applied.
Yet, there are different strategies on how the interfaces interact with Prolog,
and, thus, the interfaces often differ between implementations.

An issue that can also %
hinder portability is the large discrepancy in \emph{indexing strategies}.
Solutions so far are of a very technical nature, rather than aimed
towards a common interface,
so work is needed if this issue is to be resolved.

\subsubsection*{Conclusions}

Overall, most Prolog systems are not \emph{too} different in what they offer.
Many differences could be bridged by agreeing on certain interfaces,
or, e.g., sharing library predicates for testing, or data structures.
Differences in constraint solving capabilities are harder to reconcile,
as some solvers are of commercial nature. However, CHR's embodiment of constraints is fairly ubiquitous and permits the implementation of constraint solvers in arbitrary domains of interest.
Missing technical features, such as tabling or indexing strategies
may hinder portability or performance, but the relevance of this issue
can greatly depend on the application. It is also possible to integrate tabling with constraints à la CHR \citep{Schrijvers2004}.
As differences with ISO-Prolog usually are very small now,
the Prolog implementations' cores are very similar today.

\subsection{Influence on Other Languages}
\label{sec:influence-other-lang}
\label{pl-inspired-langs}

The concepts and ideas that have been explored during the long history
of evolution of Prolog systems have influenced and given rise to other
languages and systems, both within the LP paradigm and out to other
programming paradigms.
In the following (\cref{pl-inspired-lplangs}), we describe languages
within the LP paradigm that are heavily inspired by or emerged from
Prolog. These systems generally fail our definition of Prolog in
\cref{sec:prologdef}. 
However, sometimes they bear witness to %
useful features or extensions that have not (yet) made their way into
Prolog itself. %
Still others, such as, for example, \texttt{s(ASP)}/\texttt{s(CASP)}
or Co-Inductive LP,
are really extensions of Prolog whose support in Prolog systems could
be generalized in the future, as has happened already with constraints
or tabling. They could thus also have been listed
in~\cref{sec:extension}.
Regarding the impact that Prolog and Prolog systems have had
\emph{beyond} LP, it is outside the scope of this paper to do a full
analysis of this very interesting topic, but we briefly review in any
case in~\cref{pl-inspired-otherlangs} a few examples of other such
influences outside the LP paradigm.

\subsubsection{Influences on other Languages %
  in the LP paradigm}
\label{pl-inspired-lplangs}

\paragraph{Datalog} is a subset of Prolog, which does not allow
compound terms as arguments of
predicates. %
This topic has been worked on since the late 70s, although the term
was coined later by David Maier and David S.\ Warren in the 80s.
Datalog can be viewed as an extension of relational databases,
allowing recursion within predicates.  Datalog plays an important role
in the research field of deductive databases
\citep{ramakrishnan93survey}.  Datalog has found new applications in
many areas~\citep{huang2011datalog}, such as information extraction
\citep{shen2007}, program analysis and synthesis
\citep{DBLP:conf/aplas/WhaleyACL05,Alpuente2011,DBLP:conf/cav/JordanSS16,DBLP:journals/pacmpl/MadsenL20},
security \citep{Bonatti2010}, graph
processing~\citep{DBLP:journals/tkde/SeoGL15}, reasoning over
ontologies \citep{DBLP:journals/ws/BaumeisterS10} or natural language
processing~\citep{dahl1995treating}.

\paragraph{$\lambda$Prolog} was developed in the late 80s, by Dale Miller and Gopalan Nadathur~(\citeyear{10.5555/868509}). It was defined as a language for programming in higher-order logic based on an intuitionistic fragment of Church's theory of types.
It spawned several modern refinements and implementations~\citep{lambda-prolog-homepage}. %
Higher-order extensions have also made their way into less specialized Prologs, as we discuss in \cref{sssec:wam-based-prologs}.
Miller and Nadathur~(\citeyear{miller2012programming}) discuss uses of higher-order logic in logic programming to provide declarative specifications for a range of applications.
$\lambda$Prolog applications are still surfacing, with the main focus being on meta-programming \citep{NadathurMeta}, program analysis \citep{wang2016higherorder}, and theorem proving \citep{articleLambPro}.

\paragraph{Committed-choice Languages} As mentioned in
\cref{sec:fgcs}, the implementation complexity of combining Prolog's
backtracking with concurrency and/or parallelism led to the
development of logic languages supporting ``committed choice'' where
only the first clause whose guard succeeds is executed, instead of the
multiple execution paths supported by Prolog.  This includes GHC
(Guarded Horn Clauses)~\citep{DBLP:conf/lp/Ueda85}, KL1~\citep{ueda1990design},
Parlog~\citep{Clark1986}, and Concurrent
Prolog~\citep{shapiro83,10.5555/535468}.
Erlang~\citep{armstrong2007history} (see later) also has its origins
in this line of work.
This line of work also yielded concurrent constraint languages, such
as \texttt{cc(fd)}~\citep{DBLP:journals/lncs/HentenryckSD94},
and distributed constraint languages, such as AKL~\citep{akl-paradigms} and
Oz/Mozart~\citep{DBLP:journals/pacmpl/RoyHSS20} (see later).
The concepts brought about by the committed-choice languages, such as
guards and data-flow synchronization based on one-way unification and
constraint entailment~\citep{Maher87}, have in turn made their way back
into Prolog systems as part of extensions for concurrency and
distributed execution (e.g., in \&-Prolog/Ciao~\citep{ciao-dis-impl-prode-short,CIAO}
or ACE~\citep{ACE}).

\paragraph{Turbo-Prolog}~\citep{Hankley87} can be considered a precursor of
other strongly typed logic programming languages. It was released as a
full development environment in 1986 for PC/MS-DOS. It was strongly
typed, had support for object-oriented programming, and it compiled
directly to machine code. At that time, this pragmatic approach
provided a safe and efficient language, but lacked important dynamic
features of Prolog required in many applications. It has been
continued as PCD Prolog and Visual Prolog, focusing on supporting good
integration with Microsoft Windows APIs.

\paragraph{G\"{o}del} \citep{hill1994godel} is a logic programming language that first appeared around 1992.
It is strongly typed, with a type system based on many-sorted logic, allowing for parametric polymorphism.
It implemented a sound negation, delaying negated calls until they were ground. %
G\"{o}del also supports meta-programming, using ground representation for meta-programs, which has the advantage of having a declarative semantics.
This enabled the development of a self-applicable partial evaluator called {\sc sage} by \cite{Gurr:LOPSTR93}.
The G\"{o}del system was built on top of SICStus Prolog, employing a different syntax style. %
The development of the language came to a halt in the 1990s.

\paragraph{Curry} \citep{HanusKuchenMoreno-Navarro95ILPS} was developed
in 1995. It is a functional logic programming language that is mostly based on Haskell, but includes some features from logic programming languages such as non-determinism and constraint solving.
Curry is based on the technique of \emph{narrowing}, which is also the basis of other functional logic programming
work~\citep{10.1145/1721654.1721675}.
The language has recently been used for typesafe SQL queries~\citep{hanus2017},
for research
and for teaching both the logic and the functional paradigm~\citep{Hanus97DPLE}.

\paragraph{Oz}
Another multi-paradigm language is Oz \citep{DBLP:conf/ijcai/HenzSW93},
 incorporating concurrent logic programming, functional programming, constraints, and objects.
The design of the language started in 1991 and its
 logic programming aspects were greatly influenced by AKL, the Andorra Kernel Language (cf.\ Section~\ref{sec-andorra}).
A recent synopsis of Oz's history is available in the article by \cite{DBLP:journals/pacmpl/RoyHSS20}.
The current incarnation of Oz is available as an open source implementation called Mozart. %

\paragraph{Mercury} \citep{SOMOGYI199617} was created in 1995 as a
functional logic programming language, with characteristics from both Prolog and Haskell.
The main reasons for its development were threefold.
First, idiomatic Prolog code at the time
had short predicate and variable names,
and also lacked type annotations and comments.
This often rendered it hard for the reader to
infer the meaning, types and modes of a program.
Second, multi-mode predicates and those without static type information
could not be compiled to the most efficient WAM code.
Third, a lot of LP research was concerned with logic programs
without any impure operations, and thus were not applicable to general Prolog programs
(e.g., executing \verb|read(X), call(X)|).
Thus, Mercury features a strong, static, polymorphic type system, and a strong mode and determinism system.
It has a separate compilation step,
which allows for a larger variety of errors to be detected before actually running a program,
and for generation of faster code.
By removing non-logical features, such as \predicate{assert/1} and \predicate{retract/1},
a pure language was obtained, I/O could be implemented declaratively, and
existing research could be implemented unaltered. Some interesting research has been done in the areas of program analysis and optimization \citep{10.1007/978-3-540-77442-6_13}, as well as parallelism \citep{bone12:_contr_loops_paral_mercur_code}.

\paragraph{Assumption Grammars and Assumptive Logic Programming}~\citep{Dahl97assumptiongrammars,dahl1998assumptions,dahl2004assumptive}
 extend Prolog with hypothetical reasoning,
needed in particular for natural language processing applications ~\citep{Dahl97assumptiongrammars}. They include specialized linear logic implications, called assumptions\citep{dahl1998assumptions,dahl2004assumptive}, that range over the computation's continuation, can be backtracked upon, and can either be  consumed exactly once (linear), at most once (affine linear), any number of times (intuitionistic), or independently of when they have been made: before or after consumption (timeless). The latter is a novel form of
implication designed and first meta-programmed
by Dahl for dealing with backward anaphora.
 Their uses for abduction were also researched by \cite{dahl2004assumptive}.

\paragraph{Answer Set Programming} \label{sec:ASP}
(ASP) is arguably one of the largest successes of logic programming. It is
a logic programming paradigm that focuses on solving (hard) search problems, by reducing them to computing stable models.
Note that ASP is \emph{not} a Turing-complete programming language, but rather a language to represent aforementioned problems.
It is based on the stable models semantics and uses answer set solvers to provide truth assignments as models for programs.
The usual approach for ASP is to ground all clauses so that propositional logic techniques like SAT-solving can be applied
to find stable models for the program. Unlike Prolog's query evaluation, ASP's computational process always terminates.
The denomination ``answer set'' was first coined by \cite{Lifschitz1999}.
Its early exponents were \cite{niemela1999} and
\cite{MarekT99}; more recent developments include
Smodels~\citep{syrjanen2001smodels}, DLV~\citep{leone2006dlv}, the
Potassco toolset~\citep{clingo,gebser2008user},
and WASP \cite{AlvianoDFLR13}.
The interested reader may rely on the survey by \cite{brewka2011answer}
and on the special issue of the AI Magazine dedicated to ASP~\citep{brewka2016answer}.

\paragraph{s(ASP) and s(CASP)} \label{sec:SCASP}
\texttt{s(ASP)}~\citep{marple2017computing}
is a \emph{goal-directed, top-down} execution model which computes stable models of normal
logic programs
with arbitrary terms, supporting the use of lists and complex data
structures, and, in general, programs which may not have a finite
grounding.
It supports both deduction and abduction.
\texttt{s(ASP)} uses a non-Herbrand universe, coinduction, constructive
negation, and a number of other novel techniques.
Unlike in ASP languages, variables are (as in (C)LP) kept during
execution
and in the answer sets.
\texttt{s(CASP)}~\citep{scasp-iclp2018} is the extension of
\texttt{s(ASP)} to constraint domains, while also including additional
optimizations in the implementation.
\texttt{s(ASP)} and \texttt{s(CASP)} can be seen as Prolog extended with negation as
failure where this negation follows the stable model semantics. If 
negation is not used the behavior is as in Prolog.
Both \texttt{s(ASP)} and \texttt{s(CASP)} have been implemented in Prolog, originally in
Ciao and also ported to SWI.
For some applications this approach leads to improved performance and expressivity
compared to existing ASP systems. Another advantage is that it
naturally provides explanations. 
At the same time, for some other classical ASP applications the
grounding to propositional logic remains currently the technology of
choice.
Goal-directed evaluation of ASP was also addressed in the work of
Bonatti et
al.~\citep{DBLP:journals/jar/Bonatti01,DBLP:conf/aaai/BonattiPS08}. 
In comparison, the \texttt{s(ASP)} and \texttt{s(CASP)} work handles unrestricted
predicates and queries, with constraints.
Other early work on query-driven computation of stable models includes
that of~\cite{DBLP:journals/tkde/ChenW96} and the XNMR system within
XSB.

\paragraph{Constraint Handling Rule Grammars} (CHRG) by \cite{christiansen2002logical},
extend Prolog with sophisticated language processing capabilities because, just like Prolog does, they allow writing grammar rules that become executable. They differ from the previous grammatical default of Prolog (\emph{DCGs})
in that they work bottom up, are robust (i.e., in case of errors the recognized phrases so far are returned,
rather than silently failing by default), can inherently treat ambiguity without backtracking, and, just as Hyprolog (see below),
can produce and consume arbitrary hypotheses.
This makes it straightforward to deal with abduction, which is useful for diagnostics, %
integrity constraints, operators à la Assumption Grammars.
They can also incorporate other constraint solvers.
Applications go beyond traditional NLP, including e.g., biological sequence analysis~\citep{bavarian2006constraint}.

\paragraph{Hyprolog} \citep{christiansen2005hyprolog}
is an extension to Prolog and Constraint Handling Rules (CHR)
which includes all types of hypothetical reasoning in Assumption Grammars,
enhances it with integrity constraints, and offers abduction as well.
It compiles into Prolog and CHR through an implementation by Henning Christiansen available for SICStus, Prolog III and IV, and for SWI-Prolog.
It can access all additional built-in predicates and constraint solvers that may be available through CHR,
whose syntax can be used to implement integrity constraints associated to assumptions or abducibles.
Due to the compiled approach, which employs also the underlying optimizing compilers %
for Prolog and CHR,
the Hyprolog system is amongst the fastest implementations of abduction.

\paragraph{Co-Inductive logic programming}
\citep{simon2006coinductive} was proposed in order to allow logic
programming to work with infinite terms and infinite proofs
based on greatest fixed-point
semantics. The \emph{co-logic programming}
paradigm by \cite{10.1007/978-3-540-74610-2_4} is presented as an
extension of traditional logic programs with both inductive and
co-inductive predicates. This can be used, e.g., for model checking,
verification and non-monotonic reasoning
\citep{DBLP:conf/calco/GuptaSDMMK11}.  These concepts were implemented
based on modifying YAP's Prolog engine and are also related to 
\texttt{s(ASP)}/\texttt{s(CASP)}.

\paragraph{Probabilistic Logic Programming} (PLP) \label{sec:probabilistic-prolog}
is a research field that investigates the combination of LP with the probability theory.
A comprehensive overview on this topic is provided by \cite{riguzzi2018}.
In PLP, theories are logic programs with LPAD, i.e., logic programs with annotated disjunctions \citep{vennekens-2004}, hence they may contain facts or rules enriched with probabilities.
These may, in turn, be queried by the users to investigate not only which statements are true or not, but also under which probability.
To support this behavior, probabilistic solvers employ ad-hoc resolution strategies explicitly taking probabilities into account.
This makes them ideal to deal with uncertainty and the complex phenomena of the physical world.

From a theoretical perspective, the distribution semantics by \cite{sato1995} is one of the most prominent approaches for the combination of logic programming and probability theory.
Sato, in particular, was amongst the first authors exploiting Prolog for PLP, by building on the ideas of \cite{Poole93}.
The very first programming language laying under the PLP umbrella was PRISM, by \cite{sato-1997}, which supported not only probabilistic inference but learning as well.
Since then, many PLP solutions have been developed supporting this semantics, such as ProbLog by \cite{de-raedt-2007} and \texttt{cplint} by \cite{riguzzi-2007}.
These were often implemented on top of existing Prolog implementations.
For instance, ProbLog consists of a Python package using YAP behind the scenes, while \texttt{cplint} is based on SWI-Prolog.
They reached a considerable level of maturity and efficiency by exploiting binary decision diagrams \citep{akers1978binary} or variants of them to speed up probabilistic inference.

\paragraph{Logtalk} \citep{10.1007/978-3-642-20589-7_4}
can be considered to be an object-oriented logic programming language
as well as an extension to Prolog.
Its development started in 1998 with the goal of supporting programming in the
large.  Because it is object-oriented, Logtalk supports classes,
prototypes, parametric objects, as well as definite clause grammars,
term-expansion mechanisms, and conditional compilation.  
It addresses several issues of Prolog that have not been met with a standardized solution,
including portability of libraries and tools (e.g., for unit testing, documentation, package management and linting),
by compiling the code to a very wide range of Prolog systems.

\paragraph{Picat} \citep{zhou2015constraint} is a logic-based multi-paradigm language.
Its development started in 2012, stemming from B-Prolog \citep{zhou_2012} and
having its first alpha release in 2013 \citep{zhou2016user}.
It aims to combine the efficiency of imperative languages with the power of declarative languages.
It is dynamically typed and uses rules in which predicates, functions, and actors are defined with pattern-matching.
It also incorporates some features of imperative languages, such as arrays, assignments and loops. Its main focus of research is constraint solving \citep{DBLP:journals/corr/abs-2109-08293}.

\paragraph{Womb Grammars} (WG) \citep{dahl2012womb},
endow Prolog + CHRG with constraint solving capabilities for grammar induction,
within a novel paradigm: they automatically map a language's known grammar (the source)
into the grammar of a different (typically under-resourced) language.
This is useful for increasing the survival chances for endangered languages, with obvious positive socio-economic potential impact (over 7,000 languages are spoken in the world, of which, according to \cite{ethnologue-homepage},
2895 are endangered).
They do so by feeding the source grammar
a set of correct and representative input phrases of the target language plus its lexicon, and using the detected ''errors" to modify the source grammar until it accepts the entire corpus.
WG have been successfully used for generating the grammars of noun phrase subsets
of the African language Yor\`{u}b\'{a}~\citep{Adebara2016GrammarIA} (for which a grammar that validated the system's findings does exist)
and the Mexican indigenous language Ch'ol~\citep{dahl2020resourcing} (for which no grammar had been yet described).

\subsubsection{Some Influences on Languages and Systems Beyond LP}
\label{pl-inspired-otherlangs}

\paragraph{Theorem proving} \label{sec-theorem-proving}
Some aspects of the WAM, such as the compilation of clause heads, were
adopted by different theorem provers, such as the Boyer-Moore theorem
prover~\citep{boyer-moore-tp}, as a result of the prover team and
Prolog teams working together at MCC in the mid to late 80s. Another example of influence of these Prolog systems is the use of
Prolog technology in theorem provers, for instance, by \cite{Stickel84}
or provers directly implemented in Prolog~\citep{manthey1988satchmo,stickel1992prolog,beckert1995leantap}.

\paragraph{Java} The design of the WAM and various other aspects of
Prolog implementation influenced the design of the Java abstract
machine, since some of the designers of this machine had formerly
worked at Quintus and were Prolog implementation experts.
For instance, the semantics of type checking for Java's class files is provided as a Prolog script by \cite{lindholm2014java}.
\paragraph{Erlang}
The quite successful programming language
\emph{Erlang}~\citep{armstrong2007history} has its roots in Prolog and
the concurrent constraint languages that derived from Prolog,
and was developed with the goal of improving the development of telephony applications.
The first version of the Erlang interpreter was written in Prolog, which is the reason for syntactic similarities.
Erlang is still used nowadays by many companies, including
Cisco, Ericsson, IBM, and WhatsApp \citep{erlang-companies}.

\paragraph{Language Embeddings}
Some languages outside LP nowadays include a Prolog or logic
programming library or mode. Classical examples are the different
embeddings of Prolog in Scheme, such as Schelog~\citep{schelog-homepage} and Racket's RackLog
sublanguage~\citep{racklog-homepage}, which are generally based on the work
of~\cite{felleisen-prolog-in-scheme}
and~\cite{haynes-logic-continuations} (the former also done in part at
MCC) and~\cite{Carlsson84}. These can provide useful Prolog-like
functionality, although the performance is generally not comparable
with, e.g., native WAM-based systems.

\bigskip
Further influence outside the logic programming paradigm
is apparent in languages with inferred types and
polymorphic type systems, which sometimes include a rule system to specify
and constrain the types.  For example the \emph{concepts} of C++ 2020
\citep{C++20} %
provide predicates that form rules to statically
determine which of a set of implementations of a polymorphic function
ought to be used, according to context.
\section{Part III: The Future of Prolog}%
\label{sec:future}

While \cref{sec:current-state} establishes that some
incompatibilities between Prolog systems are not too difficult to
overcome, this section explores a different perspective: %
what are the perceived issues and potential future directions for
Prolog to grow.
In order to provide insights on the future of Prolog,
we conducted a SWOT analysis.
Its results can be found in \cref{fig:SWOT}.
In the following,
we discuss strengths (\cref{swot-S}) and opportunities (\cref{swot-O}),
followed by weaknesses of the language
(\cref{swot-W})
currently,
and external factors that may be threats to the adoption of Prolog,
its future development, or to the compatibility of Prolog systems~(\cref{swot-T}).
In \cref{sec:discussion}, we aim at providing a foundation for
community discussion
and stimulus towards future development of the language.
To this end, %
we make proposals and raise questions on
which features could be %
useful future extensions for Prolog.
Finally, in \cref{sec:nextsteps} we summarize and briefly discuss some possible next steps.
\begin{table}
    \begin{adjustwidth}{-2cm}{-3cm}
    \caption{SWOT Analysis}%
    \label{fig:SWOT}
    \begin{tabular}{|l|l|}
    \hline
    \begin{minipage}[t]{0.45\linewidth}
    {\bf Strengths} (\cref{swot-S})
    \begin{itemize}[leftmargin=*]
    \item clean, simple language syntax and semantics
    \item immutable persistent data structures, with ``declarative'' pointers (logic variables)
    \item arbitrary precision arithmetic %
    \item safety (garbage collection, no NullPointer exceptions, ...)
    \item tail-recursion and last-call optimization
    \item efficient inference, pattern matching, and unification, DCGs
    \item meta-programming, programs as data %
    \item constraint solving (\ref{section-clp}), 
        independence of the selection rule (coroutines (\ref{sect:coroutines})) %
    \item indexing (\ref{sec:indexing}), efficient tabling (\ref{sec:tabling}) %
    \item fast development, 
          REPL (Read, Execute, Print, Loop),
            debugging (\ref{sect:debugging})
    \item commercial (\ref{sec-early-commercialisation}) and open-source systems
    \item active developer community with constant new
      implementations, features, etc.
    \item sophisticated tools: analyzers, partial evaluators, parallelizers, ...
    \item successful applications
        \begin{itemize}
        \item program analysis, %
        \item domain-specific languages
        \item heterogeneous data integration %
        \item %
             natural language processing
        \item efficient inference (expert systems, theorem provers), symbolic AI
        \end{itemize}
    \item many books, courses and learning materials
    \end{itemize}
    \end{minipage}
    &
    \begin{minipage}[t]{0.45\linewidth}
    {\bf Weaknesses} (\cref{swot-W})
    \begin{itemize}[leftmargin=*]
    \item syntactically different from ``traditional'' programming languages, not a mainstream language
    \item learning curve, beginners can easily write programs that loop or consume a huge amount of resources
    \item static typing (see, however, \ref{sec:typing})
    \item data hiding %
      (see, however, \ref{sec:modules})
    \item object orientation (see, however, \ref{sec:objects})
    \item limited portability (see \ref{sec-portability-features}) %
    \item packages: availability and management
    \item IDEs and development tools: limited capabilities in some areas (e.g., refactoring; \ref{sec-dev-tools})
    \item UI development (usually conducted in a foreign language via FLI (\ref{sect:fli}))
    \item limited support for embedded or app development
    \end{itemize}
    \end{minipage}
    
    \\ 
    \hline
      
    \begin{minipage}[t]{0.45\linewidth}
    {\bf Opportunities} (\cref{swot-O})
    
    \begin{itemize}[leftmargin=*]
    \item new application areas, addressing societal challenges \ref{swot-O}:
    \begin{itemize}
        \item neuro-Symbolic AI
        \item explainable AI, verifiable AI
        \item Big Data
    \end{itemize}
    \item new features and developments
    \begin{itemize}
        \item probabilistic reasoning (\ref{sec:probabilistic-prolog})
        \item embedding ASP (\ref{sec:ASP}) and SAT or SMT solving
        \item parallelism (\ref{sec:parallelism}, \ref{subsec:parallelism}) (resurrecting 80s, 90s research)
    \item full-fledged JIT compiler
    \end{itemize}
    \end{itemize}
    \end{minipage}
    &
    \begin{minipage}[t]{0.45\linewidth}
    {\bf Threats} (\cref{swot-T})
    \begin{itemize}[leftmargin=*]
    \item comparatively small user base
    \item fragmented community with
          limited interactions (e.g., on StackOverflow, reddit), %
          see \ref{para:user-bases}
    \item active developer community with constant new implementations, features, etc. 
    \item further fragmentation of Prolog implementations, see \ref{para:implementors} %
    \item new programming languages %
    \item post-desktop world of JavaScript web-applications
    \item the perception that it is an ``old'' language
    \item wrong image due to ``shallow'' teaching of the language
    \end{itemize}
    \end{minipage}
    \\
    \hline
    \end{tabular}
\end{adjustwidth}
    
    \end{table}

\subsection{SWOT: Strengths of Prolog}%
\label{swot-S}

\subsubsection*{Ease of Expression}

Prolog is a language with a small core and a minimal, yet extremely flexible syntax.
Even though some features can only be understood procedurally, such as the cut, the semantics remains very simple. %
Combined with automatic memory management and the absence of pointers or uninitialized terms makes Prolog a particularly safe language.
Flexible dynamic typing
completes the picture by placing Prolog amongst the most high-level programming languages available to date --- a feature that makes it very close to how humans think and reason, and therefore ideal for Artificial Intelligence (AI).

The inspection and manipulation of programs at run-time also leads to faster programmer feedback and enables powerful debugging,
 in particular when coupled with Prolog's interactive toplevel
 (a.k.a.\ REPL or Read-Eval-Print-Loop, although in the case
 of Prolog the print part is richer because of multiple solutions).

Declarativity is yet another important feature of Prolog: most programs just state \emph{what} the computer should do, and let the Prolog engine figure out the \emph{how}.
For pure logic programs, the resolution strategy may also be altered, and possibly optimized, without requiring the source code of those programs to be changed.
Program analysis and transformation tools, partial evaluators and automatic parallelizers can be used for Prolog programs and can be
particularly effective for purely declarative parts.
For partial evaluation, however, the techniques have still not yet been integrated into
 Prolog compilers (the discussion by
 Leuschel and Bruynooghe (\citeyear[Section 7]{LeuschelBruynooghe:TPLP02}) is mostly still valid today), with the
 exception of Ciao Prolog's pre-processor~\citep{ciaopp-sas03-journal-scp-short} and abstract machine generator~\citep{morales05:generic_eff_AM_implem_iclp,morales15:improlog-tplp}.
 However, they have found their way into just-in-time compilers for other languages~\citep{DBLP:conf/pepm/BolzCFLPR11}.

Prolog's data structure creation, access, matching, and manipulation,
is performed via the powerful and efficiently implemented unification operation.
The logical terms used by Prolog as data
structures are  managed dynamically and efficiently.
Logical variables within logical terms can encode
``holes'' which can then be passed around arbitrarily and filled at
other places by a running program.
Furthermore, logical variables can
be bound (\emph{aliased}) together, closing or extending pointer
chains.
This gives rise to many interesting programming idioms such
as difference lists and difference structures, and in general to all
kinds of pointer-style programming, where logical variables serve as
``declarative pointers'' (since they can be bound only once).
This view of logical variables as
declarative pointers and related issues have been discussed by \cite{tutorial-pcjspecial}.
Furthermore, Prolog's automatic memory management ensures the absence of nuisances such as NullPointer exceptions or invalid pointer manipulations.
Many Prolog systems also come with arbitrary precision integers, which are used transparently without requiring user guidance.

Prolog compilers and the many program analysis and transformation
tools mentioned above are almost always written in Prolog itself,
which is an excellent language for writing program
processors. Interestingly, since the semantics of programming
languages can be easily encoded as (Constraint) Horn Clauses (CHCs), Prolog
tools can often be applied directly to the analysis of other 
languages~\cite{decomp-oo-prolog-lopstr07}. 
The survey by~\cite{anal-peval-horn-verif-tplp} in this same special
issue of the TPLP journal provides a comprehensive overview of work
using analysis and transformation of (constrained) Horn clauses and
techniques stemming from logic programming for program verification,
including those techniques most related to Prolog.

\subsubsection*{Efficiency}
In addition, Prolog is a surprisingly efficient language.
Beginners will often write very inefficient Prolog programs.
Yet, carefully written Prolog programs can build on many of the features
provided by modern implementations, such as last call optimization
(which generalizes tail recursion optimization), efficient indexing and
matching, and fine-tuned memory management with efficient backtracking.
For applications which are well-suited to Prolog, such as
program analysis~\citep{decomp-oo-prolog-lopstr07}, program verification~\citep{Le08PPDP}, or theorem proving (see Section~\ref{sec-theorem-proving}),
    this can lead to programs which are both more flexible and better-performing than counterparts written in traditional languages \citep{DBLP:conf/ifm/Leuschel20}.

\subsubsection*{Successful Applications}
Thanks to its simple foundation, the language makes it straightforward to read Prolog programs as data objects and it is almost trivial to implement meta-interpreters, as well as custom or domain-specific languages (such as Erlang \citep{armstrong2007history}, which was initially implemented in Prolog).
Therefore, it can be (and has been) used as a means to represent knowledge bases or bootstrap declarative languages in
knowledge-intensive environments.
Prolog has also been used for several successful formal methods tool
developments, such as, e.g.,~\cite{DBLP:journals/sttt/LeuschelB08,resource-verification-tplp18-short}.
Moreover, Prolog supports the implementation of novel sorts of expert systems or logic solvers, relying for instance on probabilistic, abductive, or inductive inference, which can be simply realized as meta-interpreters.
This is another reason why Prolog is well suited for symbolic AI applications.
It can also be used to integrate and reason over heterogeneous data \citep{DBLP:conf/semweb/WielemakerHOS08}.

Prolog has also been successfully used for parsing, both for computer languages and for natural languages (see also \ref{sect-metamorphosis-grammars}).
A relatively recent success story is IBM's Watson system \citep{Watson} which used Prolog for natural language processing, adapting techniques developed in logic grammars over the years for solving difficult computational linguistics problems, such as coordination ~\citep{McCord,DMc}.
Regarding parsing algorithms, Prolog's renditions of tabling admit especially succinct while elegant and efficient formulations, as demonstrated for the CYK algorithm on p. 37 of Fr\"uhwirth's book on CHR~\citep{fruhwirth2009constraint}.
Indeed,
  Prolog lends itself particularly well for grammar development and grammatical theory testing, which has potential both for compilers and for spoken and other human languages.
The simplest grammatical Prolog version, DCGs, extends context-free grammars with symbol arguments, while the first grammatical version, MGs, extends type-0 grammars with symbol arguments.
Variants that are adequate to different linguistic theories can, and have been, developed \citep{DahlGB,DahlBarriers,DahlLingConstraints,DahlRL}.
Most crucially, semantics can be straightforwardly accommodated and examined through symbol arguments, which allows for the increasingly important features of transparency and explainability to materialize naturally.
Coupled with Prolog's meta-programming abilities, grammar transformation schemes can help automate the generation of  linguistic resources that most languages in the world typically lack, as shown by \cite{dahl2012womb}.
Finally, tabling can be used for efficient parsing \citep{ChartParsing} of a wide range of grammars, even context-sensitive ones. %

Many more examples of practical applications can be found in the literature, in particular in the conference series
PADL (Practical Applications of Declarative Languages, running since 1999)
and INAP (International Conference on Applications of Declarative
Programming and Knowledge Management, since 1998).
ICLP, the premier conference in logic programming, regularly includes
papers and sessions on applications and some editions have a special
applications track.

\subsubsection*{Active Community}
As we have seen throughout this paper, there are many implementations
of the Prolog language, many of them quite mature and still being actively developed, with new features and libraries added
continuously, while new implementations keep appearing, with new aims or targeting different niches.

There are many books and tutorials on the Prolog language.
Good examples are texts by \cite{ClocksinMellish1981}, \cite{sterling1994art}, \cite{OKeefe:CraftOfProlog}, \cite{Clocksin:ClauseAndEffect}, \cite{blackburn2006learn}, \cite{Bratko2012}, or \cite{powerofprolog}.
There is also significant teaching material publicly available in the form of slides, exercises, examples, contest problems, etc., as well as plenty of topical discussions in on-line fora.
There are also some interactive learning environments and playgrounds,
e.g., GUPU~\citep{neumerkel2002gupu}, or those of SWI, Ciao, or
LogTalk, although this is certainly an area that would be well worth
improving.

A Prolog programming contest (which has now expanded to include other
LP and CLP dialects) is held every year in the context of ICLP. 

Even with Prolog being somewhat outside the mainstream of programming languages, it is taught at many universities for a simple reason: it introduces new concepts and features that are quite different from those of object-oriented as well as functional programming languages.
Accordingly, it provides computer scientists with not only a simple yet
powerful tool to understand and write elegant algorithms, but also a
new way of thinking about programming.
Getting to know the ropes of Prolog expands one's horizons and allows
programmers to significantly improve their way of solving problems.
One could easily argue that a computer scientist is not really
complete without being familiar with First-Order Logic, Resolution,
Logic Programming, and Prolog.

\subsection{SWOT: Opportunities}
\label{swot-O}

There are several opportunities to considerably improve the performance of Prolog by resurrecting earlier research on parallelism
 (\ref{sec:parallelism}, \ref{subsec:parallelism}).
A JIT compiler can also be beneficial and is provided, e.g., by SICStus Prolog. There are certainly opportunities for combining just-in time compilation with specialization to achieve even better performance.

It is very natural to integrate into Prolog features like probabilistic reasoning (\ref{sec:probabilistic-prolog}),
  ASP (\ref{sec:ASP}), or other logic-based technologies like SAT solving and SMT solving.
Making these features routinely available
  would make Prolog more appealing for a wider class of applications.

\subsubsection*{Artificial Intelligence}

\paragraph{Symbolic AI:}
Prolog is undoubtedly amongst the most impactful ideas in \emph{symbolic} AI.
However, \emph{sub-symbolic} or \emph{data-driven} AI is nowadays
attracting most of the attention and resources, mostly because of the
recent progress that has been achieved in machine and deep learning.

Despite these advances, state-of-the-art data-driven AI techniques are
far from perfect.
A common problem
that shows up in critical fields such as Healthcare
\citep{Panch2019}, Finance \citep{Johnson2019} or Law
\citep{Tolan2019} is that learning-based solutions tend to acquire
the inherent \emph{biases} of the contexts they are trained into.
This often results in decision support systems exposing sexist,
racist, or discriminatory behaviors, thus unwittingly permeating the
digital infrastructure of our societies \citep{Umoya}.

Similarly, sub-symbolic techniques have been criticized for their
inherent \emph{opacity} \citep{Guidotti19}.
In fact, while most techniques in this field (neural networks, support
vector machines, etc.) are very good at learning from data, they
easily fall short when it comes to \emph{explicitly} representing what
they have learned.
For this reason, such techniques are often described as black boxes in
the literature \citep{Lipton2018}.

\paragraph{Explainable AI:}
While all such issues are being tackled by the eXplainable AI community (XAI) \citep{darpa2016-xai} by using a plethora of sub-symbolic tricks \citep{Guidotti19}, an increasing number of works recognize the potential impact of symbolic AI models and technologies in facing these issues, such as the works by Calegari et al.~(\citeyear{lpaas-bdcc2}, \citeyear{logictech-information11}) or \cite{DBLP:conf/ijcai/Cyras0ABT21}.
It seems clear that \emph{symbolic} inferential capabilities will be crucial for transitioning into the sustainable and humanity-serving AI that is urgently needed \citep{xaisurvey-ia14}. %

Accordingly, we highlight two possible research directions where Prolog and LP may contribute further to the current AI picture.
One direction concerns the exploitation of LP either \emph{(i)} for making machine and deep learning techniques more \emph{interpretable} or \emph{(ii)} for constraining their behavior, reducing biases.
There, Prolog and LP may be exploited as a \emph{lingua franca} for symbolic rules extracted from sub-symbolic predictors \citep{xailp-woa2019,agentbasedxai-extraamas2020}, or as a means to impose constraints on what a sub-symbolic predictor may (or may not) learn \citep{SerafiniDG17}.
The other direction concerns the exploitation of sub-symbolic AI as a means to speed up or improve some basic mechanism of LP and Prolog.
For example, in the field of \emph{inductive} logic programming \citep{MuggletonR94} neural networks have been exploited to make the inductive capabilities of induction algorithms more efficient or effective %
\citep{GarcezZ99,BasilioZB01}.
For instance, in the work of \cite{FrancaZG14}, the induction task is translated into an attribute-value learning task by representing subsets of relations as numerical features, and CILP++ neuro-symbolic system is exploited to make the process faster.

Along this path, more general approaches attempt to unify, integrate, or combine the symbolic and sub-symbolic branches of AI, for the sake of advancing the state of the art.
This is the case for instance of the neuro-symbolic initiatives \citep{RaedtDMM20,LambGGPAV20,nsc4xai-woa2020,Tarau21LPandNeuro} where LP and neural networks are combined or integrated in several ways following the purpose of engineering more generally intelligent systems capable of coupling the inferential capabilities of LP with the flexible pattern-matching capabilities of neural networks.

\paragraph{Inductive Logic Programming:}
Inductive Logic Programming (ILP), first coined by \cite{MuggletonR94} is a subfield of machine learning that studies how to learn computer programs from data, where both the programs and the data are logic programs. Prolog in particular is typically used for representing background knowledge, examples, and induced theories. This uniformity of representation gives ILP the advantage, compared to other machine learning techniques, that it is easy to include additional information in the learning problem, thus enhancing comprehensibility and intelligibility. Muggleton first implemented ILP in the PROGOL system.

ILP has shown promise in addressing common limitations of the state-of-the-art in machine learning, such as poor generalization, a lack of interpretability, and a need for large amounts of training data. The first two affect quality and usability of results, and the latter affects accessibility and sustainability: the requirements of storing and processing exponentially growing amounts of data already make its processing prohibitive.
In this context, ILP shows especial promise as a tool enabling a shift from using hand-crafted background knowledge to \emph{learning} background knowledge, including learning recursive programs that generalize from few examples.
These issues, as well as future promising directions, have been recently surveyed by \cite{ijcai2020-673}.

Further developing the ideas of ILP to encompass the automated
learning of \emph{probabilistic} logic programs is key to Statistical
Relational AI (StaRAI), a successful hybrid field of AI
\citep{DBLP:series/synthesis/2016Raedt}, for which regular workshops
have been held since 2010.  Tools based on this paradigm aim to handle
complex and large-scale problems involving elaborate relational
structures and uncertainty.

\subsubsection*{Bridges to Established Research Areas}
Novel opportunities may then arise by bridging Prolog with other well-established research areas.
This is, for instance, what happened with the Multi-Agent System community, where Prolog and LP have been extensively exploited in the last decades as the technological or conceptual basis for tens of agent-oriented technologies \citep{lptech4mas-jaamas35}.
Similarly, the Prolog language has been proposed within the scope of Distributed Ledger Technologies (a.k.a. Blockchains), by \cite{blockchainlp-woa2018}, or as a means to provide more declarative sorts of \emph{smart contracts}, as suggested by  \cite{autonomoussc-paams2019} and \cite{blockchainmas-applsci10}.
There, LP pursues the goal of making smart contracts declarative, hence easing their adoption, and increasing their expressiveness. More generally, begining with the  Prolog implementation of the British Nationality Act, Prolog has been used extensively for applications to computational law, and is proabably still the dominant approach. For example, the logic programming language in Oracle Policy Automation%
\footnote{\url{https://documentation.custhelp.com/euf/assets/devdocs/cloud19b/PolicyAutomation/en/Default.htm\#Guides/Welcome/Welcome.htm}}
was originally implemented in Prolog.

In data science, several data sources need to be cleaned and combined
before applying statistical analysis and machine learning.  This
preprocessing step is often the most labor intensive phase of a data
science project.  Prolog, particularly when extended with tabling
support, is a suitable tool for data preprocessing.  It can
transparently access data from different sources without needing to
repeatedly import all the data, e.g., from a relational database
management systems. %
Subsequently, a view on the data can be established from small
composable predicates that can be debugged separately and
interactively.  These ideas have been explored in SWISH
datalab~\citep{bogaard2016swish} which provides a web frontend for
cooperative development of both Prolog data preprocessing steps and
subsequent statistical analysis using R, and used in applications in
Biology using large
data~\citep{DBLP:journals/corr/abs-1909-08254}. Other Prologs, such as
Yap, include support for dealing with large data
sets~\citep{DBLP:conf/padl/Costa07,DBLP:journals/tplp/CostaV13}.

Summarizing, it might prove very useful for the community to anticipate
what features may attract programmers %
for ends such as improving AI, the Internet, programming languages, or knowledge intensive systems in general.

\subsection{SWOT: Weaknesses}
\label{swot-W}

Prolog is syntactically quite different from traditional mainstream programming languages.
One could argue that this is a necessary side effect of many of its strengths, and that it is possibly a reason why Prolog has a place in many computer science curricula.
However, it also means that Prolog can appear strange or unfamiliar to many beginners or even seasoned programmers.
Mastering Prolog also implies a considerable learning curve and it certainly takes a while for a beginner to become truly productive in what is often a radically new language and paradigm.

While Prolog's dynamic typing can be an advantage for rapid prototyping and meta-programming, it is also often considered a weakness.
Optional static typing (see \ref{sec:typing}) would definitely help in catching many bugs early on.

Data hiding is also more difficult in Prolog.
In the de-facto standard module system used by most Prologs one can decide which predicates are exported, but not which data types (i.e., functors and constants) are exported.
One can typically not prevent another module from manipulating internal data structures (see, however, \ref{sec:modules}).
This issue goes hand-in-hand with the limited support for object orientation (see, however, \ref{sec:objects}).

Another issue is the limited support for user interface (UI) development within Prolog.
There was more attention to this issue in the past (e.g., BIM-Prolog
had the Carmen library for this purpose), and there were interesting approaches in
which declarativeness and/or constraints were exploited for this purpose, 
but nowadays UIs are usually developed in a foreign language via FLI (\ref{sect:fli}).
Similarly, there is limited support for developing mobile applications
or embedded software with Prolog, even if a number of Prologs can run
on small devices and Prolog program optimization techniques have been
shown to be up to the task~\cite{carro06:stream_interpreter_cases}. 

Portability of Prolog code can also be nontrivial, despite the ISO standard (see \ref{sec-portability-features}).
Also, the Prolog community does not have a standard package manager, making it more difficult to distribute libraries or modules.
(On the upside, Prolog also does not have all the  version management hell and security issues prevalent in other languages.) %

Finally, the support for Prolog in some integrated development environments can be less mature than for mainstream languages like Java (see \ref{sec-dev-tools}).
In particular, the available support for refactoring is often quite limited.

\subsection{SWOT: Threats}
\label{swot-T}

Some threats to Prolog come from competing programming languages.
Indeed, Prolog may be perceived as an ``old'' language, and new programming languages such as previously Java, or now Rust or Go, may be more appealing to new generations of programmers.
Also, for some application domains, like web-based applications, which have obviously become increasingly important in recent years, other languages like JavaScript are much more popular and easier to deploy than Prolog, although this is being addressed by current Prolog implementations.

Another threat to the perception of Prolog comes from the fact that,
when teaching it, if the presentation does not go deep enough to
convey the real power and elegance of the language and the programming
paradigm, it may instead leave the incorrect impression of being a
shallow or just academic tool.

On the other hand, as mentioned before, Prolog
systems are still very actively developed, with new features and new
implementations appearing continuously. While this is obviously positive,
some threats we see for Prolog stem precisely from a resulting divergence of
implementations (which tends to fragment the community into several,
changing user camps) and a lack of strong stewardship (which, in turn,
may fuel further divergence of systems).

We discuss the aforementioned issues in more detail in the following
two sections.

\subsubsection{Fragmentation of the Community}%
\label{subsec:fragmented-community}

A large threat to the future of Prolog as an accessible programming language
is further fragmentation of the community.
This manifests itself in two dimensions:

\paragraph{User Bases}
\label{para:user-bases}
Today, the Prolog language lacks a strong, united presence on the Web,
especially when compared to other languages, such as
the \cite{python-homepage}, \cite{rust-homepage},
\cite{r-homepage}, \cite{clojure-homepage} or \cite{julia-homepage}.
Hence, there is limited exchange between users of Prolog, in particular of different Prolog systems.
Only a few shared platforms exist, such as the venerable
\href{https://groups.google.com/g/comp.lang.prolog}{comp.lang.prolog} newsgroup,
the Prolog subreddit, or StackOverflow.
They have in common that they are not very active and are mostly used for announcements
or beginner questions.
There are some exceptions, such as the SWI-Prolog discourse forum, hosting active
discussions on the implementation, but, again, it is a barrier since
it is essentially local to this system.
The lack of an overarching presence, thus, lowers Prolog's visibility,
hinders development of shared libraries, and
might be an entry barrier for new Prolog users.
It may even deter programmers from adapting LP technology due to outdated or wrong information
on Wikipedia pages on Prolog or Logic Programming in general,
or due to missing FAQs, API documentation, and tutorials.

The current fragmentation of the Prolog community
may be seen as a threat to the language standardization effort and,
thus, the advancement of Prolog.
Compilation of standardization documents takes a very long time,
and is currently driven by a few volunteers.
While only experts can evaluate the %
impact of changes in the standard on existing Prolog systems,
the community may assist such efforts by pinpointing differences
between systems, prioritizing features that should be considered next,
or by providing test cases.
However, the community is hard to reach in a unified way and such contributions
are often limited to motivated individuals and remain scarce.

\paragraph{Implementors}
\label{para:implementors}
While implementors of Prolog systems meet at the CICLOPS workshops and other conferences,
they do not have a good shared infrastructure to revise syntax,
discuss libraries, tools and technical questions,
or to offer existing code or tests.
Without such workflows and regular discussion of future directions,
Prolog systems may further diverge in features, libraries,
and possibly even from the ISO standard if no concerted effort is made to find common best solutions.
New developers need a forum to ask questions and benefit from lessons learned.
Otherwise, implementation work may be duplicated,
no common Prolog tooling will be developed, etc.

\subsubsection{Lack of Strong Stewardship}%
\label{stewardship}

The largest single threat we see for the future compatibility of Prolog systems
and attractiveness of the language in the long term,
is the lack of a strong stewardship.
All implementors we were able to reach for comments
agree with the need for compatibility,
are willing to discuss issues and work on their systems to address them.
However, the most lacking resource is time, as
compatibility work diverts efforts from research and development.

A dedicated entity, that acts as a steward, calls meetings on a
regular basis to ensure progress, and oversees open issues is missing
and necessary, but establishing it is not without challenges.
Sufficient financial and/or institutional backing to motivate
implementors and to fund at least one steward position would be an
asset.  Additional human resource positions may be needed to address
specific aspects, such as interoperability, source code compatibility,
website maintenance, etc.

Two major collective efforts, the ISO process and the Prolog Commons group,
have already been established that can be
regarded as such entities and persist with varying success:

\paragraph{The ISO Process}
The ISO working group has the strong mission to provide a robust and concise
standard that makes Prolog attractive for the industry,
giving it a competitive advantage.
The core standard was a huge leap in the right direction,
providing a strong basis for compatibility.
However, further progress has been rather slow,
which may be due to the nature of ISO as well as
the voluntary nature of the work of the participants
and their aspirations for stability and high-quality work.
Unfortunately, only a few of the actual system implementors
are currently active in the ISO standardization efforts.
Also, the process is complex
and may be too slow despite steady progress,
since even more and more features and libraries are developed independently.

\paragraph{Prolog Commons}
The Prolog Commons project started as a series of informal implementor meetings
to improve the portability of Prolog code,
in a more agile and interactive setting than the ISO process.
This impetus pushed many participating systems in converging
directions, producing changes in their documentation systems, improved
compatibility/emulation layers, and plans for further integrations.
The reason that the project has not
fully materialized into a common code base is, again,
the lack of stewardship, meeting schedules, and deadlines, combined
with the available time of the implementors.

\subsection{Improving Prolog}%
\label{sec:discussion}

In preparation of this article, we reached out to researchers from different application areas of Prolog
in order to gauge the most pressing topics,
as well as to Prolog implementors in order to assess the potential for convergence of Prolog systems.
Based on the results of this survey and on the SWOT analysis in Sections \ref{swot-S}--\ref{swot-T},
 we now discuss the issues and areas of improvements which we feel are most pressing.

\subsubsection{Portability of Existing Features}
\label{sec-portability-features}

The differences between the various Prolog implementations, %
either in the set of features provided (cf. Table~\ref{tbl:feature-overview})
or in the way the features work,
lead to a \emph{code portability problem} between Prolog implementations. Circumventing this problem rather than solving it results in fragmenting the community into
smaller, non-compatible subgroups. This makes it hard for users to find support and stay interested.

Strong and universally accepted standards for available features may raise interest
from programmers and industry.
Yet, if some combination of them are missing in many Prolog systems,
it will negatively impact the perception of Prolog.
One of the non-standard offers of Prolog is constraint logic programming.
Yet, choosing a Prolog system requires certain trade-offs concerning the available features.
Many of these features concern performance, such as
tabling,
efficient coroutining via block annotations,
multi-argument indexing.
Some programmers may not want to give up what they are used to from other programming languages,
like
multi-threaded, concurrent and distributed programming,
standard libraries for formatting and pretty printing,
efficient hash map data structures and
universally available data structures such as AVL trees or sets.
Finally, some features are needed to embed Prolog as a component in a larger software system,
e.g.,
non-blocking IO,
interfaces to other programming languages or
(de-)serialization support, such as fastwrite/fastread, XML, JSON, YAML.

\subsubsection{Improved Development Tools} \label{sec-dev-tools}
An important area of future improvement is in the development tools
available for the language.  The dynamic nature and complex control of
Prolog raises new challenges in the implementation of some of these
features, but its clear logical foundations provide an advantage for
others.  We see a potential and a need for the following improvements
to the Prolog tooling ecosystem:

\begin{itemize}
 \item Increased capabilities in debuggers (e.g., constraint debugging or graphical interaction),
     performance profiling tools, and testing frameworks. The
     combination of static and dynamic verification, debugging, and
     testing of Ciao is relevant here \citep{ciaopp-sas03-journal-scp-short,verifly-2021-tplp-short}.
 \item Better experience using interactive shells: Prolog could also
    profit from being integrated within notebooks such as Jupyter, as
    has been done in
    SWISH~\citep{DBLP:journals/tplp/WielemakerRKLSC19}.
  \item More regular, user-friendly and standardized ways of
    interfacing with other languages.  Features which should be
    accounted for include non-determinism, convenient and safe data
    types, and memory management co-existence.
 \item More capable IDEs, in particular with good refactoring tools,
    which help the programmer to safely reorder or change the arguments
    of predicates, rename or move predicates to other modules, etc.,
    in the line of SICStus' SPIDER~\citep{CarlssonM12}.
    Prolog itself may be used as the GUI programming language in
    developing the IDE. %
 \item Linters that enforce community-sourced coding guidelines
     based on, e.g., the proposals by \cite{covington_bagnara_okeefe_wielemaker_price_2012}
        and \cite{nogatz2019}. %
 \item Improved code location facilities, such as Ciao's \emph{Semantic Search}~\citep{deepfind-iclp2016}.
 \item Other static program analysis tools, enabling for instance the inspection of
   dependencies amongst modules, and the existence of call hierarchies amongst predicates. Also,
   dynamic process inspection tools aimed at visualizing call
   stacks, choice points queues, proof trees, etc. (see also~\cref{sec:state-inspection}, State Inspection).
\end{itemize}

\subsubsection{Application- and Domain-Specific Needs}\label{sec:app-specific}

In this section, we consider a few domains, that we think may influence Prolog in the future,
and try to anticipate future developments that are needed to satisfy their needs.

\paragraph{Parsing}
The parsing domain has implications both for computing sciences, in that it can be applied to compilers or other program transformations, and for  sciences and the arts, in that many kinds of human languages (e.g., written, spoken, or those of molecular biology or music) can be computationally processed for various ends through parsing.

As far as pure parsing is concerned, one can of course easily write
top-down recursive descent parsers in Prolog using DCGs.  However,
encoding deterministic parsing with lookahead requires the careful use
of the cut, which is tedious and error-prone.  Ideally, the cuts could
be automatically generated by standard compiler construction
algorithms (Nullable, First, Follow) and a Prolog-specific parser
generator.  Bottom-up parsers are also easy to write using, for
instance, CHR.  Another way to improve Prolog's parsing capabilities is
through memoing, which avoids the infinite loops that left-recursive
grammars are prone to, thus needing to resort less to the cut, and
avoids recomputations of unfinished constituents in the case of
alternative analyses where one analysis is subsumed by another.  This
is done by storing them in a table so they need not be recomputed upon
backtrack (e.g., in ``Marie Curie discovered Polonium in 1898'', the
partial analyses of the verb phrase as just a verb or as a verb plus
an object are stored in a table for reuse rather than disappearing
upon failure and backtrack).  Memoing can be easily implemented in
Prolog, e.g., through assumptions, as discussed by
\cite{christiansen2018natural}, or through CHR, as discussed by
\cite{fruhwirth2009constraint}, or using tabling as in XSB Prolog.

 \paragraph{NLP and Neural Networks}
 Neural-network approaches to NLP use word embedding strategies to generalize from known facts to plausible ones (e.g., BERT \citep{devlin-etal-2019-bert}, GPT-3 \citep{GPT3}).
 They typically train only on form, in that they retrieve those responses that are statistically pertinent, with no regard to meaning. Their results are unstable, since they rely on ever larger and changing internet-mined data.
 While this approach  has achieved noteworthy performance milestones in machine translation,
 sentence completion, and other standard benchmarking tasks,
 it offers a priori no way to learn meaning \citep{bender-koller-2020-climbing}
 and relies on undocumented, un-retraceable or otherwise partial (and therefore unaccountable) data, which tends to perpetuate harm without recourse~\citep{BIRHANE2021100205}.
 It is also resource-intensive, both computationally and energetically, and prone to spectacular failure \citep{Marcus2020}. It seems that overcoming such drawbacks will need inferential programming capabilities, which integrations with deductive reasoning might help achieve. We anticipate that  efforts in that direction, which are already happening \citep{sun2021faithful}, will be increasingly needed.

 Another NLP area which we anticipate will require much attention and that Prolog can be ideal for is that of under-resourced human languages. Very few of the over 7,000 languages in existence have at their disposal the computational tools that are needed for their adequate processing. Since texts on the Web are also overwhelmingly in mainstream languages, and the machine learning approaches that are in vogue typically rely on mining massive volumes of text, when these do not exist (or as soon as the existing ones become more protected) we need more logical approaches, such as grammatical inference by grammar transformation.

 \paragraph{Problem Solving, Solvers}
Constraint programming blends nicely within Prolog (see Sect. \ref{section-clp}), in the form of CLP($\fd$), CLP($\cal B$), CLP($\cal R$), or CLP($\cal Q$), and also in the CHR form, which lets users define constraint solvers for their own domain of interest.
However, the binding to SAT, SMT, or ASP solvers is often still quite awkward.
In particular, ASP with its Prolog syntax could be made available as a seamless extension to Prolog.
Similarly, SAT and SMT solving could also be linked in a seamless way to Prolog facts or clauses.
In an ideal world, one could even link various solvers via shared variables and coroutines.

 \paragraph{Visualization and GUIs}
Visualization is nowadays most often performed via the foreign language interface or by exporting data to external tools (e.g., dot
text files for GraphViz).
While BIM-Prolog (cf. Sect.~\ref{sections-bim-prolog}) had a declarative graphical toolkit, unfortunately Prolog systems
have moved away from this approach. Other communities, however, are discovering the advantages of a ``declarative approach''
to visualization and user-interface design (e.g., React~\citep{gackenheimer2015introduction} in the JavaScript world).
Maybe it is time again to implement visualizations or user interfaces within Prolog itself.

\paragraph{State Inspection} %
\label{sec:state-inspection}
Prolog 0 already included primitives to programmatically inspect the
computation state.
These can be useful for example in debugging or in parsing, in order to implement context-sensitive grammars.
Current Prologs allow different degrees of state inspection, including
for example the classic facilities that enable meta-programming.
Such primitives could acquire new relevance under the increasing needs
for further transparency and inspectability in AI applications.

\subsubsection{Prolog Aspects That Need Joint, Public and Earnest Discussion}

There are a number of issues that would need early discussion
in the process of giving a new impetus to Prolog standardization.
While we raise questions here,
we cannot speak on behalf of the entire community.
Thus, we think that a visible (in the sense of commonly-known) platform for public discussion between implementors and users is required.
The Prolog language should eventually evolve on results of discussions
and needs of (potential) programmers.
For some issues we discuss below, several solutions have been offered by the research community.
However, the Prolog community has to discuss what features shall be adapted as standard.
Further, one might also consider whether purely declarative implementations are preferable or even feasible, since some features, such as assert and retract, do not even have a declarative, logical semantics.
The following points are some examples:

\paragraph{Types, Modes,  and Other Properties}
Often, during development, a Prolog program may terminate %
without finding a solution, get stuck in an infinite loop,
backtrack unexpectedly, etc.
Often, this sort of situation is due to type errors in
the code.  Should a type (and mode, etc.)  system be part of an
improved Prolog, allowing for more powerful static analysis?  If so,
what should it look like?  What would have to be changed for a useful
gradual static type system that allows one to progressively add types
to existing code?  Should more general properties than classical types
and modes be supported?  Should static types be combined with dynamic
checks?  With testing?  The logic programming community has been
pioneering in these areas with research, solutions, and systems well
ahead of other languages, but it has not yet seen widespread use.
Ciao's comprehensive approach to this overall topic could shed some
light here.
\paragraph{Reactivity}
A rule in Prolog can be viewed as a part of a definition that defines
the predicate in the head of that rule in terms of the predicates in
the body of the rule.  In contrast, most rules in imperative languages
are reactive rules that perform actions to change state, as in the
case of condition-action rules in production systems and
event-condition-action rules in active database systems.  Extensions
of logic programming to include reactive rules have been developed in
CHR (see Chapter 6 in \citep{fruhwirth2009constraint}) and in the
Logic-based Production System Language LPS \citep{KowalskiSadri2015,
  DBLP:journals/tplp/WielemakerRKLSC19}.

\paragraph{Module System}
As mentioned before,
the second part of the ISO standard regarding modules was universally
ignored and most Prolog systems settled for a Quintus-inspired module
system. However, the implementations incorporate some deviations to support
new features.  Rules regarding visibility of operators, predicates,
and perhaps atoms should be reconsidered. This is addressed to some
extent, for example, in the Ciao and SWI module systems but, again,
with some differences, which should however not be too difficult to
bridge.
Systems support some of the legacy code loading methods such as
\texttt{consult/1} and \texttt{ensure\_loaded/1} for backwards
compatibility, but \texttt{use\_module/\{1,2\}} is recommended,
specially for large-scale applications.
New solutions must address errors and inconsistencies that have already been uncovered by~\cite{10.1007/11799573_6} and
later by \cite{10.1007/978-3-642-02846-5_3}.

\paragraph{Objects} \label{sec:objects} An aspect which is closely
tied to the module system is that of integrating logic programming
with object-oriented features.  This has been an elusive goal but,
nevertheless, several effective proposals have been put forward which
achieve ways of doing whole-program composition more in line with the
Logic Programming paradigm.  \cite{DBLP:conf/iclp/MonteiroP89} present
the basic concepts of Contextual Logic Programming
while~\cite{DBLP:journals/jlp/BugliesiLM94} offers an extensive
discussion on this topic.  A more advanced design, in which modules
(called \emph{units}) may take parameters and where the context itself
becomes a first-class object is provided by
GNU~Prolog/CX~\citep{DBLP:conf/iclp/AbreuD03}, which appears as both
an object-oriented extension and a dynamic program-structuring
mechanism.  An application of this system is described by
\cite{DBLP:conf/inap/AbreuN05}.

Another example is the O'Ciao model, which implements a source-to-source approach to objects as a
conservative \emph{extension} of the Prolog module system~\citep{pineda02:ociao}.
Logtalk offers a different approach, also based on source-to-source
transformations, that adds a separate object-oriented layer on top of
many Prolog systems. Thus, it regards objects as a \emph{replacement} for 
a module system.

\paragraph{Interfaces}
Beyond modules,
programming in the large can be supported by mechanisms that allow
describing and enforcing separation between interfaces and
implementations.
Such interfaces are typically sets of signatures,
specifying types and other properties, to be matched by all
implementations,
and thus allow stating what a given implementation must meet, and
making implementations interchangeable in principle.
They should include at least constructs to express the external
connections of a module, not only in terms of the predicates it should
expose, but also the types, modes, and other characteristics of their
arguments, and a compiler/interpreter capable of \emph{enforcing}
those interfaces at compile- or loading-time.
Ciao's assertion language, preprocessor, and interface definitions
offer a possible solution in this area.

\paragraph{Syntactical Support for Data Structures}
One of Prolog's strengths is a minimalistic syntax.
SWI-Prolog's
syntactic support for strings and dictionaries
responds to a demand for interfacing with prevalent protocols and exchange formats,
e.g., XML and YAML.
Other systems have other extensions such as feature terms~\citep{HASSAN93,hassan-wildlifetutorial}.

Many questions need to be answered:
Should syntactical support for data types such as associative data
structures (feature terms) or strings be standardized?
Would the current syntax be affected (e.g., a prevalent syntax for maps, \verb|{"key": "value"}|, might break DCGs)?
What are trade-offs the community is willing to take?
Should other containers, such as linear access arrays, sets and character types, be included?
How should unification of variable elements work?

\paragraph{Library Infrastructure and Conditional Code}
As Moura's efforts concerning Logtalk (cf.~\cref{iso}) showed,
it is possible to support large portable applications across almost all Prolog systems.
Yet, a non-trivial amount of conditional code for abstraction libraries
is needed to provide support for each new Prolog system.
Is the community willing to develop libraries supporting more than a couple of systems?
Is a stronger standardized core library required to attract programmer?
Is the language that emerges from the portability layer still Prolog?

\paragraph{Macro System}

Many Prolog systems support the source-to-source transformation of terms and goals,
via the so-called \emph{term expanders}.
It is a powerful feature that is not part of the standard
and in some cases can be clunky to use and error-prone
since term expanders are often not local to a module,
they are order dependent
and every term or goal, relevant or not,
is fed into the transformation predicate.
Ciao shows how module locality and priorities can be used in this context.
Or should a more lightweight macro system be made available for code transformation?

\paragraph{Functional Programming Influences}
Curry and Mercury (cf. \cref{pl-inspired-langs}) are well-known for
combining the functional and logic programming paradigms. Some Prolog systems
such as Ciao have strong support for functional syntax and higher
order.
Most Prolog systems provide support for meta-predicates such as
\verb|call| and some derivatives such as \verb|maplist|,
\verb|filter|, \verb|reduce|, etc.  Should a notation for lambda
expressions be introduced as done in, e.g., Ciao's predicate
abstractions?

\paragraph{Standard Test Programs Beyond  ISO}

While test suites for ISO Prolog exist,
many features outside the core language are prevalent in a lot of Prolog systems.
Unstandardized features, such as modules, constraint programming or DCGs,
are provided by almost all modern Prolog systems,
yet implementors rarely share test programs.
Obviously, creation of and agreement on tests is challenging,
especially since some systems might want to maintain their features as they are.
Availability of such shared tests can guarantee that systems behave
similarly, allow implementors to test compatibility with other
systems, and foster standardization of these features.  As a positive
example in this line, Logtalk's Prolog test suite is quite comprehensive,
covering not only ISO- and de-facto standard features, but many other
tests of predicates and notable features such as modules, threads,
constraints, coroutining, Unicode support or
unbounded integer arithmetic.

\subsection{Summary and Next Steps}\label{sec:nextsteps}

Despite differences and incompatibilities between Prolog systems and
thus user code, the ensuing divergences are not fundamental.  It is
our belief that the best initiatives that were put in place in the
past, with some tweaks that we hope to have covered, can be re-applied
to make much stronger progress in the future.

The necessary implementor involvement seems attainable: most surveyed
implementors and users agree that efforts for a common ground should
be made, namely by meetings and the sharing of ideas, implementations
and even common infrastructure.
It is important that the entire community be involved in creating and
maintaining the existing core standard as well as debating desired
features, in order that upcoming standards take hold and be
respected.  %
Since support tools are a must for all modern programming languages,
it is important to pool resources.  In particular, mature debuggers,
as well as developing and testing environments are often required by
Prolog programmers.  %

The diversity that is unique to Prolog's ecosystem is a strength which
should be taken advantage of.  In short, Prolog systems need to be
useful and usable so they can be more universally employed.

Accordingly, to coordinate converging efforts, an organization,
perhaps a \thegang{}, could be
established, either independently or as an initiative within, for
instance, the ALP.
With converging systems and tool infrastructures, possibilities in
community participation and user documentation can vastly improve,
rendering Prolog much more attractive to new users both in academia
and industry, thereby building a more unified community.

Standard procedures already exist for Prolog improvement, such as the
ISO standardization.  We feel these would benefit from being
complemented by less formal ones, such as the Prolog Commons.

A time- and cost-efficient first step could be to create a web
platform to publicly share feature extensions and modifications and
propose new ones.  It could also provide the community with a forum
for public debate, and gather pointers to previous and complementary
efforts.  It could be linked to from the ALP website.

Another productive step would be to define a structured workflow aimed
at tracking, supporting, and coordinating the evolution of proposals,
efforts, and further initiatives.

Teaching Prolog to children could be most influential: they would be
learning logic and computing at the same time, and Prolog literacy
would spread much more rapidly through them.  There was much activity
in this area in the 1980s with the Prolog Education Group (PEG) with
its annual workshops.  More recently, ``Logical English''
\citep{kowalski2020logical}, although focused primarily on legal
applications, including smart contracts, is simply sugared syntax for
``pure'' Prolog, making it suitable for teaching to children (and
others).

Finally, it could be useful for the community to analyze all the
relevant Wikipedia pages and other web content in search of
misinformation, outdated information, or in general inaccurate content
on Prolog or logic programming, in the aim of correcting it.

\section{Conclusions}
\label{sec:conclusions}

We have provided an analysis of the Prolog language and community from
the historical, technical, and organizational perspectives.

\medskip
The first contribution of the paper (\cref{sec:history}) has been a
historical discussion tracking the early evolution of the Prolog
language and the genealogy of the many implementations available
nowadays.
This discussion stemmed from a general definition of what constitutes
a ``Prolog system'' and covered the most notable implementations and
milestones in chronological order.

We have seen how Prolog implementations, and even the language itself,
have experienced %
significant evolution and extensions over the years.  Despite the common roots, the
maintenance of the Prolog name throughout these transformations, and
the rich exchange of ideas among system implementers, inevitably some
parts of each system have progressed somewhat independently.
This has naturally led to some divergence. %

\medskip
In order to assess the current situation, the second contribution of
the paper (\cref{sec:current-state}) has consisted in a technical
survey of the current status of the main Prolog systems available
today, comparing them on a per-feature basis with the objective of
identifying commonalities and divergences.  In
addition,
we have also tried to identify the core characterizing aspect(s) of
each Prolog system, i.e., what makes each system unique.

We have observed that there is widespread adherence to the ISO
standard and its extensions.
More differences understandably appear in the added features that lie
beyond ISO.
At the same time we have seen that many of these extensions are common
to many Prolog systems, even if there are often differences in the
interfaces.
We have also observed that some Prologs offer unique characteristics
which would be very useful to have in other systems.
Such differences in general affect portability and make it harder for
a programmer to write Prolog programs that work and behave
consistently across systems.

However, we have observed that the divergences between Prolog
systems are not fundamental.
It seems that a good number of the differences could be easily bridged
by agreeing on interfaces and/or standard implementations that can be
easily ported to all systems.

\medskip
\cite{wenger2002cultivating} suggest that communities reach one or
more crucial points where they either fade away, or, alternatively,
they find new impetus to start further cycles.
We believe that our community has reached such crucial points several
times to date, at junctures such as the loss of interest in
parallelism in the early 90s (only to come back in full swing, over a
decade later), the end of the Fifth Generation project, the advent of
Constraints, the long AI winter (also with a mighty comeback), the
appearance of the Semantic Web,  the
advent of ASP, etc.
In all previous occasions the community has been able to adapt to
these environmental changes and incorporate new scientific and
technological advances as extensions to the language and the different
implementations, and has continued to produce truly state-of-the-art
programming systems, with advanced characteristics that often appeared
ahead of many other paradigms.

We believe we are now at one of these crucial points where action is
needed. The present Prolog systems, and Prolog as a whole,
continue being important and unique within the programming
languages landscape,
and it is encouraging to see that after all this time new Prolog
implementations, new features, and new uses of Prolog appear
continuously.

However, at the same time there is some risk of losing unity and
strength in the community.
This is partly due to its very success (in that it has spawned new
communities or successfully complemented existing ones) and partly to
community fragmentation and divergences in systems due to new
functionalities.

Thus, the last contribution of the paper (\cref{sec:future}) has been
to provide an analysis, including strengths, weaknesses, opportunities
and threats, as well as some proposals for the LP community aimed at
addressing the most relevant issues concerning the Prolog ecosystem,
in a coordinated way.
We argue that a joint, inclusive, and coordinated effort involving
Prolog implementers and users is necessary.
The great initiatives consisting of a small group of experts, such as
the ISO working group or the Prolog Commons project, with some tweaks
that we hope to have covered, can be re-invigorated to make stronger
progress in the future.  However, they may not be enough to bring
together a user community that contributes to the language and its
growth.  Accordingly, to apply the lessons learned from the past,
we stress the need for both community involvement and
stewardship, including perhaps the ALP, for ensuring that the effort
is kept moving forward consistently. 

Good initiatives are already being undertaken 
with the declaration of 2022 as the ``Year of Prolog''%
\footnote{\url{http://www.prolog-heritage.org/en/Prolog_50.html}}: its celebrations
will include an Alain Colmerauer Prize awarded by an international jury 
for recent application-oriented work of Prolog-inspired technology and its potential for more human-oriented computing in the future, and a
``Prolog School Bus'' that will travel to reintroduce declarative programming 
concepts to the younger generation.  It is expected that both initiatives 
will continue into the future, with a yearly prize also forseen for the ``nicest''
Prolog program written by a student participating in the school caravan.

\section*{Acknowledgments}

The authors would like to thank
Egon B\"{o}rger,
Ugo Chirico,
Bart Demoen,
Maarten van Emden,
John Patrick Gallagher,
Sebastian Krings,
John Lloyd,
Michael Maher,
Fernando Pereira,
Lu\'\i{}s Moniz Pereira,
Zoltan Somogyi,
German Vidal,
David H.D.\ Warren,
David S.\ Warren,
and Isabel Wingen
for their input and fruitful discussions.
They also thank Juan Emilio Miralles for proofreading.
The authors are particularly grateful for the effort made by the group
of people (more than 120 individuals) that drafted the ISO-Prolog
Standard and piloted its later evolution, and specially to Jonathan
Hodgson, Roger Scowen, Tony Dodd, Pierre Deransart,
Ali Ed-Dbali, Paulo Moura, and Ulrich Neumerkel, the current editor.
The authors would also like to thank the anonymous reviewers for their
thoughtful criticism and suggestions, which gradually helped improve
this paper.
Apologies to the many that are not mentioned.
The authors also endorse Paul McJones' efforts
to maintain a historical archive on Prolog, at
\url{http://www.softwarepreservation.org/projects/prolog/},
and thank all contributors.

\paragraph{Competing interests:} The authors declare none.

\bibliographystyle{acmtrans}
\bibliography{paper}

\end{document}